\documentclass[draft=false,twocolumn,showpacs,preprintnumbers]{aa}
\usepackage{natbib,twoopt}
\usepackage{graphicx,units,multirow,subfigure,color,amsmath,amssymb,epsfig}
\usepackage{courier}
\usepackage[varg]{txfonts}
\usepackage[usenames,dvipsnames,svgnames,table]{xcolor}
\usepackage[absolute,overlay]{textpos}

\bibpunct{(}{)}{;}{a}{}{,} 

\definecolor{light-gray}{gray}{0.95}
\definecolor{dark-gray}{gray}{0.4}

\def\pbar{$\overline{p}$}

\usepackage[pdftex,             
            breaklinks=true,%
            colorlinks=true,%
            pdfauthor={Derome et al.},%
            pdftitle={Fitting B/C cosmic-ray data in the AMS-02 era: A cookbook}%
           ]{hyperref}
\makeatletter
\newcommandtwoopt{\citeads}[3][][]{\href{http://adsabs.harvard.edu/abs/#3}{\def\hyper@linkstart##1##2{}\let\hyper@linkend\@empty\citealp[#1][#2]{#3}}}
\newcommandtwoopt{\citepads}[3][][]{\href{http://adsabs.harvard.edu/abs/#3}{\def\hyper@linkstart##1##2{}\let\hyper@linkend\@empty\citep[#1][#2]{#3}}}
\newcommandtwoopt{\citetads}[3][][]{\href{http://adsabs.harvard.edu/abs/#3}{\def\hyper@linkstart##1##2{}\let\hyper@linkend\@empty\citet[#1][#2]{#3}}}
\newcommandtwoopt{\citealpads}[3][][]{\href{http://adsabs.harvard.edu/abs/#3}{\def\hyper@linkstart##1##2{}\let\hyper@linkend\@empty\citealp[#1][#2]{#3}}}
\newcommandtwoopt{\citealtads}[3][][]{\href{http://adsabs.harvard.edu/abs/#3}{\def\hyper@linkstart##1##2{}\let\hyper@linkend\@empty\citealt[#1][#2]{#3}}}
\newcommandtwoopt{\citeyearads}[3][][]{\href{http://adsabs.harvard.edu/abs/#3}{\def\hyper@linkstart##1##2{}\let\hyper@linkend\@empty\citeyear[#1][#2]{#3}}}
\newcommandtwoopt{\citeadsstar}[3][][]{\href{http://adsabs.harvard.edu/abs/#3}{\def\hyper@linkstart##1##2{}\let\hyper@linkend\@empty\citealp*[#1][#2]{#3}}}
\newcommandtwoopt{\citepadsstar}[3][][]{\href{http://adsabs.harvard.edu/abs/#3}{\def\hyper@linkstart##1##2{}\let\hyper@linkend\@empty\citep*[#1][#2]{#3}}}
\newcommandtwoopt{\citetadsstar}[3][][]{\href{http://adsabs.harvard.edu/abs/#3}{\def\hyper@linkstart##1##2{}\let\hyper@linkend\@empty\citet*[#1][#2]{#3}}}
\newcommandtwoopt{\citeyearadsstar}[3][][]{\href{http://adsabs.harvard.edu/abs/#3}{\def\hyper@linkstart##1##2{}\let\hyper@linkend\@empty\citeyear*[#1][#2]{#3}}}
\newcommandtwoopt{\citeauthoradsstar}[3][][]{\href{http://adsabs.harvard.edu/abs/#3}{\def\hyper@linkstart##1##2{}\let\hyper@linkend\@empty\citeauthor*[#1][#2]{#3}}}
\newcommandtwoopt{\citepthesis}[3][][]{\href{http://tel.archives-ouvertes.fr/docs/#3}{\def\hyper@linkstart##1##2{}\let\hyper@linkend\@empty\citep[#1][#2]{#3}}}
\newcommandtwoopt{\citetthesis}[3][][]{\href{http://tel.archives-ouvertes.fr/docs/#3}{\def\hyper@linkstart##1##2{}\let\hyper@linkend\@empty\citet[#1][#2]{#3}}}
\makeatother

\newcommand{\beq}{\begin{equation}}
\newcommand{\eeq}{\end{equation}}
\newcommand{\bea}{\begin{eqnarray}}
\newcommand{\ena}{\end{eqnarray}}

\newcommand{\usine}{{\sc usine}}
\newcommand{\minuit}{{\sc minuit}}
\newcommand{\minos}{{\sc minos}}
\newcommand{\Rmean}{{\ensuremath{R_{\rm mean}}}}
\newcommand{\Rmeansqrt}{{\ensuremath{R_{\rm mean}^{\times}}}}
\newcommand{\Rmeanarit}{{\ensuremath{R_{\rm mean}^{+}}}}
\newcommand{\Rmeanpt}{{\ensuremath{R_{\rm mean}^{\times,+}}}}
\newcommand{\Rmin}{{\ensuremath{R_{\rm min}}}}
\newcommand{\Rmax}{{\ensuremath{R_{\rm max}}}}

\begin{document}

\begin{textblock*}{5cm}(15.8cm,2.8cm) 
   LAPTH-015/19
\end{textblock*}

\input epsf
\title{Fitting B/C cosmic-ray data in the AMS-02 era: a cookbook}
\subtitle{\large Model numerical precision, data covariance matrix of errors,\\
cross-section nuisance parameters, and mock data}
\titlerunning{Fitting B/C data in the AMS-02 era (a cookbook)}

\author{L. Derome\inst{1}\thanks{\url{laurent.derome@lpsc.in2p3.fr}}
  \and D. Maurin\inst{1}\thanks{\url{david.maurin@lpsc.in2p3.fr}}
  \and P. Salati\inst{2}\thanks{\url{pierre.salati@lapth.cnrs.fr}}
  \and \\ M. Boudaud\inst{3}\thanks{\url{boudaud@lpthe.jussieu.fr}}
  \and Y. G\'enolini\inst{4}\thanks{\url{yoann.genolini@ulb.ac.be}}
  \and P. Kunz\'e\inst{1}
}

\institute{
LPSC, Universit\'e Grenoble Alpes, CNRS/IN2P3, 53 avenue des Martyrs, 38026 Grenoble, France
\and LAPTh, Universit\'e Savoie Mont Blanc \& CNRS, 74941 Annecy Cedex, France
\and LPTHE, Sorbonne Universit\'e \& CNRS, 4 Place Jussieu, 75252 Paris Cedex 05, France
\and Service de Physique Th\'eorique, Universit\'e Libre de Bruxelles, Boulevard du Triomphe, CP225, 1050 Brussels, Belgium
}

\date{Received / Accepted}

\abstract
{AMS-02 on the International Space Station has been releasing data of unprecedented accuracy. This poses new challenges for their interpretation.} 
{We refine the methodology to get a statistically sound determination of the cosmic-ray propagation parameters. We inspect the numerical precision of the model calculation, nuclear cross-section uncertainties, and energy correlations in data systematic errors.} 
{We used the 1D diffusion model in \usine{}. Our $\chi^2$ analysis includes a covariance matrix of errors for AMS-02 systematics and nuisance parameters to account for cross-section uncertainties. Mock data were used to validate some of our choices.} 
{We show that any mis-modelling of nuclear cross-section values or the energy correlation length of the covariance matrix of errors biases the analysis. It also makes good models ($\chi^2_{\rm min}/{\rm dof}\approx1$) appear as excluded ($\chi^2_{\rm min}/{\rm dof}\gg1$). We provide a framework to mitigate these effects (AMS-02 data are interpreted in a companion paper).}
{New production cross-section data and the publication by the AMS-02 collaboration of a covariance matrix of errors for each data set would be an important step towards an unbiased view of cosmic-ray propagation in the Galaxy.}


\maketitle

\section{Introduction}

Particle physics detectors in space have opened a new era for the study of Galactic cosmic rays (GCRs). The Alpha Magnetic Spectrometer (AMS-02) instrument on the International Space Station (ISS) provides the best data to date for leptons and nuclei \citepads{2013PhRvL.110n1102A,2014PhRvL.113v1102A,2014PhRvL.113l1102A,2015PhRvL.114q1103A,2015PhRvL.115u1101A,2016PhRvL.117w1102A,2018PhRvL.120b1101A,2018PhRvL.121e1103A}, with an uncertainty of a few percent on a large energy range. Its measurements will probably remain unrivalled for at least the next decade in the GeV-TeV energy range.

In principle, high-precision data can be used to constrain different propagation scenarios or candidates in the context of dark matter indirect detection. However, promises of high-precision cosmic-ray (CR) physics can only be fulfilled if the various sources of uncertainties, data and model, are fully accounted for.
\begin{itemize}

   \item Data uncertainties: AMS-02 systematic uncertainties are diverse in origin and dominate the error budget of measured fluxes and ratios overall. In experiments measuring spectra, correlations in adjacent energy bins may be introduced at the data analysis stage. These correlations could wash out or mimic spectral features in the data. In principle, the best approach to automatically account for such effects is to fold the model prediction to the instrument full response and directly compare with the number of events. However, because the AMS-02 instrument response is not available, the next best approach is to incorporate a correlation matrix of errors or use nuisance parameters when comparing a model to the data. Such a matrix is also not available; however, we can rely on educated guesses to derive it and inspect the consequences on the model parameters.

   \item Input ingredient uncertainties: the dominant source of uncertainty in the modelling is from nuclear cross sections \citepads{2001ApJ...563..172D,2010A&A...516A..67M}. Propagating these uncertainties to the model parameters have already been investigated in several studies \citepads{2015JCAP...09..023G,2015A&A...580A...9G,2017PhRvD..96j3005T,2018JCAP...01..055R}, and we revisit this question here in more details with the use of mock (simulated) data. Controlled data were introduced in a CR propagation context in \citetads{2012A&A...539A..88C} to study possible biases on model parameters from using primary and secondary CR data of very different accuracy. In this study, mock data are used to characterise the bias on reconstructed model parameters when accounting for cross-section uncertainties, and more importantly, to assess how well nuisance parameters on cross sections allow one to recover unbiased values of these parameters.

   \item Model numerical precision: to be able to test various model hypotheses at the required level of the data, the model calculation must be at a much higher precision than the data uncertainty. The diffusion equation is a second order differential equation in space and momentum, and we discuss the impact of energy boundary conditions here, along with other effects that could prevent reaching the desired precision.

\end{itemize}

Except for the numerical precision, studying how to best deal with data and cross-section uncertainties is mostly independent of the propagation model and of the specific quantity studied. For practical purpose, our methodology is illustrated on the B/C ratio, which is one of the most frequently discussed quantity in the CR literature (e.g. \citealtads{2001ApJ...555..585M}).

This is a methodology paper, and the full analysis of AMS-02 data and the interpretation of the model parameters are left for the companion paper \citepads{2019arXiv190408917G}. The paper is organised as follows: Sect.~\ref{sec:setup} briefly introduces the 1D model and its parameters, and details two configurations used for the analysis. Sect.~\ref{sec:precision} discusses how to ensure a good precision at various stages of the calculation: boundary conditions, numerical stability, and comparison of the model to the data. Sect.~\ref{sec:XS} shows two different parametrisations of nuisance parameters for nuclear cross sections, and mock data are used to fully characterise the impact of cross-section uncertainties. In Sect.~\ref{sec:cov}, we detail the systematics of AMS-02 data---from which a covariance matrix of error is built---, and we show the impact of the correlation length (of the covariance matrix) on the best fit results and errors on the model parameters. Cross sections and covariance matrix effects are then dealt with together in Sect.~\ref{sec:xs_and_covmatrix}, to assess which one dominantly impacts the analysis of AMS-02 B/C data. We summarise our findings and give some recommendations in Sect.~\ref{sec:conclusions}.

For the sake of readability, some more technical and detailed discussions are postponed in the appendices: Appendix~\ref{app:RtoEkn} shows that the AMS-02 B/C data conversion from $R$ to $E_{k/n}$ as presented in \citetads{2016PhRvL.117w1102A} induces an energy-dependent bias ($~\sim 3\%$); App.~\ref{app:chi2} details how the covariance matrix of errors and nuisance parameters are included in the $\chi^2$ analysis;App.~\ref{app:boundary} gathers several boundary coefficients that can be implemented for the numerical solution of the discretised second-order differential diffusion equation; App.~\ref{app:stability} presents a thorough analysis of the stability of the latter solution (for the 1D model), also determining a stability criterion for the Crank-Nicholson solution with a vanishing second-order term ($V_a\rightarrow 0$); App.~\ref{app:xs_impact} shows a more detail view of how specific cross-section reactions impact the calculated B/C ratio.

\section{Model and parameters}
\label{sec:setup}

Many modellings of propagation models are possible, with different geometries, more or less involved spatial dependences, possible time-dependence, etc. For simplicity, and because most of our results and conclusions should not depend on this modelling, we use throughout the paper a 1D diffusion model, as implemented in the {\sc usine} package \citepads{2018arXiv180702968M}\footnote{A specific release, \usine{}~{\sc v3.5} (to appear), was developed for this analysis, improving on the first public version {\sc v3.4}, allowing for more nuisance parameters, more displays, etc.}. The equations and solutions can be found for instance in \citetads{2011A&A...526A.101P}, and for conciseness, we do not repeat them here. Such a model also proves useful, because it can be compared, in some simple cases, to compact analytical solutions (see, e.g. App.~\ref{app:stability}).
\subsection{1D propagation model}
\label{sec:prop_model}

In this model, sources and gas are in a thin disc of half-height $h$ in which energy losses and reacceleration occur. Particles diffuse in an homogeneous region of height $L$ above and below the disc. The free parameters of the model that we vary in this study are:
\begin{itemize}
  \item the rigidity dependence, $R\!=\!pc/(Ze)$, of the homogeneous and isotropic spatial diffusion coefficient $K(R)$, see Sect.~\ref{sec:model_params};

  \item the Alfvénic speed $V_a$ of scatterers mediating the strength of the diffusion in momentum $K_{pp}(R)$, the latter being related to the spatial diffusion coefficient and depends on the geometry of the turbulence \citepads{2002cra..book.....S};

  \item the strength of $V_c$, a constant Galactic wind perpendicular to the thin disc.
\end{itemize}
Other model parameters (halo size $L$, source spectrum) are irrelevant for this study. For instance, changing $L$ would merely lead to a rescaling of the above parameters \citepads[e.g.][]{2011A&A...526A.101P}, so we fix it to 10~kpc for practical purpose (and we fix $h$ to 100~pc). The B/C value is insensitive to the value of the universal spectral index of sources \citepads[e.g.][]{2002A&A...394.1039M,2015A&A...580A...9G}, so it is fixed in the analysis.

\subsection{Parameters for Model~A and Model~B}
\label{sec:model_params}

The impact of various uncertainties on transport parameters depends on the exact number and nature of the free parameters considered. We rely on two configurations, which correspond to two extreme cases of a more generic parametrisation of the diffusion coefficient, further detailed in the companion paper. They are denoted Model~A and Model~B in the rest of the paper:

\begin{itemize}
   \item Model~A: diffusion-convection-reacceleration model with
   \begin{equation}
      K(R) = \beta^{\eta_t} K_0 \left(\frac{R}{\rm 1~GV}\right)^\delta \times K_{\rm HE}(R)\,,
      \label{eq:K_std}
   \end{equation}
   where $\eta_t$ allows for a sub-relativistic change of the diffusion coefficient as parametrised in \citetads{2010A&A...516A..67M}, and where $K_{HE}=(1+(R/R_h)^{\delta_h/s_h})^{-s_h}$ is a high energy break whose parameter values are taken from \citetads{2017PhRvL.119x1101G}. This configuration has 5 free transport parameters: $K_0$, $\delta$, $\eta_t$, $V_a$, and $V_c$.

   \item Model~B: pure diffusion model (no $V_c$, no $V_a$) with a double broken power-law, at both high and low energy:
   \begin{equation}
      K(R)=\beta K_0\left(\frac{R}{\rm 1~GV}\right)^{\delta}{\left(1+\left(\frac{R_l}{R}\right)^{(\delta+\delta_l)/s_l}\right)^{s_l}} \times K_{HE}(R)\;.
      \label{eq:K_damped}
   \end{equation}
   This configuration has 4 free parameters: $K_0$, $\delta$, $R_l$, and $\delta_l$. The smoothness parameter $s_l$ has only a minor impact on the results, so it is fixed to 0.05 (quick transition) to speed up the fitting procedure.

\end{itemize}
Following \citetads{1994ApJ...431..705S}, we take, for both models, the diffusion coefficient in momentum
\begin{equation}
  K_{pp}(R)\times K(R)=\frac{4\,(V_a\,\beta\, E)^2}{3\delta(4-\delta^2)(4-\delta)}\,.
  \label{eq:kpp}
\end{equation}

\section{Model precision: general considerations}
\label{sec:precision}

The requirements for any analysis is (i) that the model calculation can be enforced at a precision much better than the precision of the data which will constrain it, and (ii) to ensure that the model and the data compared refer to the exact same quantity.

We focus on the model calculation in this section, but at the data level, biases can also sometimes be introduced, as exemplified in App.~\ref{app:RtoEkn} with the conversion of B/C data from $R$ to $E_{k/n}$ proposed in \citetads{2016PhRvL.117w1102A}.

\subsection{Energy boundary conditions and numerical stability}

The transport equation of CRs \citepads[e.g.][]{2018arXiv180702968M} is a second order differential equation in space and in momentum (if reacceleration is present), whose solution depends on the boundary conditions. These conditions should in principle be fixed by physics requirements, but there is no consensus as to what they are. We underline that spatial and momentum boundary conditions are generally not dealt with on the same footing: while they are usually clearly indicated in publications and are recognised as being part of the properties of a model for the former, this is not the case for the latter (which is discussed below).

The transport equation can be cast into a generic conservation equation on the energy co-ordinate $x=\ln{E_{k/n}}$:
\beq
 u + \alpha(x)_{\,}{\displaystyle \frac{dJ_E}{dx}} = u^{0} \quad\text{and}\quad J_E= \beta(x)_{\,} u - \gamma(x)_{\,}{\displaystyle \frac{du}{dx}}\;, \label{eq:crank_0}
\eeq
with $u$ is the cosmic-ray differential density in energy, $u^0$ a source term, $J_E$ a current (convection and diffusion), and $\alpha$, $\beta$, and $\gamma$ are coefficients related to energy losses, diffusion in momentum, etc.\footnote{For instance, in the case of the 1D diffusion model described in Sec.~\ref{sec:prop_model}, the above coefficients are given by \citetads{2010A&A...516A..66P}:
$\alpha(x)=2h/(E_k {\cal A})$, $\beta(x)=b_{\rm losses}(E)$, $\gamma(x)=\beta^2K_{pp}/E_k$, and $u^0(x)=2hq/{\cal A}$, with ${\cal A}=2h\,nv\sigma_{inel}+V_{c} \, + \,K S \coth \left( {{S L}/{2}} \right)\;$ and $S = \left(V_{\rm c}^2/K^2 + 4 \, \Gamma_{\rm rad}/K  \right)^{1/2}$.}

\paragraph{Boundary conditions}

At very high energy, for nuclei, the timescale of energy losses and gains becomes very large compared to the escape time, suggesting the condition $u(x_{\rm max})=u^0(x_{\rm max})$, which is seemingly used in all propagation codes.

At low energy, different boundary conditions have been used in the literature, corresponding to different physical situations:
\begin{itemize}
   \item No curvature in the spectrum, as introduced in \citetads{2001ApJ...563..172D} for antiprotons: $\partial^2 u/\partial x^2|_{x_{\rm min}}=0$.

   \item No energy flow, that is a vanishing current $J_E(x_{\rm min})=0$. Physically this means that, at $x_{\rm min}$, the outward current from energy losses balances exactly the inward reacceleration current. This condition thus depends on the coefficients $\alpha$, $\beta$ and $\gamma$ of the equation. From Eq.~(\ref{eq:crank_0}), one can infer that for very small values of $\gamma$, this will create a strong gradient of $u$ to maintain $J_E=0$.

   \item No density gradient in momentum $\partial f/\partial p|_{x_{\rm min}=0}$ \citepads{2017JCAP...02..015E}, which translates into $\partial/\partial p (u/(pE))|_{E_{\rm min}}=0$.
\end{itemize}

\paragraph{Numerical solution and precision}

In the semi-analytical models implemented in \usine{} and used here, as for most propagation codes, the second order differential equation in energy is solved numerically, by solving the equation on a grid. In practice, we make use of the semi-implicit Crank-Nicholson method and the resolution proceeds via the inversion of a tridiagonal matrix (see App.~\ref{app:boundary}). The first and last energy bins are set by the boundary conditions, as reported in Table~\ref{tab:coeff}.

\begin{figure}[t]
  \begin{center}
    \includegraphics[width=0.5\textwidth]{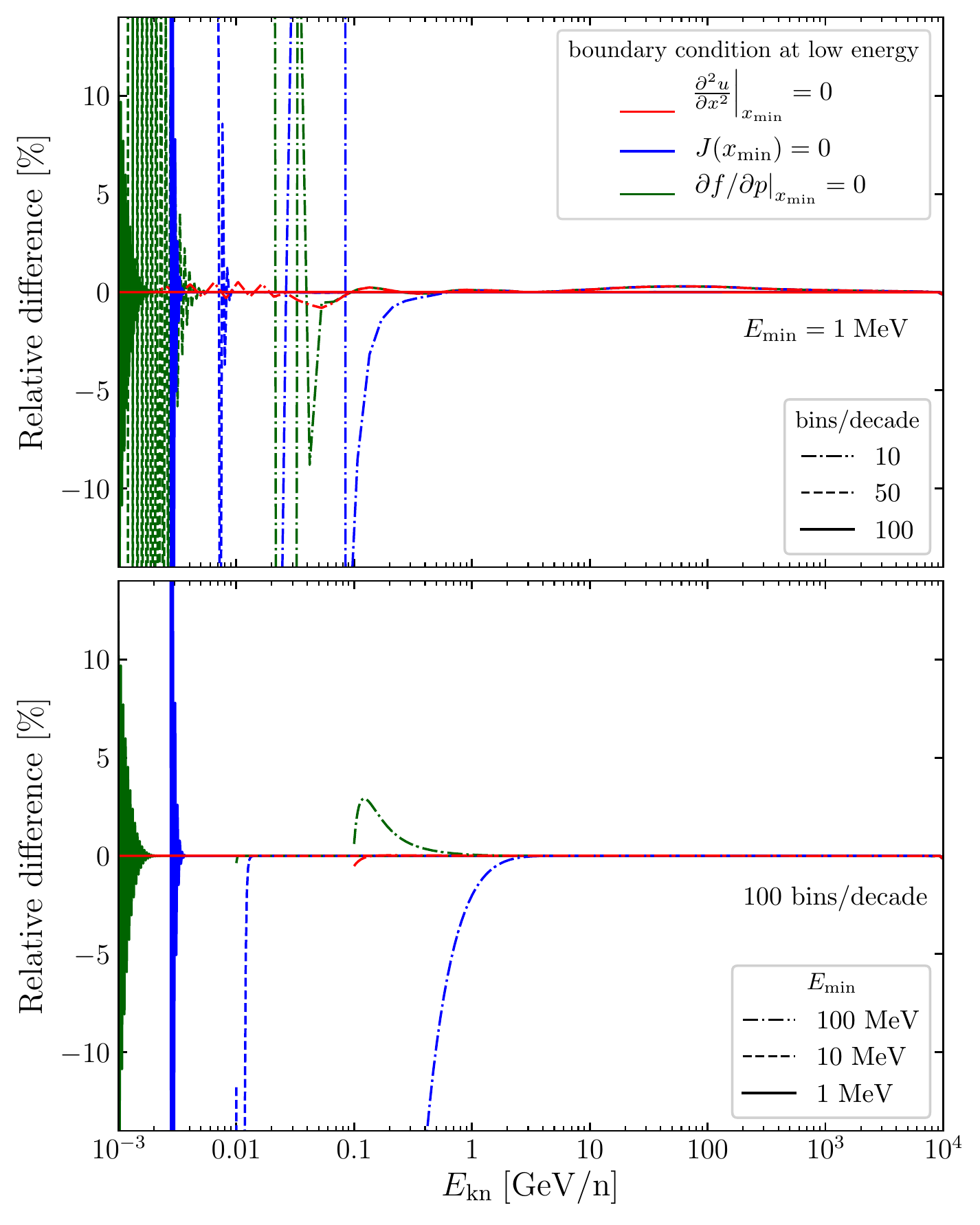}
  \end{center}
  \caption{Relative difference for the B/C computed with different boundary conditions at LE, varying the number of energy bins per decade (top panel) and the minimal energy of the grid (bottom panel). The reference B/C has been computed using 5000 energy bins per decade and setting $E_{\rm min}=1 \, {\rm MeV}$.}
    \label{fig:boundary_conditions}
\end{figure}

The precision of the numerical inversion depends on the number of points on the grid. We find that the boundary conditions at low energy described above yield similar results, provided that the number of energy bins is large enough and that the lower energy $E_{\min}$ is far below the range of interest. However, they differ with respect to the minimum number of energy bins required to reach the same level of precision. 
The B/C ratio obtained from each of these conditions are compared in  Fig~\ref{fig:boundary_conditions} to a reference ratio obtained setting $E_{\rm min} = 1 \, {\rm MeV/n}$ and using 5000 points per energy decade. We checked that the reference ratio does not depend, at the per mille level, on the chosen boundary condition. The red, blue and green curves are obtained for the first, second and third boundary conditions at low energy, respectively. In the top panel, we change the number of points in the energy grid (10, 50 and 100 bins per decade) whereas in the bottom panel we move the low-energy bound $E_{\rm min}$ from 100 MeV/n down to 1 MeV/n.
We obtain similar results for each low-energy boundary condition, provided that the number of points per energy decade is larger than 10 and $E_{\rm min}$ is much lower than the energy range of interest. This is probably the reason why no boundary condition at low energy is specified in the numerical codes GALPROP~\citep[][p.35]{Galprop_manual} and PICARD~\citepads[][p.5]{2014APh....55...37K}.
However, the choice of the boundary condition is important for the convergence.
We find that the first condition (no curvature in the spectrum) converges better than the others and yields a precision at the percent level even using only 10 bins per decade when starting from $E_{\rm min} = 1 \, {\rm MeV/n}$. We therefore recommend to make use of the no curvature boundary condition at low energy, with $E_{\rm min} = 1 \, {\rm MeV/n}$ and using 50 bins per decade, to ensure numerical systematics lower than the percent level.

Finally, this method can be unstable for $\gamma \rightarrow 0$, in which case Eq.~(\ref{eq:crank_0}) boils down to a first order differential equation. A simple prescription to obtain the solution keeping the same solver is to enforce a non-null but small $\gamma$, with a non-vanishing yet small $V_a$. We checked that setting a lower limit on the Alfv\'enic speed such that ${V_{a}^{2}}/({K_{0}\Delta x^2}) \geq 10^{-1} \; {\rm Myr^{-1}}$ where ${\Delta x}$ is the spacing of the energy grid, stabilises the $V_a=0$ solution at the per mille level.

\subsection{`Model vs data' error using \Rmean{} or of bin range estimate}

To our knowledge, models are always calculated at a single point and then compared to data measured over a bin range. This can lead to a systematics we quantify here.

\paragraph{Meaning of $R_{\rm mean}$ in data}
All data measurements are based on a number of events in the detector (corrected for the acceptance, efficiencies, etc.) per unit area, solid angle, unit time, and energy bin. To be practical, let us assume that, like AMS-02, the experiment provides data per rigidity bin $[\Rmin,\,\Rmax]$\footnote{The reasoning would be the same if we were to take $E_{k/n}$ instead.}. Usually, for display purpose, the central bin reported in the experiments is the geometric or more rarely the arithmetic mean:
\begin{equation}
   \Rmeansqrt\equiv\sqrt{\Rmin\Rmax}
   \quad{\rm or}\quad
   \Rmeanarit\equiv(\Rmin+\Rmax)/2.
   \label{eq:Rmean_def}
\end{equation}
Neither of those means are satisfactory to represent the measured flux $F$, because the correct central bin $R_{\rm mean}^{\rm exact}$ should be calculated from the condition
\begin{equation}
   F(R_{\rm mean}^{\rm exact})=F_{[\Rmin,\,\Rmax]} \equiv\frac{\int_{\Rmin}^{\Rmax} F(R) dR}{\Rmax-\Rmin}.
\end{equation}
This is a well known issue \citepads{1995NIMPA.355..541L}: providing the exact $R_{\rm mean}$ from the data requires the knowledge of the spectral shape of the flux that the experiment is actually trying to measure!

\paragraph{$F(\Rmeanpt{})$ vs $F(R_{\rm mean}^{\rm exact})$ for fluxes and ratios}
In the context of minimisation studies, the above discussion should be irrelevant as the theoretical flux is known: it can be integrated over the bin range and compared with the data without approximation. However, standard practice in the literature is to fit the data using \Rmeanpt{}. We quantify below the relative difference between the exact and approximate calculation for a flux,
\begin{equation}
   \mathcal{E}_F \,[\%] = \frac{\left(F_{\Rmean}-F_{[\Rmin,\,\Rmax]}\right)}{F_{[\Rmin,\,\Rmax]}}\times 100.
   \label{eq:reldiff}
\end{equation}
For practical calculations, we assume power-law fluxes $F = F_0 R^{-\alpha}$, so that the normalisation $F_0$ simplifies in Eq.~(\ref{eq:reldiff}). We can also estimate the relative difference for B/C, taking the ratio of two power-law fluxes of indices $\alpha_{\rm num}$ and $\alpha_{\rm den}$:
\begin{equation}
   \mathcal{E}_R \,[\%] = \frac{\displaystyle 
     \left(\frac{F^{\alpha_{\rm num}}}{F^{\alpha_{\rm den}}}\right)_{\Rmean}
       - \frac{F^{\alpha_{\rm num}}_{[\Rmin,\,\Rmax]}}{F^{\alpha_{\rm den}}_{[\Rmin,\,\Rmax]}}
     }{\displaystyle \frac{F^{\alpha_{\rm num}}_{[\Rmin,\,\Rmax]}}{F^{\alpha_{\rm den}}_{[\Rmin,\,\Rmax]}}}
     \times 100.
   \label{eq:reldiff_ratio}
\end{equation}

\paragraph{Model error on fluxes}
\begin{figure}[t]
\begin{center}
\includegraphics[width=\columnwidth]{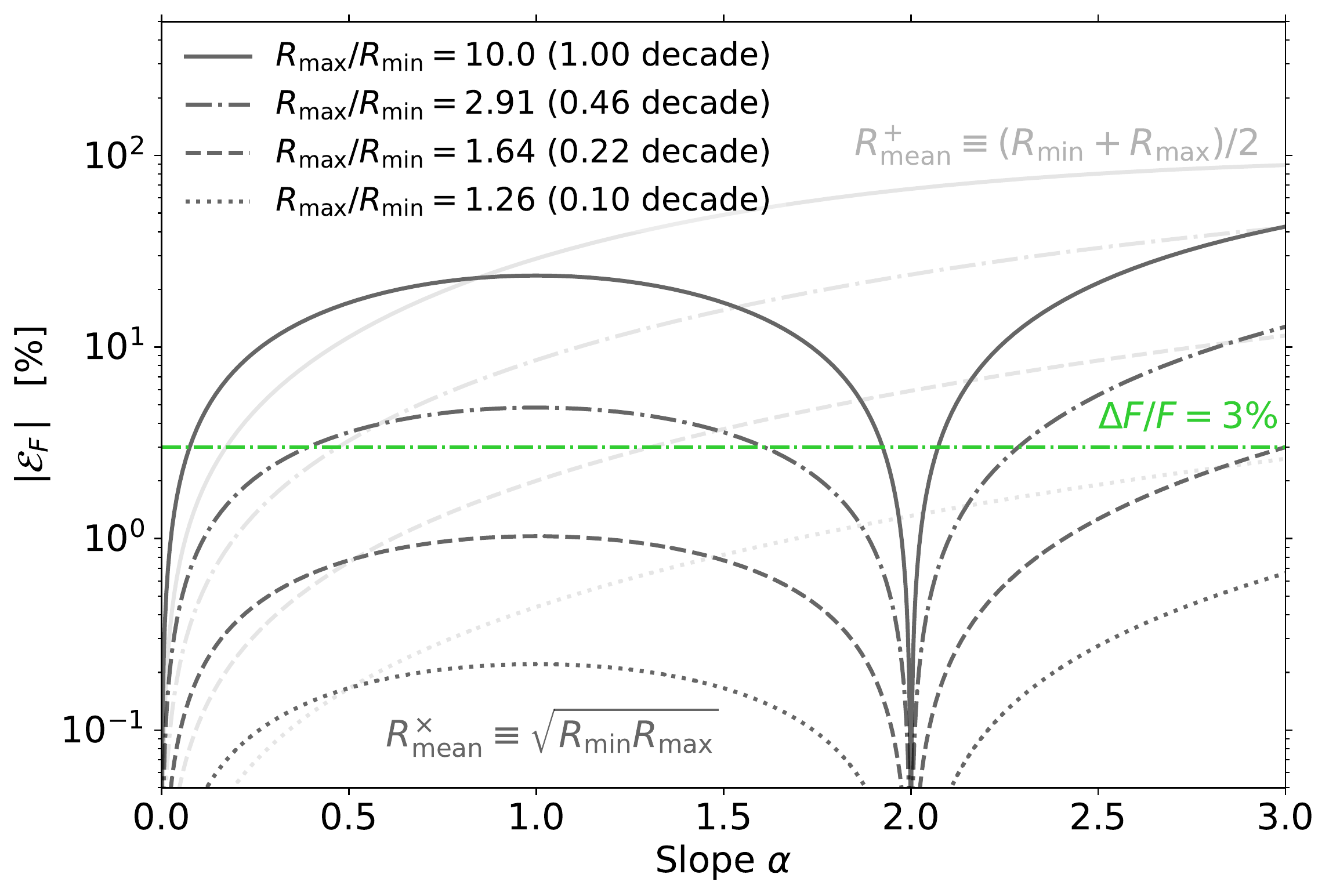}
\caption{Absolute value $|\mathcal{E}_F|$ of the relative difference between the exact and approximate flux calculation in a bin, Eq.~(\ref{eq:reldiff}), as a function of the flux spectral index $\alpha$. Two mean rigidity definitions are compared, \Rmeansqrt{} in dim grey and \Rmeanarit{} in light grey, see Eq.~(\ref{eq:Rmean_def}), for different bin size ranges (from 1 decade, solid line, down to 0.1 decade, dotted line). To guide the eye, the green dash-dotted line shows when the relative difference is $3\%$. See text for discussion.}
\label{fig:meanvsbin_flux}
\end{center}
\end{figure}
We show in Fig.~\ref{fig:meanvsbin_flux} the module $|\mathcal{E}_F|$ as a function of $\alpha$. The different line styles correspond to different bin widths, and the larger the bin width, the larger $|\mathcal{E}_F|$. Also, $F(\Rmeansqrt{})$ is a better approximation than $F(\Rmeanarit{})$, except for $\alpha\sim 0$ (thick dim vs thin light grey lines). $F(\Rmeansqrt{})$ is exact for $\alpha=0$ and $\alpha=2$ and the relative difference is positive above $\alpha=2$ and negative below $\alpha=2$.
To guide the eye, the green dash-dotted line shows the typical $3\%$ uncertainty on the AMS-02 data: all values of $\alpha$ and $\Delta R$ above this line, using $F(\Rmeanpt{})$ in model calculations produces a bias larger than $3\%$. We do not study fluxes here, but primary and secondary fluxes have slopes of $\sim 2.8-3.1$, for which the bias is maximal. Nevertheless, the maximal rigidity bin width in AMS-02 data for proton and He flux is $R_{\rm max}/R_{\rm min}\approx1.6$, for which the approximate calculation is smaller than the data uncertainty.

\begin{figure}[t]
\begin{center}
\includegraphics[width=\columnwidth]{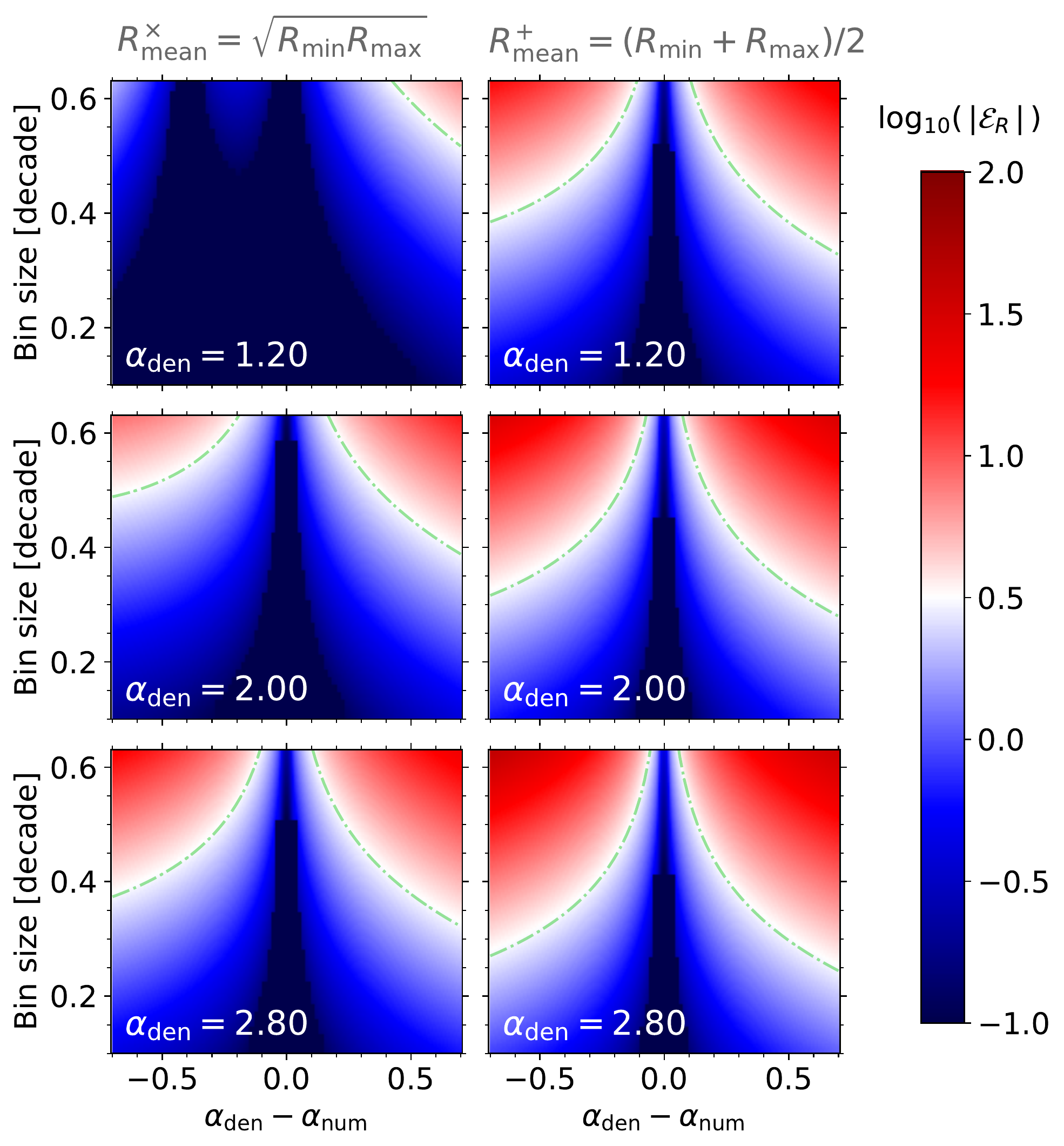}
\caption{Colour-coded relative difference $\log_{10}(|\mathcal{E}_R|)$ between the exact and approximate ratio calculation in a bin, Eq.~(\ref{eq:reldiff_ratio}), as a function of the numerator and denominator spectral difference, $\alpha_{\rm den}-\alpha_{\rm num}$ ($x$-axis), and as a function of the rigidity bin width, $\log_{10}(\Rmax/\Rmin)$ ($y$-axis). For instance, a value of $2$ (red) corresponds to a relative difference of $100\%$, and a value of $-1$ (blue) to a relative difference of $0.1\%$. The left and right panels correspond to calculations using \Rmeansqrt{} and \Rmeanarit{} respectively, see Eq.~(\ref{eq:Rmean_def}). The rows correspond to three different values of the denominator spectral index $\alpha_{\rm den}$. In each panel, the dash-dotted green contour delimits regions in which the relative difference is above or below $3\%$. See text for discussion.}
\label{fig:meanvsbin_ratio}
\end{center}
\end{figure}

\paragraph{Model error on ratios}
We repeat the analysis for the ratio of two power laws of slope $\alpha_{\rm num}$ and $\alpha_{\rm den}$, see Eq.~(\ref{eq:reldiff_ratio}). Figure~\ref{fig:meanvsbin_ratio} shows colour-coded values of $\log_{10}(\mathcal{E}_R)$. The $x$-axis is $\alpha_{\rm den}-\alpha_{\rm num}$: for the B/C ratio, this difference ranges from $\sim 0.2$ to $0.7$ at high energy, while negative values mimic the decreasing ratio below a few GV. We show our results for three $\alpha_{\rm den}$ values, where the lower value (top panels) mimics the flattening of the Carbon spectrum at low energy, while the higher value (bottom panels) corresponds to the high-energy slope: $\mathcal{E}_R$ grows with increasing $\alpha_{\rm den}-\alpha_{\rm num}$ and with the data bin size ($y$-axis); the three rows show a growing relative difference (from top to bottom) with $\alpha_{\rm den}$. Also, as already observed for fluxes, the approximation ${\rm Ratio}(\Rmeansqrt{})$ is always better than ${\rm Ratio}(\Rmeanarit{})$: for any given $x-y$ position in Fig.~\ref{fig:meanvsbin_ratio}, $\mathcal{E}_R$ is always smaller in the left panel than in the right panel. To guide the eye, the green dashed line delimit the contour for which $\mathcal{E}_R\equiv 3\%$, with larger uncertainties in reddish regions and smaller uncertainties in blueish ones.

For B/C from AMS-02 data, the bin range size goes from 0.08 decade at low rigidity to 0.3 decade for the last few rigidity bins where the systematic uncertainty reaches $10\%$ \citepads{2016PhRvL.117w1102A,2018PhRvL.120b1101A}\footnote{In experimental data, larger bins are used to limit statistical uncertainties whenever smaller number of events in the detectors are measured. For AMS-02 data, this happens at high energy (because of the power-law behaviour) but also at low energy (because of the geomagnetic rigidity cut-off).}. From Fig.~\ref{fig:meanvsbin_ratio}, this translate into negligible values of $\mathcal{E}_R$ for almost all rigidities. However, for the highest rigidity bins, the model error from using ${\rm Ratio}(\Rmeanpt{})$ is systematic (same sign) and reaches a maximum of $-3\%$ for $\alpha_{\rm num}-\alpha_{\rm den}\gtrsim 0.7$. In a region where power-law breaks are usually fitted, the exact calculation is recommended\footnote{In \usine{}, the keyword {\tt IsUseBinRange} in the parameter file allows to calculate the full bin content value assuming a power law for each isotope within the bin range.}.

\section{Handling cross-section uncertainties}
\label{sec:XS}

Nuclear cross sections are measured by `external' experiments, and these measurements can be incorporated as a distribution of probability in the $\chi^2$ minimisation via nuisance parameters (see App.~\ref{app:chi2}): cross sections far from their most probable values must be penalised in the minimisation, see Eq.~(\ref{eq:chi2_nuis}).

The difficulty lies in the characterisation of the uncertainties, the choice of the nuisance parameters, and assessing the robustness of the procedure. We start by characterising the impact of cross-section uncertainties on the B/C ratio (\S~\ref{sec:bc_impact}). We then present two different strategies for the choice of nuisance parameters (\S~\ref{sec:xs_nuis}). To assess the successfulness of these two strategies, we have to rely on the analysis of mock data for many configurations (\S~\ref{sec:xs_mock}). We then discuss how well these configurations capture and propagate all cross-section uncertainties to the transport parameter level (\S~\ref{sec:xs_mock_results}).

\subsection{Quantifying the impact on B/C ratio}
\label{sec:bc_impact}
Cross section data uncertainties are typically at $\sim 5-10\%$ level for inelastic cross sections, and $15-25\%$ level for production cross sections \citepads{2018PhRvC..98c4611G}. However, because the data are sometimes scarce, old, not always consistent with one another, and sometimes even missing for some reactions, several parametrisation of the whole network of reactions exist. A conservative estimate of the impact of cross-section uncertainties on the B/C calculation can be based on the scatter observed from using several of these parametrisations (see \citealtads{2018PhRvC..98c4611G} for more details):
\begin{itemize}
  \item Inelastic cross sections, $\sigma_{\rm inel}$: we use below B94~\citepads{BarPol1994}, W96~\citepads{1996PhRvC..54.1329W}, T99~\citepads{1996NIMPB.117..347T,1999NIMPB.155..349T}, and W03~\citepads{2003ApJS..144..153W}. Except for T99, the scaling $\sigma_{\rm He}/\sigma_{\rm H}$ is taken from \citetads{1988PhRvC..37.1490F}.

  \item Production cross sections, $\sigma_{\rm prod}$: W98 \citepads{1998ApJ...508..940W,1998ApJ...508..949W,1998PhRvC..58.3539W}, S01\footnote{Same dataset as in \citetads{2003ApJS..144..153W}, but fitted by Aimé Soutoul (private communication).}, W03~\citepads{2003ApJS..144..153W}, and G17~\citepads{2001ICRC....5.1836M,2003ICRC....4.1969M}.
\end{itemize}

\begin{figure}[t]
\begin{center}
\includegraphics[width=0.49\columnwidth]{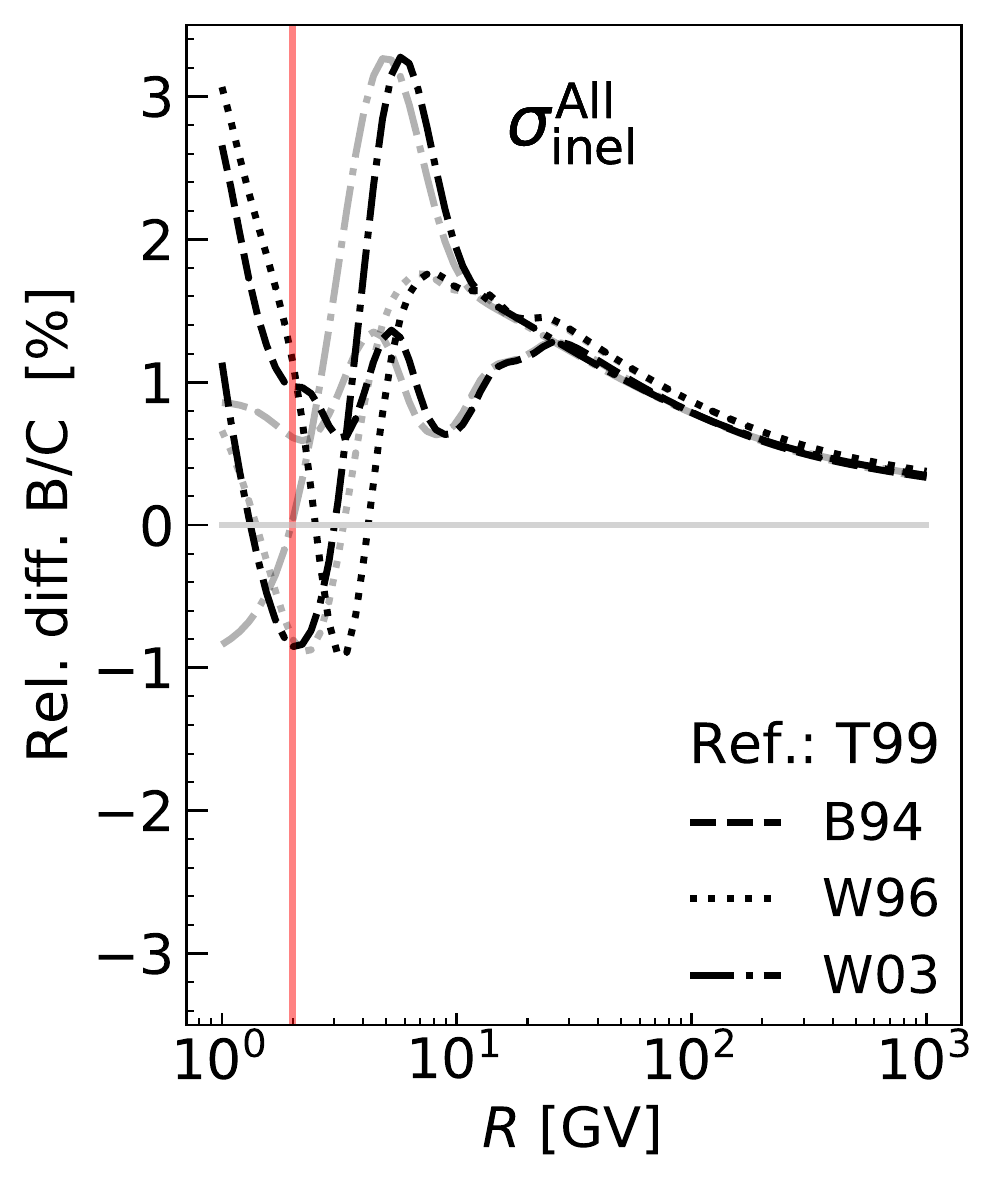}
\includegraphics[width=0.5\columnwidth]{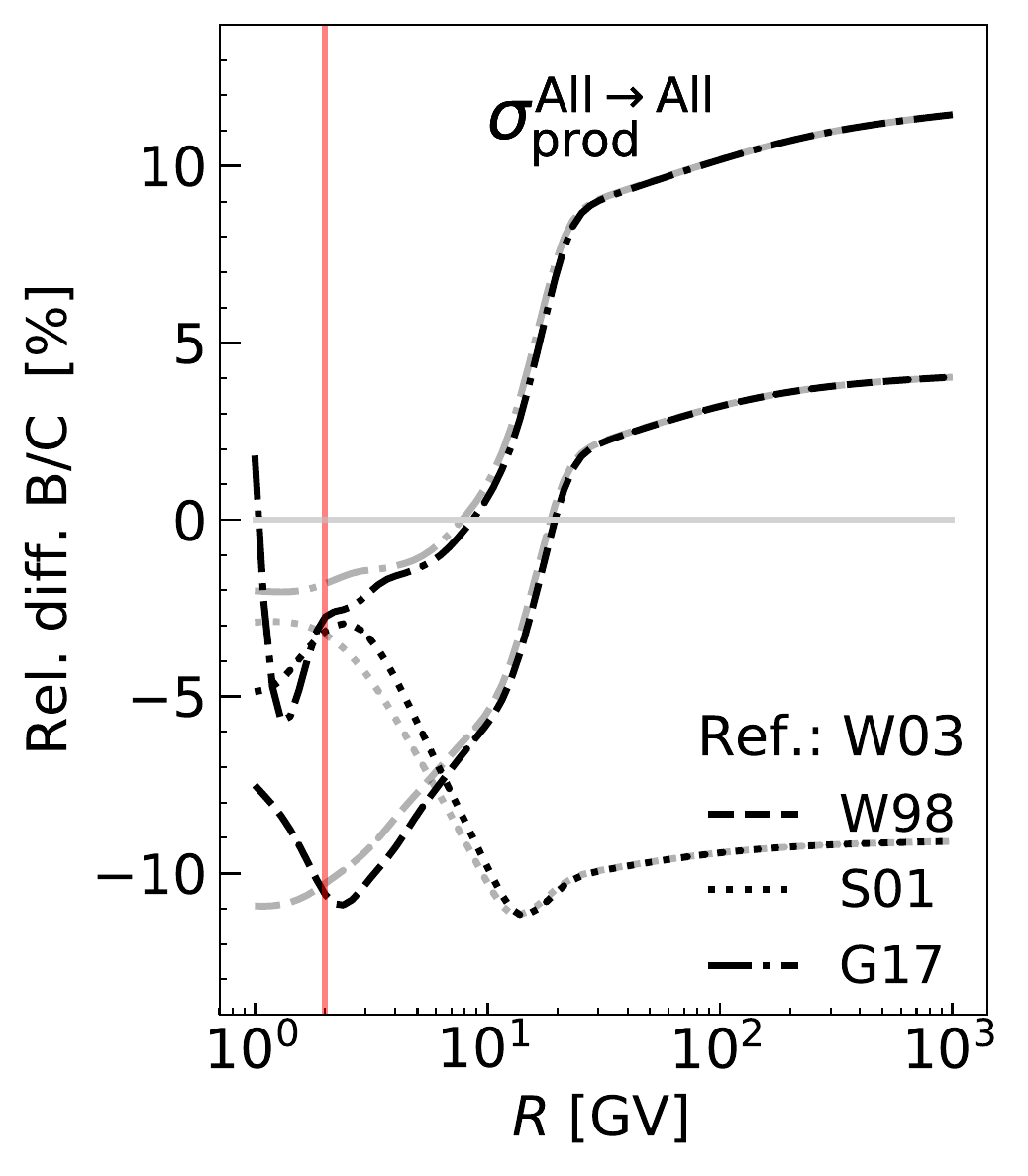}
\caption{Impact of different cross-section parametrisations on B/C flux calculation as a function of rigidity, w.r.t. a reference (denoted {\tt Ref.} in the legend). The left and right panels are for inelastic and production cross sections respectively (see Sect.~\ref{sec:bc_impact}). The thick (resp. thin) lines are for the interstellar (resp. solar-modulated at $\phi_{\rm Force-Field}=800$~MV) calculation. The red vertical line highlights the first rigidity point of AMS-02 B/C data.}
\label{fig:xs_val_all}
\end{center}
\end{figure}

For a given set of propagation parameters, we calculate the B/C ratio for various parametrisations, and we plot in Fig.~\ref{fig:xs_val_all} the relative variation with respect to a reference T99 for $\sigma_{\rm inel}$ and W03 for $\sigma_{\rm prod}$). The maximum impact of inelastic cross sections is $\lesssim 3\%\footnote{We wrongly reported a 10\% impact in \citetads{2018PhRvC..98c4611G} because of an error in the inelastic cross section on He for two parametrisations (the faulty files have been corrected in \usine{}~v3.5).} $at $\sim 5$~GV (left panel). It slowly decreases to zero at higher $R$, because the escape time from the Galaxy decreases with $R$ while the destruction time remains constant (see, e.g. Fig.~\ref{fig:taux_H_NEW_model}). The maximum impact of the production cross sections is $\lesssim 10\%$, and it is equally seen at low and high rigidity (right panel) because the Boron flux is directly related to its production cross section. These results are similar for Models~A and~B (not shown) and are independent of the solar modulation level (compare the black and grey lines in Fig.~\ref{fig:xs_val_all}).

Actually, a huge network of reactions is involved in the calculation of any given secondary cosmic ray, making the modelling of uncertainties for each individual reaction a daunting task. Obviously, this full network is taken into account to calculate B/C, but we model and incorporate cross-section uncertainties for the most relevant reactions only: for the production, $^{16}$O and $^{12}$C make $\sim 70\%$ of the Boron flux via $^{10,11}$B \citepads{2018PhRvC..98c4611G}, and for inelastic interaction, $^{16}$O, $^{12}$C, and $^{11}$B are the most relevant. The results of App.~\ref{app:xs_impact} show that the variation seen on the B/C ratio from cross-section uncertainties almost completely originate from these dominant reactions.

\subsection{From uncertainties to nuisance parameters}
\label{sec:xs_nuis}

To estimate uncertainties on selected reactions, one could use a parametric formula to fit the cross-section data and extract the best-fit parameters and uncertainties to propagate as nuisance parameters. This is the strategy followed by \citetads{2018JCAP...01..055R} for production cross sections. However, as already said, the reliability of the data is not always clear, with many inconsistent data points, from a mixture of very old and more recent experiments with probably underestimated systematics \citepads{2018PhRvC..98c4611G}. All these unknowns most certainly break down the statistical meaning of the $\chi^2$ values and uncertainty determination of these nuclear data. We assume here that the cross-section uncertainties are fully captured by existing parametrisations (Sect.~\ref{sec:bc_impact})\footnote{These parametrisations are based on fits to the same inhomogeneous sets of cross-section data, but the authors used different approaches and assumptions to fit them. For instance, the GALPROP parametrisation is renormalised to data whenever available, whereas other data sets are based on semi-empirical formulae designed to give an good fit over all reactions. The former parametrisation is expected to better represent the data, but by construction it sometimes shows non-physical energy dependences (step-like behaviour), whereas the other parametrisations do not.}.
It is difficult to argue which of the two above approaches gives the most realistic description. Indeed, the values and uncertainties on the cross-section data and the models could be fully, partly, or not at all correlated, which would increase or decrease the uncertainty on the calculated B/C ratio (see the discussion in \citealtads{2018PhRvC..98c4611G}). Without new cross-section data, the degree of belief one can have in the modelling of cross-section uncertainties can hardly be improved.

Our approach probably allows for larger uncertainties than those taken in \citetads{2017PhRvD..96j3005T} and \citetads{2018JCAP...01..055R}, which also focus on a subset of reactions only. However, we recall that the uncertainties on this subset must `emulate' the total uncertainties from the whole network of reactions (see the discussion in App.~\ref{app:xs_impact}).
To go beyond qualitative arguments, we inspect in Sect.~\ref{sec:xs_mock} whether the degrees of freedom used to model cross-section uncertainties are conservative enough not to bias the determination of the transport parameters. Before doing so, the next two paragraphs discuss two ways to model cross-section uncertainties as nuisance parameters.

\subsubsection{Normalisation, scale, and slope (NSS)}

\begin{figure}[t]
\begin{center}
\begin{tabular}{p{0.45\columnwidth}p{0.45\columnwidth}}
{\tiny \hspace{1.1cm} Inelastic cross sections} &  {\tiny \hspace{1.1cm} Production cross sections}\\[-0.3cm]
\vspace{0pt}\includegraphics[width=0.49\columnwidth]{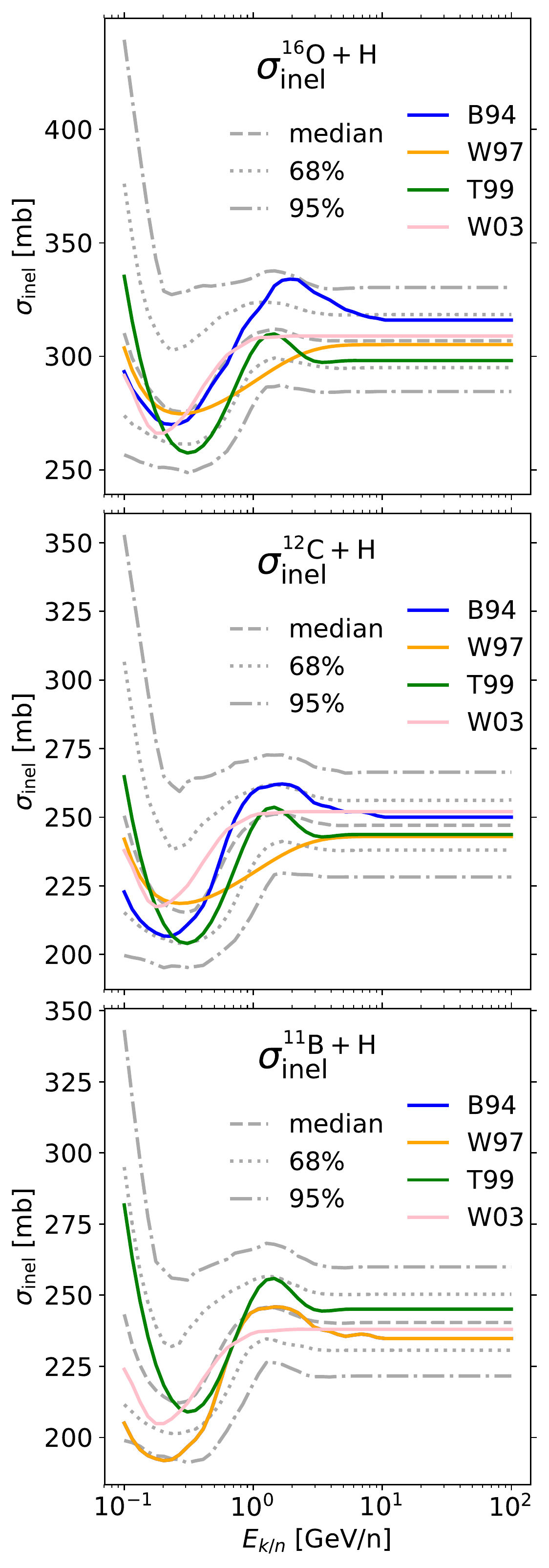} &
\vspace{0pt}\includegraphics[width=0.49\columnwidth]{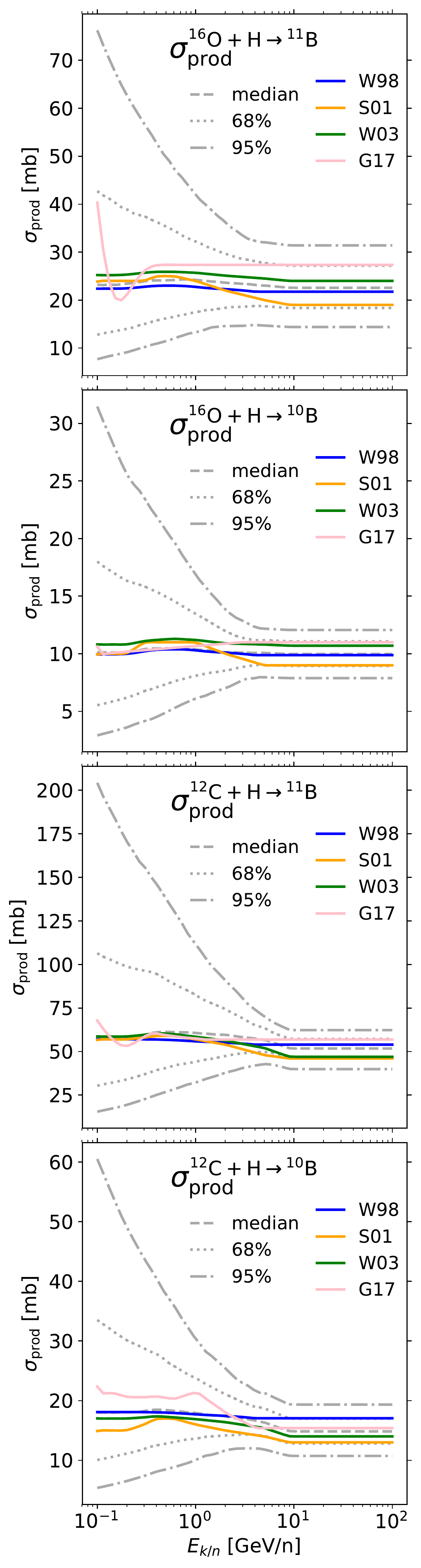}
\end{tabular}
\caption{Models (as listed in Sect.~\ref{sec:bc_impact}) and range of cross sections (quantiles corresponding to median, $68\%$, and $95\%$ CLs) generated from a reference cross section biased by Gaussian distributed nuisance parameters (as gathered in Table~\ref{tab:xs_nuis_pars}). The reactions showed correspond to the dominant ones discussed in Sect.~\ref{sec:bc_impact}, for inelastic (left column) and production (right column) cross sections.}
\label{fig:xs_nuis}
\end{center}
\end{figure}

Technically, how to choose nuisance parameters so that they enable to move from one parametrisation to another? The latter are shown (solid lines) in Fig.~\ref{fig:xs_nuis}, with $\sigma^{\rm inel}$ (resp. $\sigma^{\rm prod}$) in the left (resp. right) panels, and there is no obvious scaling formula between the curves. A possibility is to start from a reference cross section and apply several simple (uncorrelated with some non-commutative) transformations:
\begin{align}
   \quad-\; &
      \textrm{Normalisation:}\;\; \sigma \rightarrow \sigma \times {\rm Norm.}
      \label{eq:nss_n}\\
   \quad-\; &
      \textrm{Scale:}\;\; E_{k/n} \rightarrow E_{k/n} \times {\rm Scale}
      \label{eq:nss_s1}\\
   \quad-\; & \textrm{Slope:}\;\; \sigma(E_{k/n}) \rightarrow
      \label{eq:nss_s2}
          \begin{cases}
            \sigma(E_{k/n})     & \!\!\!\!{\rm if}~E_{k/n}\geq E_{k/n}^{\rm thresh.}\\
            \displaystyle\sigma\times \left(\frac{E_{k/n}}{E_{k/n}^{\rm thresh.}}\right )^{\rm Slope}\!\!\!\!\! & {\rm otherwise}.
          \end{cases}
\end{align}

\begin{figure}[t]
\begin{center}
\includegraphics[width=\columnwidth]{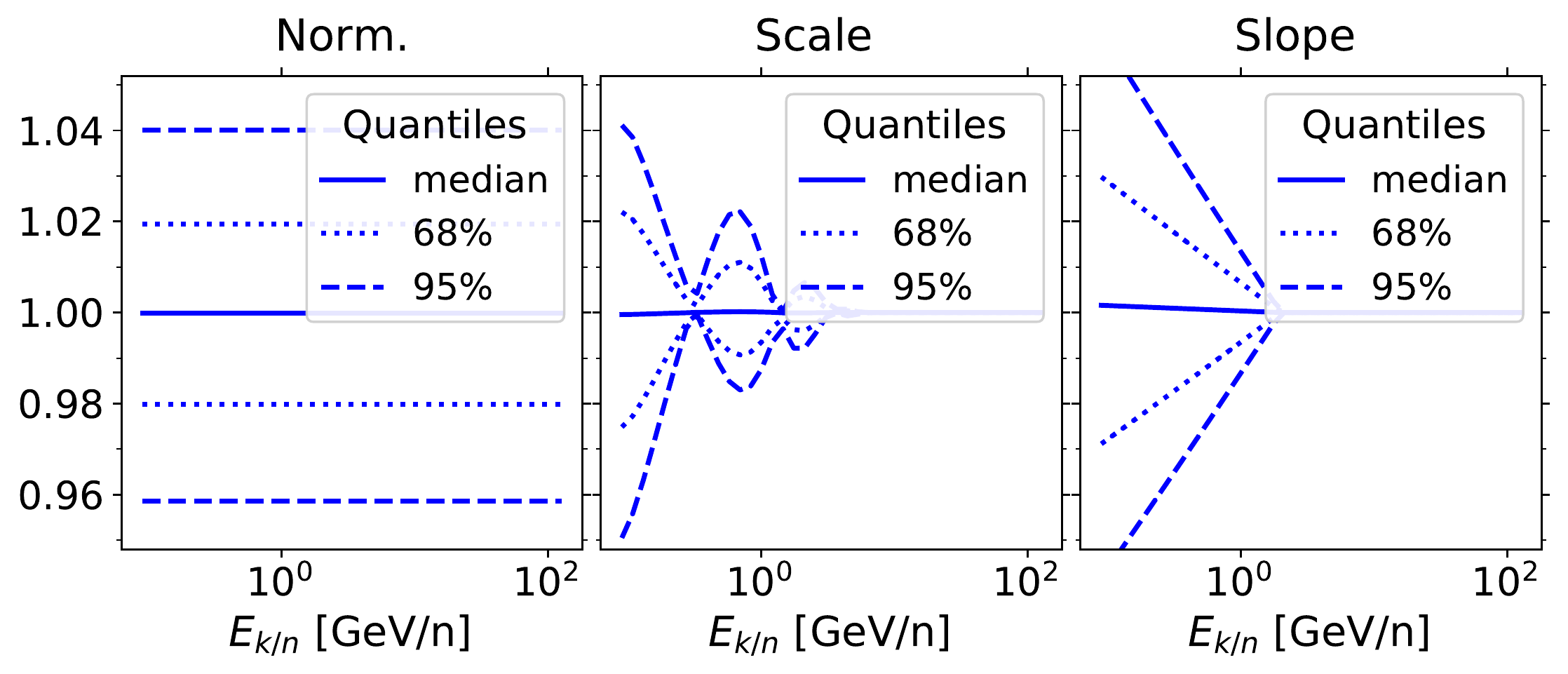}
\caption{Median, 1-$\sigma$, and 2-$\sigma$ for the distribution of $\sigma^{\rm NSS}/\sigma_{\rm ref}(E_{\rm k/n})$ values, from 1000 Gaussian distributed normalisation values (\ref{eq:nss_n}), energy scale (\ref{eq:nss_s1}), and slopes (\ref{eq:nss_s2}) with respective mean and variance $(\mu,\sigma)^{\rm Norm}=(1,0.05)$, $(\mu,\sigma)^{\rm Scale}=(1,0.2)$, and $(\mu,\sigma)^{\rm Slope}=(0,0.02)$.}

\label{fig:xs_gen}
\end{center}
\end{figure}

This set of transformations is denoted NSS in the following. It is our first option to generate nuisance parameters for cross-section uncertainties. To better visualise how NSS change cross sections, we draw 1000 values for each of the three uncorrelated NSS parameters (Norm., Scale, and Slope), and then show in Fig.~\ref{fig:xs_gen} the median, $1\sigma$, and $2\sigma$ contours from the associated 1000 realisations of $\sigma^{\rm NSS}/\sigma^{\rm ref}$. Whereas the normalisation and the low-energy slope changes are independent of any reference cross section (left and right panels), the energy scale bias is strongly dependent on it (middle panel). Indeed, the reference in Fig.~\ref{fig:xs_gen} is an inelastic cross section whose energy dependence has a low-energy peak, a dip, and a second smaller peak (see Fig.~\ref{fig:xs_nuis}): shifted to the right-hand side and divided by the unscaled one, it gives a series of three bumps.

For each reaction, the NSS nuisance parameters are chosen so that $\sigma^{\rm NSS}/\sigma^{\rm ref}\pm 1\sigma$---calculated from Gaussian distributed samples of $(\mu,\sigma)^{\rm Norm,~Scale,~Slope}$---encompasses the various cross-section parametrisations. This is shown in Fig.~\ref{fig:xs_nuis} for inelastic (left panels) and production (right panels) cross sections, with grey lines showing the median (dashed), $1\sigma$ (dotted), and $2\sigma$ (dash-dotted) envelopes of $\sigma^{\rm NSS}/\sigma^{\rm ref}$; solid coloured lines are the literature parametrisations. The corresponding NSS parameters are gathered in Table~\ref{tab:xs_nuis_pars}, and they serve as nuisance parameters in the analysis below\footnote{See \usine{}~v3.5 documentation for the syntax of the NSS parameters.}. In order to keep as fewest nuisance parameters as possible (not to slow down too much the minimisation procedure), only a normalisation and energy scale is applied to inelastic cross sections, whereas a normalisation and a slope suffices to capture the range covered by production cross-section parametrisations.

\begin{table}[t]
\centering
\caption{Values of $\mu$ and $\sigma$ for Gaussian distributed nuisance parameters for cross sections listed: top rows are for inelastic cross-section parameters defined w.r.t. T99, while bottom rows are for production cross sections defined w.r.t. W03. For information purpose, the numbers in parenthesis correspond to the maximum estimated uncertainties the given reaction has on B/C (as read from Fig.~\ref{fig:xs_var_inel} and \ref{fig:xs_var_prod}). From left to right, the nuisance parameters correspond to a normalisation, energy scale, and a modification of the slope below an energy threshold $E_{k/n}^{\rm thresh.}$. Unused parameters are indicated by `-`.}
\label{tab:xs_nuis_pars}
\begin{tabular}{lcccc}
\hline\hline
Reaction (max.                &    Norm.      &   Scale    &  Slope  & $\!\!E_{k/n}^{\rm thresh.}\!\!$ \\
impact on B/C)              &               &            &         & {\small \!\![GeV/n]\!\!}      \\
\hline
                        &$\mu$ | $\sigma$& $\mu$ | $\sigma$ &$\mu$ | $\sigma$& $\mu$ | $\sigma$ \\[0.2cm]
\!\!$^{16}$O+H \;(1\%)             & \!\!\!1.030\,|\,0.04  & \!\!0.7\,|\,0.5\!\! &    -    &  -  \\
\!\!$^{12}$C+H \;(3\%)             & \!\!\!1.015\,|\,0.04  & \!\!0.8\,|\,0.5\!\! &    -    &  -  \\
\!\!$^{11}$B+H \;(2\%)             & \!\!\!0.980\,|\,0.04  & \!\!0.7\,|\,0.4\!\! &    -    &  -  \\[0.2cm]
\!\!$^{16}$O+H$\rightarrow^{11}$B \;(15\%)\!\!&  0.96\,|\,0.18  &  -  &\!\!0.00\,|\,0.15\!\!& 5\,|\,0.\\
\!\!$^{16}$O+H$\rightarrow^{10}$B \;(9\%) \!\!&  0.93\,|\,0.10  &  -  &\!\!0.00\,|\,0.15\!\!& 5\,|\,0.\\
\!\!$^{12}$C+H$\rightarrow^{11}$B \;(12\%)\!\!&  1.10\,|\,0.12  &  -  &\!\!0.03\,|\,0.15\!\!& 8\,|\,0.\\
\!\!$^{12}$C+H$\rightarrow^{10}$B \;(14\%)\!\!&  1.07\,|\,0.15  &  -  &\!\!0.00\,|\,0.15\!\!& 5\,|\,0.\\
\hline
\end{tabular}
\tablefoot{Parameters with $\sigma\!=\!0$ amount to fixed parameters.\vspace{-0.2cm}}
\end{table}

\subsubsection{Linear combination (LC)}
Our second and more straightforward option is to define cross sections as a linear combination of the available cross-section parametrisations,
\begin{equation}
   \sigma^{\rm LC} = \sum_i C_i \times \sigma_i,
   \label{eq:LC}
  \end{equation}
where the index $i$ runs on the parametrisations shown in Fig.~\ref{fig:xs_nuis}. The sum of the $C_i$ coefficients must be close to 1, so that we naturally recover each parametrisation when only one $C_i$ is non-null. This LC allows to combine the different shapes (energy dependences) of the various cross sections, which is key for the determination of the transport parameters (see next section). To allow for possible normalisation systematics, we apply a loose constraint on the sum of the $C_i$ coefficients,
\begin{equation}
    \sum_i C_i= \mu_C^{\rm user} \pm \sigma_C^{\rm user}.
   \label{eq:LC_constraint}
\end{equation}
To be able to compare this approach to the NSS approach above, we set $(\mu_C,\sigma_C)^{\rm inel}=(1,0.04)$ and $(\mu_C,\sigma_C)^{\rm prod}=(1,0.15)$ to match the spread set on normalisation parameters in Table~\ref{tab:xs_nuis_pars}. The constraint~(\ref{eq:LC_constraint}) is accounted for as a penalty in the minimisation, that is an additional term in the $\chi^2$,
\begin{equation}
     \chi^2_{\rm LC-penalty}= \left( \frac{\mu_C-\sum_i C_i}{\sigma_C}\right)^2.
   \label{eq:LC_penalty}
\end{equation}
The $C_i$ parameters are taken to be flat in $[-0.5,1.5]$ and are forbidden to wander outside this range.

\subsection{Mock data: generation and configurations}
\label{sec:xs_mock}

We are almost ready to address some important questions related to cross-section uncertainties. How do they propagate to transport parameter uncertainties? Can we recover the true values of the transport parameters using `wrong' values for the cross sections? However, the only way to answer these questions is to analyse controlled data, whose input ingredients and parameters are known.

\subsubsection{Mock data generation}
To generate simulated data as close as possible to real data, we proceed as follows. Firstly, select the model (e.g. 1D model here) and its input ingredients (e.g. a specific cross-section dataset). Secondly, select a dataset to fit (e.g. B/C AMS-02 data) and perform the fit. Thirdly, use the best-fit model to simulate data close to the real data in the following way: (i) interpolate model values at data energies, $y_k^{\rm model}(e_k^{\rm data})$; (ii) draw at each $e_k^{\rm data}$ a value from a normal distribution ${\cal G}(0,1)$, and use the latter and $\sigma_k^{\rm data}$ (data error) to form the mock data values
        \[
          y_k^{\rm mock} = y_k^{\rm model} \times \left(1 + {\cal G}(0,1)\frac{\sigma_k^{\rm data}}{y_k^{\rm data}}\right)\,;
        \]
(iii) repeat as many times as necessary to obtain the desired number of simulated data. Lastly, analyse the mock data using the same setup or varying the input ingredients (depending on what is studied, see below).

\subsubsection{Mock data configurations}
We list and label the many cases considered below: in the forthcoming figures, each configuration is associated with a unique colour- and line-style, as recapped in Table~\ref{tab:configs}.

\begin{figure}[t]
\begin{center}
\includegraphics[width=0.95\columnwidth]{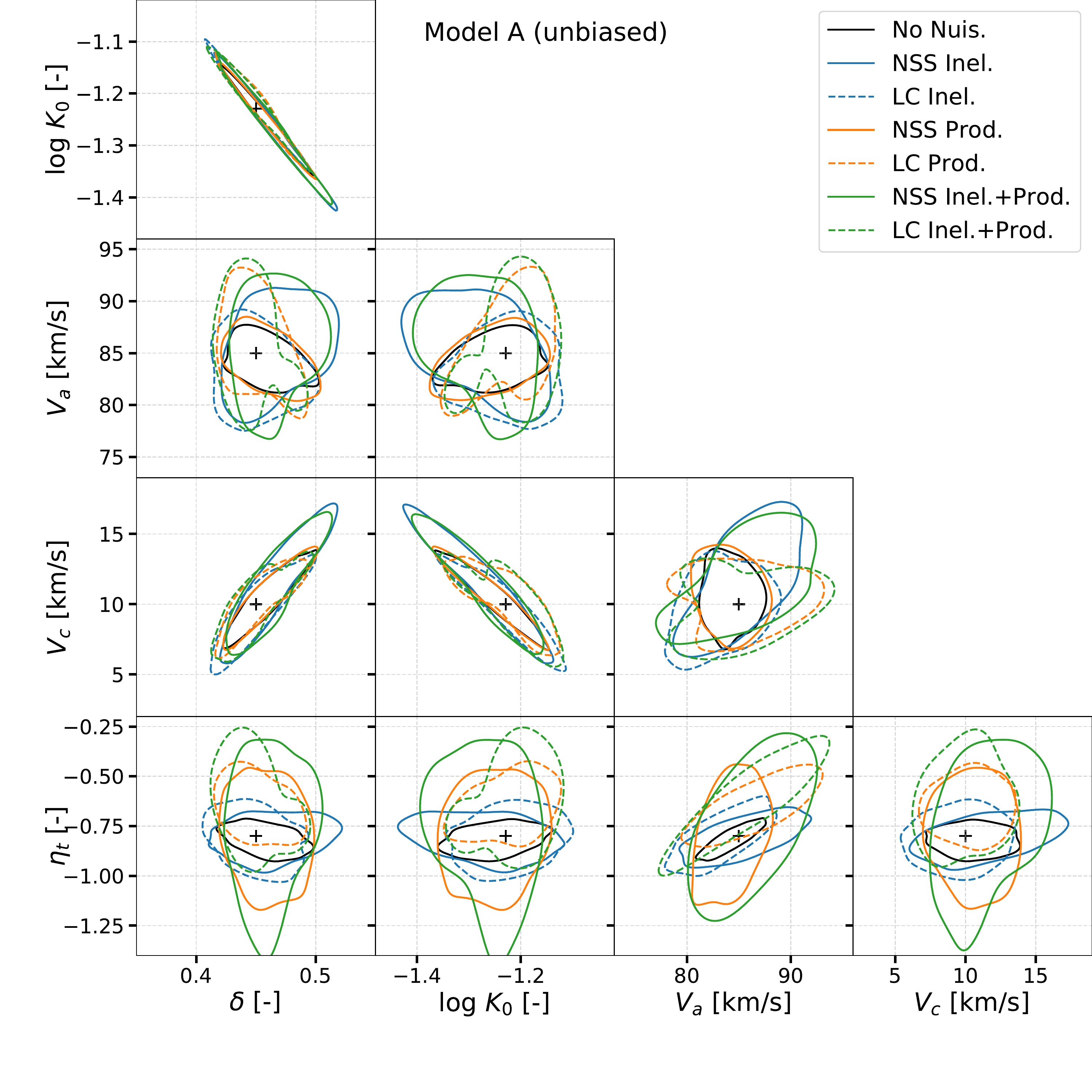}
\includegraphics[width=0.95\columnwidth]{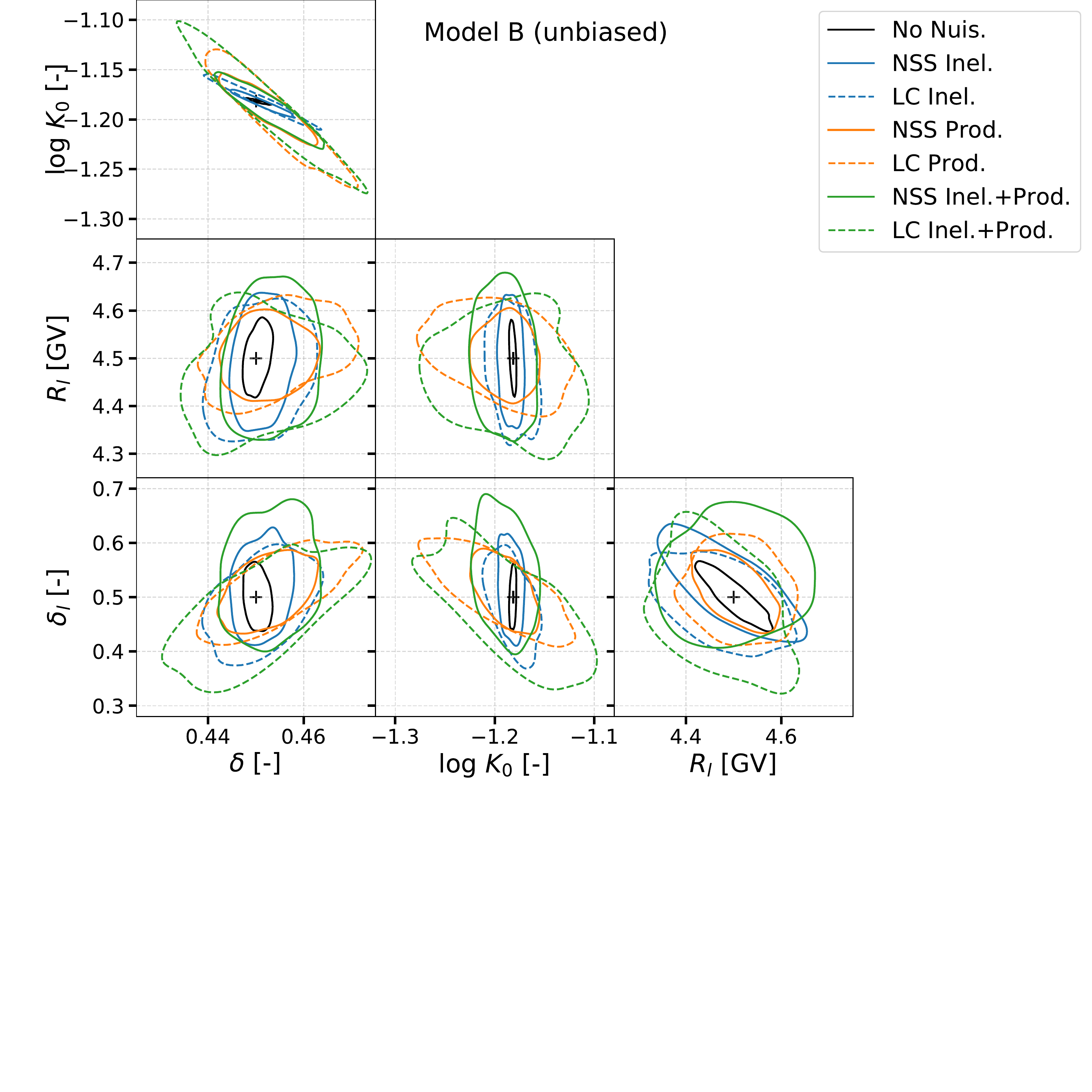}
\vspace{-2.2cm}
\caption{Reconstruction of $1\sigma$ contours (68\% confidence level) from the analysis of 1000 mock data for Model~A (top panel) and Model~B (bottom panel), for the {\em unbiased} case (cross sections for the analysis are the same as the cross sections used to generate mock data). For display purpose, the 2D probability distribution functions are estimated using a Gaussian Kernel (default method to define the bandwidth of {\tt gaussian\_kde} in {\tt scipy} python library); the irregular shapes are related to statistical fluctuations. The colour code is related to the nuisance parameters used and the line style to the type of nuisance parameters (see Table~\ref{tab:configs}). The `+' symbols represent true values. See text for discussion.}
\label{fig:xs_nuis_2D}
\end{center}
\end{figure}

\begin{itemize}
  \item{\em Propagation parameters ($\times2$):}\footnote{Reference values are those of Sect.~\ref{sec:cov}. They are used for all mock configurations to allow for a more compact presentation of the results in Figs~\ref{fig:xs_nuis_2D} and \ref{fig:xs_nuis_1D}. Thus, mock B/C data do not always represent the measured one (not shown), but this does not impact our conclusions.} each analysis is repeated for Model~A (free parameters $K_0$, $\delta$, $\eta_t$, $V_a$, and $V_c$) and Model~B (free parameters $K_0$, $\delta$, $R_l$, and $\delta_l$), see Sect.~\ref{sec:model_params}. We do so because the two models have different correlations between their parameters and cross-section uncertainties may impact them differently. For instance, reacceleration smooths spectral features, and it is present in Model~A only.

  \item{\em Cross sections to generate and fit mock data ($\times2$):} In the {\em unbiased} case, the cross sections in the propagation model used to generate and fit mock data are the same (T99 for $\sigma^{\rm inel}$ and W03 for $\sigma^{\rm prod}$). In the  {\em biased} case, mock data are still fit with T99 and W03, but they were generated from a different set of cross sections (W96 for $\sigma^{\rm inel}$ and G17 for $\sigma^{\rm prod}$)\footnote{We emphasise that cross-section values for all reactions, not just the dominant ones (see Sect.~\ref{sec:bc_impact}), are taken from the other parametrisations.}.

  \item{\em Type of nuisance parameters to fit mock data ($\times2$):} we either use `normalisation, scale, and slope' (NSS) nuisance parameters, that is $\mu|\sigma^{\rm Norm,\,Scale,\,Slope}$ values prescribed in Table~\ref{tab:xs_nuis_pars}, or linear combination (LC) nuisance parameters, that is $C_i$ coefficient weighting various cross-section parametrisations with normalisation uncertainties $\sigma_{C_i}=\sigma^{\rm Norm}$ (see previous section and Eq.~\ref{eq:LC}).

  \item{\em Nuisance parameters to fit mock data ($\times4$):} we have four types of runs to assess the impact of adding more and more nuisance parameters in the analysis, labelled {\em No nuis.} (transport parameters only), {\em Inel.} (free transport parameters and nuisance for $\sigma^{\rm inel}$), {\em Prod.} (same but $\sigma^{\rm prod}$ instead), and {\em Inel.+Prod.} (combined).
\end{itemize}

\begin{table}[t]
\centering
\caption{Summary of mock data configurations used to test cross-section nuisance parameters in Sect.~\ref{sec:xs_mock}. The $1^{\rm st}$ column lists configuration names. The $2^{\rm nd}$ column provides keys (for each configuration tested) used in legends of Figs.~\ref{fig:xs_nuis_2D}, \ref{fig:xs_nuis_1D}, and \ref{fig:xs_nuis_violon}; keys represented with specific line styles (solid or dashed) and colours (black, orange, blue, and green) are highlighted in parenthesis. The $3^{\rm rd}$ column gives synthetic information related to keys (see main text for details).}
\label{tab:configs}
{\small
\begin{tabular}{lll}
\hline\hline
  Configs    & Key       & Parameters or description \\
\hline
\multirow{2}{*}{Propag.}
    & Model~A    & $K_0$, $\delta$, $R_l$, and $\delta_l$\\
    & Model~B    & $K_0$, $\delta$, $\eta_t$, $V_a$, and $V_c$\\[0.5cm]
\multirow{2}{*}{Mock\,\&\,fit$^\star$}
    & Unbiased  & $\sigma^{\rm T99\;(W03)}_{\rm \,inel\;(prod)}$ for mock and fit\\[0.2cm]
    & Biased    &  $\sigma^{\rm W97\;(G17)}_{\rm \,inel\;(prod)}$ mock, $\sigma^{\rm T99\;(W03)}_{\rm \,inel\;(prod)}$ fit\\[0.5cm]
\multirow{6}{*}{Nuisance$^\star$}
    & \multirow{3}{*}{NSS$^\dagger$ {\em (solid)}}  & $\mu|\sigma^{\rm Norm,\,Scale}$ for $\sigma_{\rm inel}^{(^{16}\rm O,\,^{12}{\rm C},\,^{11}{\rm B})+{\rm H}}$\\
    &     & \hspace{1.7cm}{\em and}\\
    &     & $\mu|\sigma^{\rm Norm,\,Slope}$ for $\sigma_{\rm prod}^{(^{16}\rm O,\,^{12}{\rm C})+{\rm H}\rightarrow ^{11,10}{\rm B}}$\\[0.2cm]
    & \multirow{3}{*}{LC$^\ddagger$ {\em (dashed)}} & $C_{\rm T99,\,W97}$ for $\sigma_{\rm inel}^{(^{16}\rm O,\,^{12}{\rm C},\,^{11}{\rm B})+{\rm H}}$\\
    &     & \hspace{1.7cm}{\em and}\\
    & & $C_{\rm W03,\,G17}$ for $\sigma_{\rm prod}^{(^{16}\rm O,\,^{12}{\rm C})+{\rm H}\rightarrow ^{11,10}{\rm B}}$\\[0.5cm]
\multirow{4}{*}{Fit config.}
     & No nuis. (\textcolor{black}{\em black})   & Transport parameters only\\
     & Inel. (\textcolor{blue}{\em blue})      & Transport + $\sigma_{\rm inel}$ nuisance \\
     & Prod. (\textcolor{orange}{\em orange})  & Transport + $\sigma_{\rm prod}$ nuisance \\
     & Inel.+Prod. (\textcolor{OliveGreen}{\em green})\!\!\!\! & Transport + $\sigma_{\rm inel,\,prod}$ nuisance \\
\hline
\end{tabular}
}
{\small \\
$^\dagger$ See Eqs~(\ref{eq:nss_n}-\ref{eq:nss_s2}) and Tab.~\ref{tab:xs_nuis_pars}.\\
$^\ddagger$ See Eq.~(\ref{eq:LC}).
\vspace{-0.2cm}
}
\tablefoot{$^\star$To generate mock data and analyse them (Mock \& fit), the cross-section values in the model are set to the indicated parametrisations for all reactions in the network. For nuisance parameters (Nuisance), only the cross-section values for the specified reactions are modified.}
\end{table}

\subsection{Results of the mock data analysis}
\label{sec:xs_mock_results}

The analysis starts with the generation of 1000 mock data (see above), based on given values of the transport parameters, and a given choice of cross-section parametrisations. We then perform a $\chi^2$ analysis on each mock data and store the best-fit parameters and the associated $\chi^2_{\rm min}$ value\footnote{Statistical uncertainties only (taken from real data) are used in our analysis, in order to disentangle the issue of cross-section uncertainties and more involved data uncertainties (discussed in Sect.~\ref{sec:cov}).}. By construction, $\chi^2_{\rm min}/{\rm dof}\sim 1$ for our mock data, so that from the one-dimensional distribution of the parameter values and their correlations, we can reconstruct $1\sigma$ and $2\sigma$ confidence intervals. Also, comparing the parameter distribution and their `true' input value allows to assess the successfulness of our procedure.

\subsubsection{Unbiased case: Sanity check}
Figure~\ref{fig:xs_nuis_2D} shows the $1\sigma$ contours (68\% confidence level) from the 2D probability distribution functions of the transport parameters, with and without nuisance parameters in the fit. We underline a few features in these plots:
\begin{itemize}
  \item \textit{Contours from statistical errors only:} black solid lines are $1\sigma$ contours from the analysis without nuisance parameters. Their size and shape depend on the level and energy dependence of the data statistical error (blue line of Fig.~\ref{fig:AMS_errors}). Tight correlations are seen on $K_0$ and $\delta$ for both models. As expected for the {\em unbiased} analysis (i.e. same cross sections used to generate and fit mock data), the contours encompass the true value of the parameters (`+' symbols).

  \item \textit{Fit with $\sigma^{\rm inel}$ and $\sigma^{\rm prod}$ nuisance parameters:} blue lines (resp. orange lines) show the $1\sigma$ contours from the fit with $\sigma^{\rm inel}$ (resp. $\sigma^{\rm prod}$) as nuisance parameters. The size of the contours is not too strongly impacted by the cross-section nuisance parameters, because the minimum of $\chi^2$ is left unchanged (not shown) and the additional degrees of freedom provided by the nuisance parameters are `unused'. Nevertheless, the contours are deformed differently, because inelastic and production cross sections impact differently the B/C ratio---see for instant the blue ($\sigma^{\rm ine}$) and orange ($\sigma^{\rm prod}$) contours for $K_0$ vs $R_{\rm lo}$ in Model~B. The fit with combined nuisance parameters (green lines) gives contours that encompass both the previous ones. The fine details depend on the Model (A or B) and the type of nuisance (NSS in solid and LC in dashed lines) used.
\end{itemize}

From these results alone ({\rm unbiased} case), one cannot conclude on the impact of cross-section uncertainties on the transport parameters. In fact, the {\em unbiased} case is just an elaborate sanity check. It confirms that, in a scenario where cross sections are perfectly known, adding cross-section uncertainties has only a marginal effect.

\begin{figure}[t]
\begin{center}
\begin{tabular}{l}
{\tiny \hspace{1.2cm} Model A (\textcolor{OliveGreen}{biased}) \hspace{2.0cm} Model B (\textcolor{OliveGreen}{biased})}\\
\includegraphics[width=0.49\columnwidth]{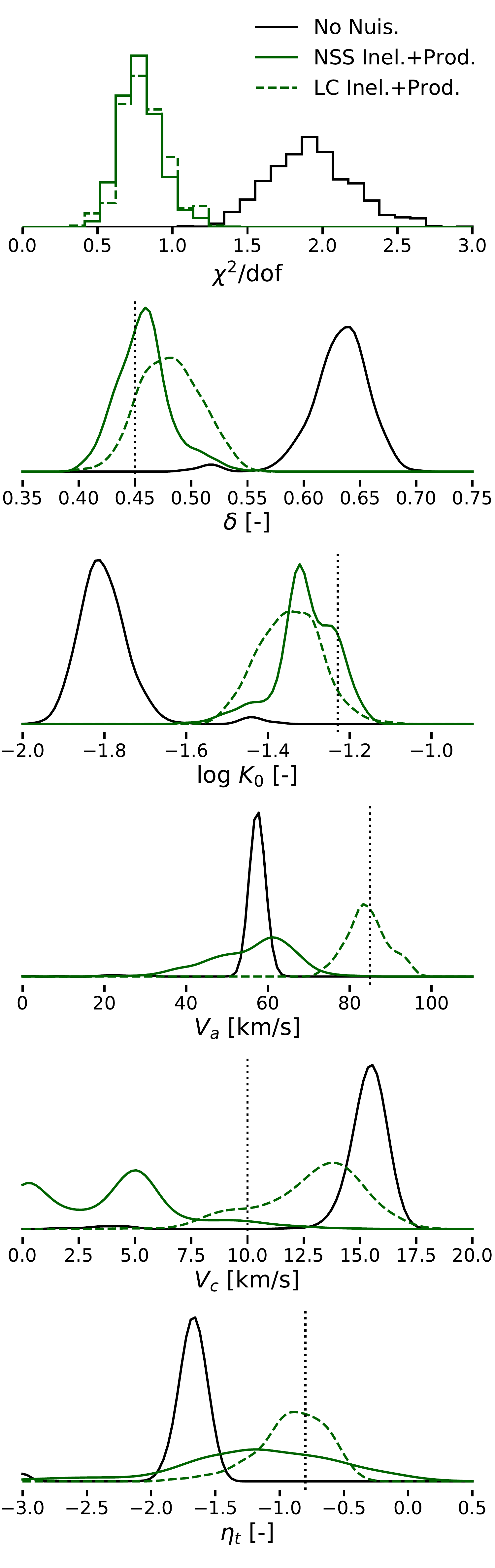}
\includegraphics[width=0.49\columnwidth]{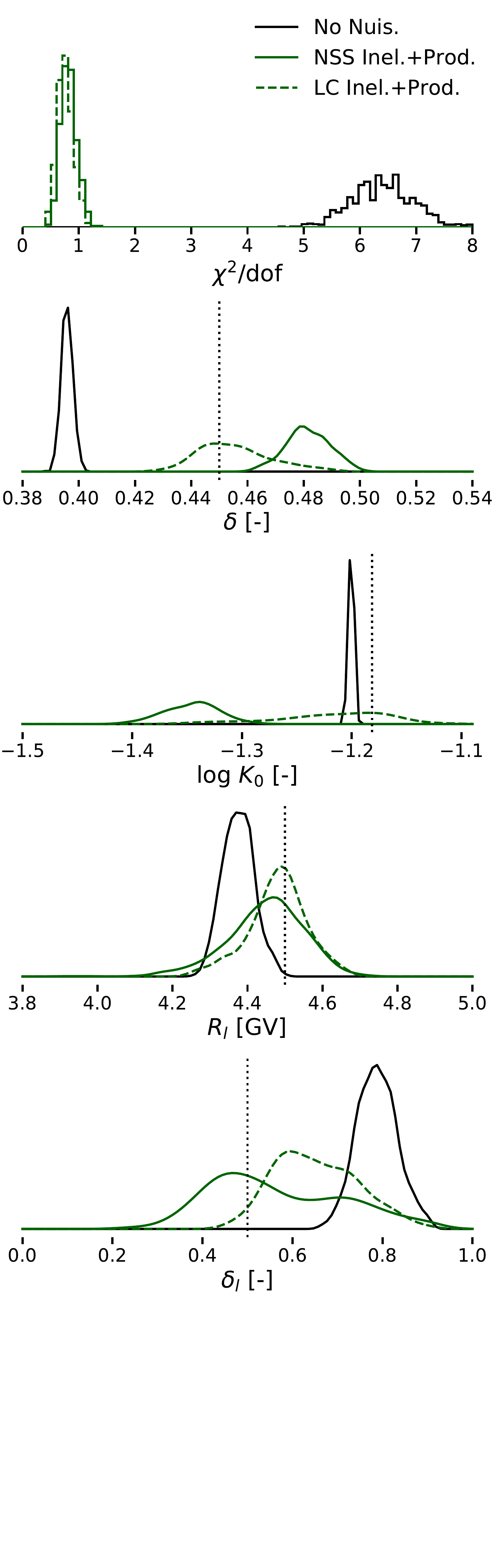}
\end{tabular}
\caption{Distribution of best-fit values ($\chi^2_{\rm min}$, then transport parameters) from the analysis of 1000 mock data for Model~A (left panel) and~B (right panel), for the {\em biased} case (cross sections for the analysis differ from those used to generate mock data). For display purpose, 1D probability distribution functions are estimated using a Gaussian Kernel. The line style and colours indicate the type and configuration of nuisance parameters used (see Table~\ref{tab:configs}). The vertical dashed lines represent true values. See text for discussion.}
\label{fig:xs_nuis_1D}
\end{center}
\end{figure}

\subsubsection{Biased case: uncertainties and biases on transport parameters}

To go further, we repeat the analysis fitting mock data with cross sections that differ from the ones used to generate them, that is {\em biased} case (see Table~\ref{tab:configs}). We show in Fig.~\ref{fig:xs_nuis_1D}, from top to bottom, the $\chi^2_{\rm min}/{\rm dof}$ distribution and the 1D probability distribution function of all transport parameters. For readability, we only show the results for the {\em No nuis.} (black lines) and {\em Inel.+Prod.} (green lines) cases. In this figure, solid (dashed) lines correspond to NSS (LC) nuisance type.

\begin{itemize}
  \item \textit{Impact on goodness of fit (top panels):} the black lines, which correspond to a fit with the transport parameters only ({\em No nuis.}), show that using wrong cross sections can lead to $\chi^2_{\rm min}/{\rm dof}$ values larger than one. Taken at face value, one would conclude that the model is excluded. Adding cross-section nuisance parameters---which encompass the true cross-section values at $1\sigma$---allows to recover $\chi^2_{\rm min}/{\rm dof}\sim 1$ (green lines). The LC nuisance parameters (green dashed lines) fare slightly better than NSS ones (green solid lines): this is understood as the `true' cross-section values can be reached in the LC case, whereas they can only be approached in the NSS case.

  \item \textit{Biased transport parameters (remaining panels):} without nuisance parameters (black lines), the transport parameters are strongly biased, up to several $\sigma$ away from their true value (vertical dashed line). Using nuisance parameters (black vs green lines) has two effects: it enlarges the probability distribution function of the transport parameters, and it shifts the distribution towards the true value. Overall, the two schemes allow to recover unbiased parameters. A mismatch is observed for the strongly correlated $\delta$ and $K_0$ parameters when using NSS in Model B. The latter is particularly sensitive to any small energy-dependent difference in the cross-section values as it directly reflects on the calculated B/C. On the other hand, in Model~A, this difference can be smoothed out by reacceleration.

\end{itemize}

We finally comment on the fact that the LC case does not recover fully unbiased transport parameters. Whereas nuisance parameters enable the cross sections to match their `true' values (the one used to generate the data), they can only do so for the selected four production cross sections and three inelastic reactions. The remaining ones are different from those used to generate the mock. This `reaction network' effect explains the observed residual biases.

\begin{figure}[t]
\begin{center}
\begin{tabular}{c}
{\tiny Model A}\\[0.1cm]
\includegraphics[width=0.45\columnwidth]{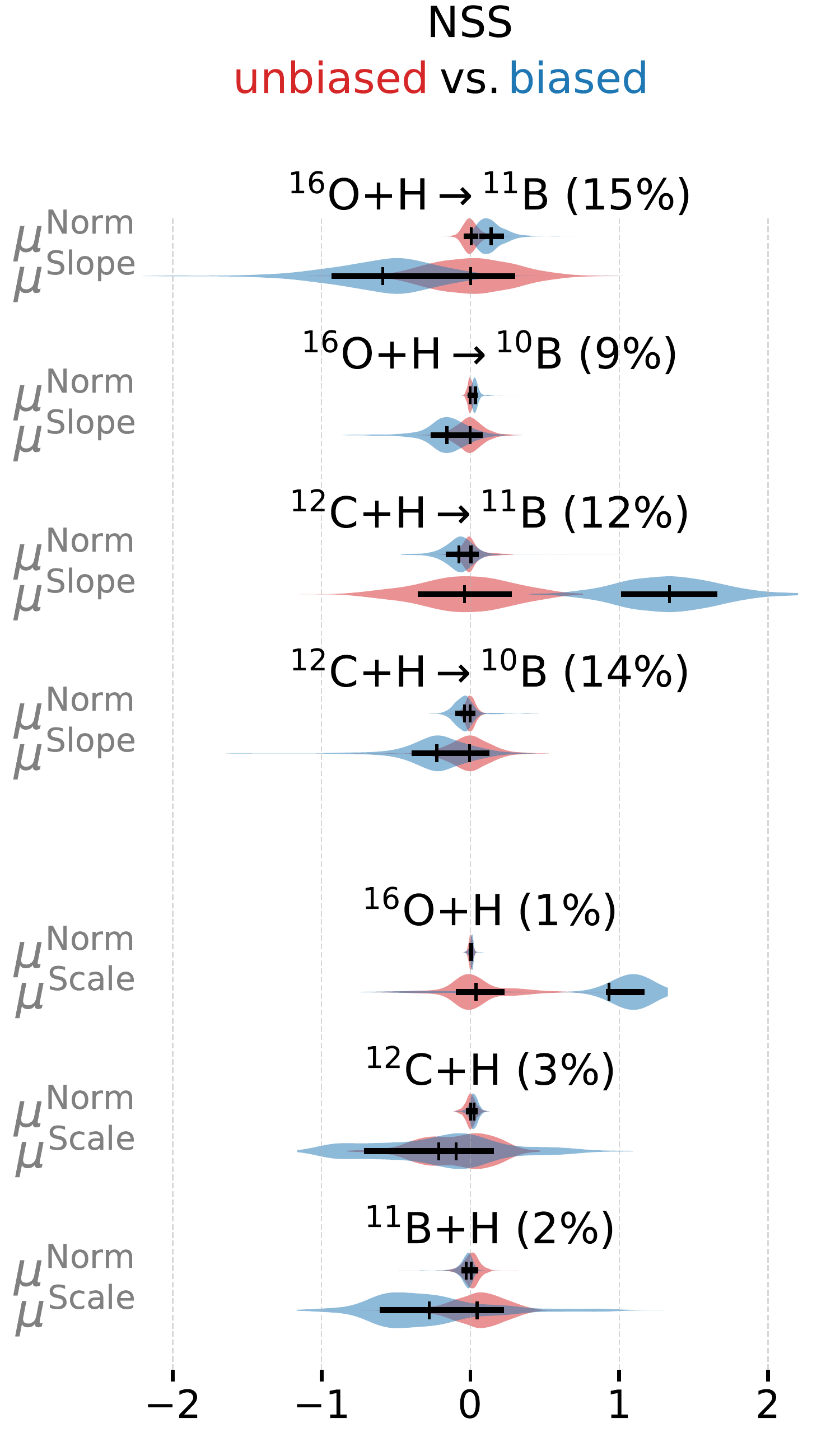}
\hspace{2mm}
\includegraphics[width=0.45\columnwidth]{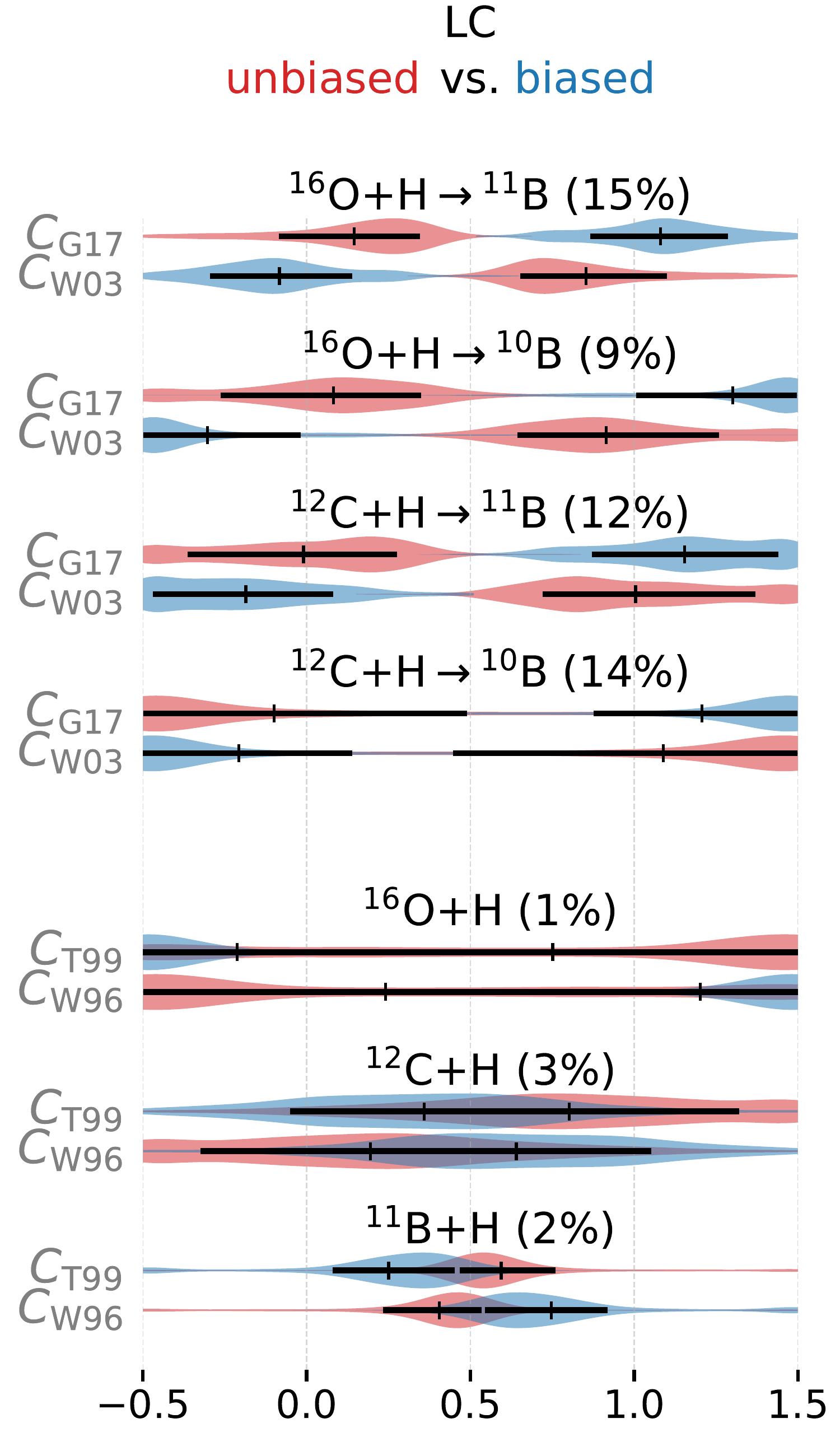}\\[0.5cm]
{\tiny Model B}\\[0.05cm]
\includegraphics[width=0.45\columnwidth]{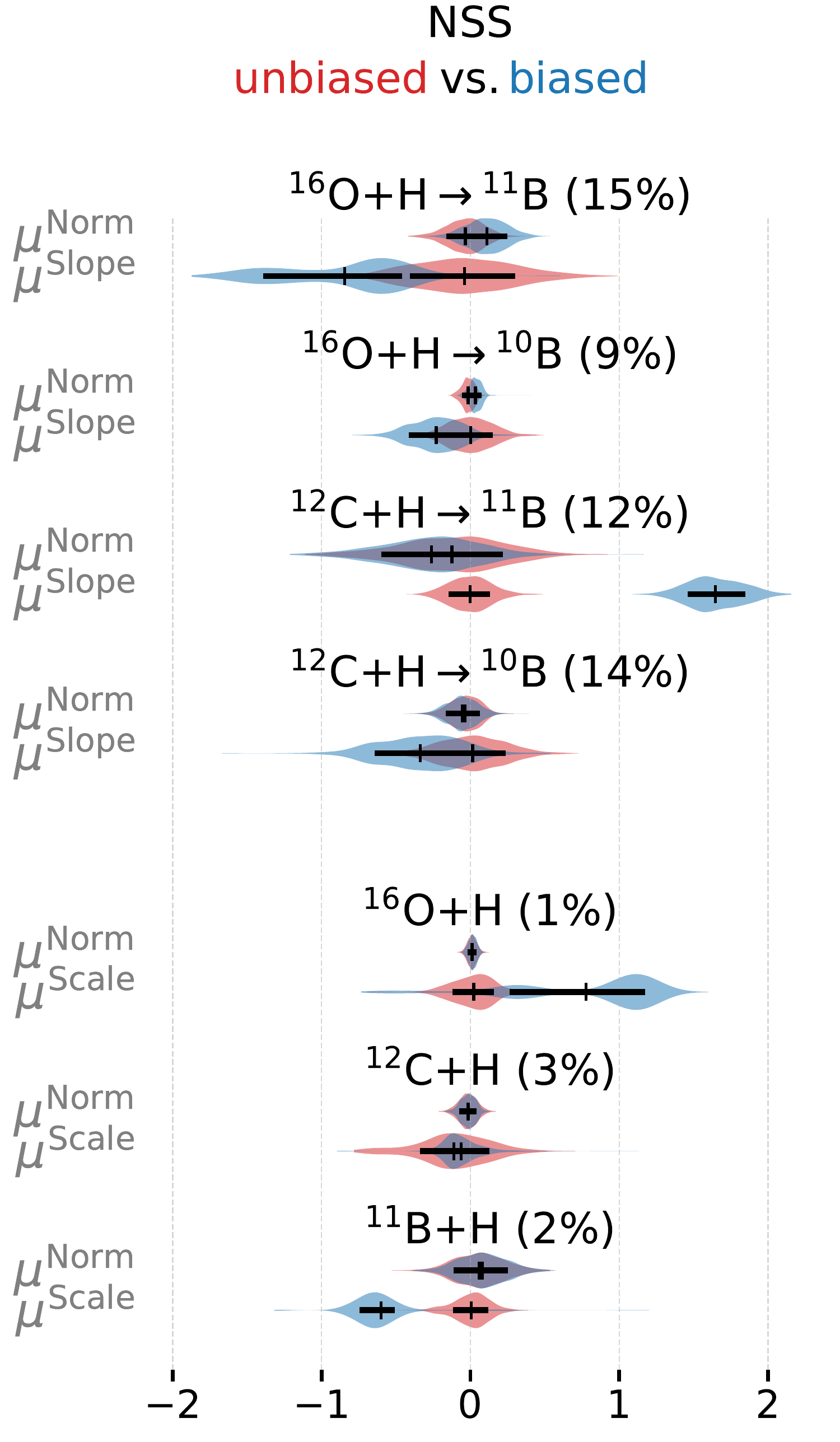}
\hspace{2mm}
\includegraphics[width=0.45\columnwidth]{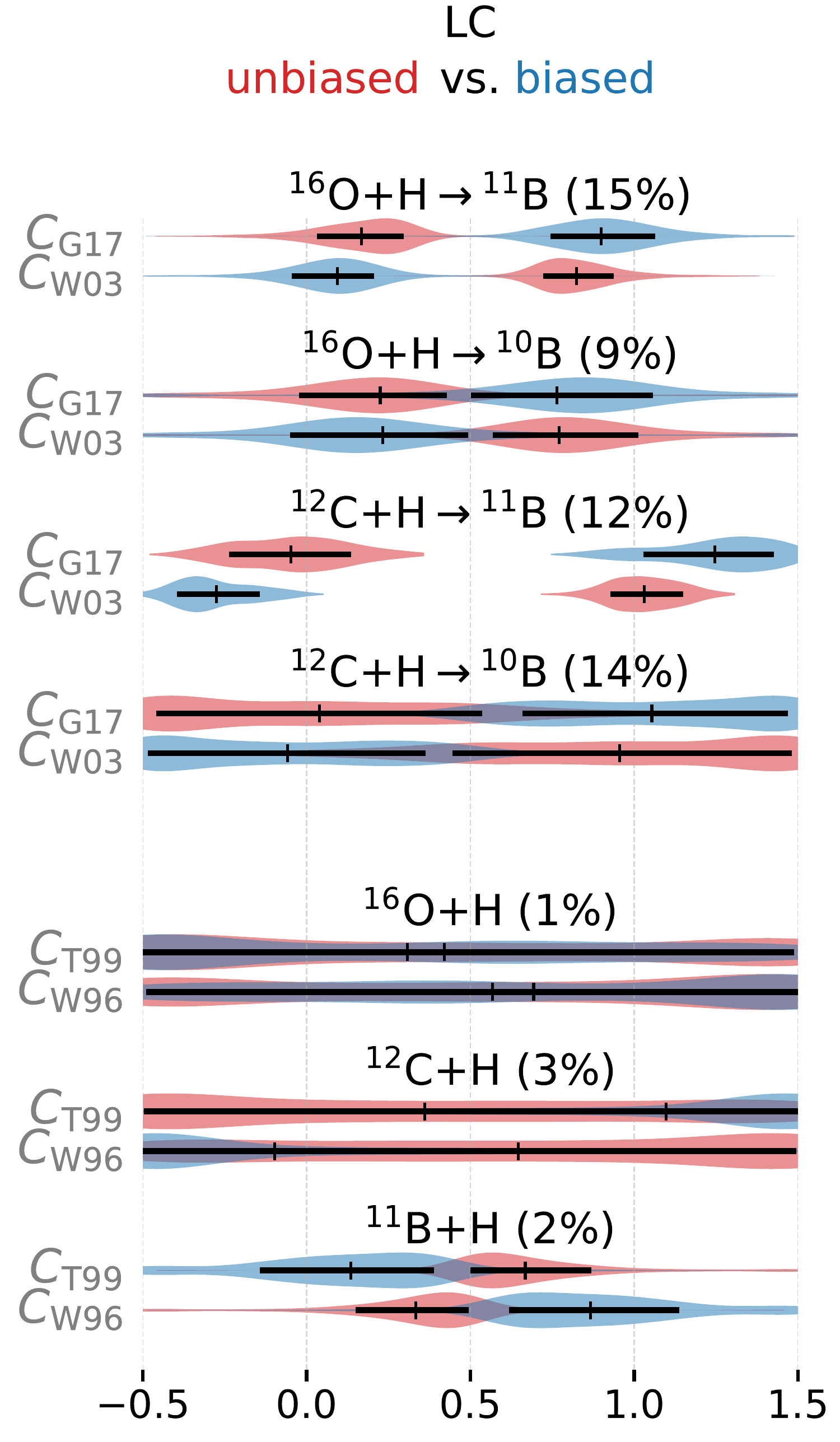}
\end{tabular}
\caption{Nuisance parameters from the analysis of 1000 mock data for Model~A (top panels) and Model~B (bottom panels). The nuisance parameters are shown reaction by reaction (rows: production then inelastic cross sections), each reaction having two nuisance parameters (see list in Table~\ref{tab:configs}). The number in parenthesis beside the reaction corresponds to its overall impact on the B/C calculation, as reported in Table~\ref{tab:xs_nuis_pars}. We show for the NSS analysis (left panels) normalised and centred parameters, whereas LC analysis (right panels) are shown between zero and one. We use violin plots to highlight the probability density of the parameter ($y$-axis) at different values ($x-$axis), and our violin plots include a marker for the median (black `+' symbol) along with the $1\sigma$ range of the parameter (thick black line). In order to directly compare the distribution of the nuisance parameters for the {\em unbiased} and {\em biased} analyses, we superimpose them on the same line in red and blue respectively. See text for discussion.}
\label{fig:xs_nuis_violon}
\end{center}
\end{figure}

\subsubsection{Biased case: posterior on nuisance parameters}
We show in Fig.~\ref{fig:xs_nuis_violon} violin plots for the nuisance parameters obtained after the fit. We only show these parameters for the {\em Inel.+Prod.} configuration, that is the analysis in which we fit mock data with transport parameters plus production and inelastic cross-section nuisance parameters (see Table~\ref{tab:configs}). The rows in Fig.~\ref{fig:xs_nuis_violon} show, for the four production and three inelastic reactions (see Sect.~\ref{sec:bc_impact}), the values of the associated nuisance parameters (two per reaction). It is interesting to show the results for the {\em unbiased} (resp. {\em biased}) case, in red (resp. blue), corresponding to the use of the same (resp. different) cross sections to generate and fit mock data. Let us comment on these distributions.

\paragraph{NSS analysis (left panels):} the nuisance parameters for the NSS analysis are a normalisation and energy scale parameters for production cross section, and a normalisation and low-energy slope for inelastic cross sections (see Eqs.~\ref{eq:nss_n}-\ref{eq:nss_s2} and Table~\ref{tab:xs_nuis_pars}). The parameters are centred and normalised (to their $\sigma$ value) so that on the $x-$axis, unbiased parameters are expected to be centred on zero and between -1 and +1 ($1\sigma$ range).
  \begin{itemize}
      \item {\em Unbiased} analysis (red): as expected, the distributions are centred on zero and have overall a very small width.
      \item {\em Biased} analysis (blue): the distributions are now offset, because nuisance parameters are used to improve the fit. Almost all parameters fall within their $1\sigma$ value, indicating that the range of variation for these nuisance parameters was well calibrated in Sect.~\ref{sec:xs_nuis}.
   \end{itemize}

Actually, we tried different values of the nuisance parameters (not shown): with a smaller range, the `true' cross sections cannot be recovered as the nuisance parameters would need to be several $\sigma$ away from their central value, which would penalise the $\chi^2$. In turn, this biases the transport parameters (discussed in the previous section): the mild bias that was observed for the NSS case (w.r.t. to the LC case) is related to the fact that $\mu^{\rm slope}$ is almost $2\sigma$ away in Fig.~\ref{fig:xs_nuis_violon}.

\paragraph{LC analysis (right panels):} nuisance parameters for the LC analysis are the coefficients $C_i^j$ (see Eq.~\ref{eq:LC}), where the index $i$ runs on the cross-section parametrisation enabled in the analysis, and $j$ runs on the various reactions considered. These coefficients typically vary from 0 (if the parametrisation is unused) to 1 (if the cross section is dominant in the LC). In principle, we should have as many coefficients as the number of available cross-section parametrisations. However, the inspection of Fig.~\ref{fig:xs_val_all} shows that some parametrisations are, up to a normalisation, very close to some others. As a result, when more than two parametrisations are considered as nuisance (not shown), there are several possible combinations to reproduce the original cross section (used to generate the mock): the $\chi^2$ function has several local minima and \minuit{} has difficulties to find the true minimum\footnote{There are more involved methods to find it, but this would further complicate the analysis for no obvious gain on the results. What matters is the capability to recover the true cross section with one combination of $C_i$, not with several.}. That is why we chose only two parametrisations in our analyses (as listed in Table~\ref{tab:configs}) and the distribution of values of the nuisance parameters are shown in the right panels of Fig.~\ref{fig:xs_nuis_violon}:
  \begin{itemize}
      \item {\em Unbiased} analysis (red): cross sections used to generate and analyse mock data are T99 for $\sigma_{\rm inel}$ (resp.~W03 for $\sigma_{\rm prod}$), so that we should recover $C_{\rm T99}\approx 1$ (resp. $C_{\rm W03}\approx 1$) for all inelastic (resp. production) cross sections and 0 for the other $C_i$. This is what is observed for all reactions. The parameters for inelastic cross sections (three bottom rows) display overall very broad distributions: we recall that the latter only have a small impact on the B/C calculation ($\lesssim 3\%$), so that a fit with data with similar uncertainties will not be sensitive to them.

      \item {\em Biased} analysis (blue): now, the starting cross sections to generate mock data, for all reactions in the network, are W96 for $\sigma_{\rm inel}$ and G17 for $\sigma_{\rm prod}$, and we observe $C_{\rm W96}\approx 1$, $C_{\rm G17}\approx 1$, and $C_{\rm others}\approx0$, as expected.
   \end{itemize}

\subsubsection{Conclusions on the impact of cross-section uncertainties}
\label{sec:xs_concl}

We have seen that assuming wrong cross sections can strongly bias the model fit, and thus bias the deduced transport parameters. Starting from the wrong cross-section values, we showed that nuisance parameters on a limited number of reactions allow to mostly recover the true values of the transport parameters. However, the procedure is not perfect owing to `reaction network` effects, that is the fact that we only use as nuisance a small, though representative, sample of all the reactions involved. The LC parameters fare slightly better than NSS parameters, but this is only true because LC parameters always contain `true' cross sections of the analysis. In real life, we do not know what are the real cross sections, and there is no guarantee that the LC approach would still fare better than the NSS one.

We want to stress that the above procedure is not even as straightforward as presented. There was, to some extent, some fine tuning done on the range chosen for the nuisance parameters to best recover our mock data. Although we were guided by the spread between the cross sections (see Fig.~\ref{fig:xs_nuis}), we had somehow to extend the range of some parameters in successive tests. In particular, for NSS, the low-energy slope was taken larger than what was strictly required from the inspection of the spread (see Fig.~\ref{fig:xs_nuis}). However, a posteriori, this made sense, because not only the normalisation but also the slope of the cross-section reaction matters in the analysis (especially for Model~B), and a larger slope parameter was needed to reconcile the energy-dependences of W03 and G17 parametrisations. Even for LC, which looks less problematic, there were some issues. We already underlined the pitfalls (in terms of minimisation with \minuit{}) of having too similar cross-section parametrisations in the linear combination. It was not mentioned earlier, but the allowed range set for the LC coefficients also matters: using $[0,1.5]$ instead of $[-0.5,1.5]$ affects the distribution of the nuisance parameters shown in Fig.~\ref{fig:xs_nuis_violon}, although the transport parameters are only very mildly affected.

Furthermore, the reader should keep in mind that regarding (i) the importance of cross-section uncertainties and (ii) which reactions should be used as nuisance, the conclusions strongly depend on the data uncertainties assumed. Indeed, the above analysis was based on AMS-02-like statistical uncertainties, that is an extreme and too conservative situation for the data. This was chosen in order to demonstrate the proof of principle of our approach. Adding systematics, which are dominant over most of the energy range in AMS-02 data, will obviously make cross-section uncertainties less impacting and the residual biases on transport parameters less severe. Accordingly, with larger uncertainties, the number of reactions to include as nuisance is also decreased: there is no gain in adding cross sections whose impact on B/C is smaller than the data uncertainties, only issues. Indeed, unnecessary reactions increase the run time of minimisations, and worse, these reactions create multiple minima that are harder to deal with. Part of these issues would be alleviated by using more evolved sampling engines, like a Markov Chain Monte Carlo \citepads[e.g.]{2014PDU.....5...29P}, but it remains better to use as few reactions as possible.

To assess the realistic impact of cross sections on the B/C analysis, we could repeat the above analysis with mock data accounting for statistical and systematic uncertainties. However, it is more interesting to illustrate how real data analysis should proceed. This is presented in Sect.~\ref{sec:xs_and_covmatrix}, but before doing so, we have to discuss how to handle systematic uncertainties in the B/C analysis.

\section{Handling systematics from experimental data}
\label{sec:cov}

Almost all, if not all CR phenomenological studies, account for data uncertainties as the quadratic sum of statistical and systematics uncertainties. Doing so ignores any possible energy correlations for the systematic errors. This has two important consequences on the model best-fit analysis. For instance, considering two extreme cases, fully uncorrelated and fully correlated uncertainties, corresponds to adding quadratically the uncertainties or to allow for a global normalisation of the data (or more precisely to an energy-dependent normalisation related to the energy dependence of the uncertainty). Starting from the same uncertainties, a $\chi^2$ analysis on the two different cases would lead to a smaller $\chi^2_{\rm min}$ in the former than in the latter case, and possibly to different values for the best-fit parameters of the model.

A better approach is to use the correlation matrix of error in the $\chi^2$ analysis (see App.~\ref{app:chi2}). However, the AMS-02 collaboration does not provide this matrix, and we have to rely on the provided information to build one. We then inspect how sensitive the analysis is on our choices.

\subsection{Origin of B/C systematic errors}

\begin{figure}[t]
\begin{center}
\includegraphics[width=\columnwidth]{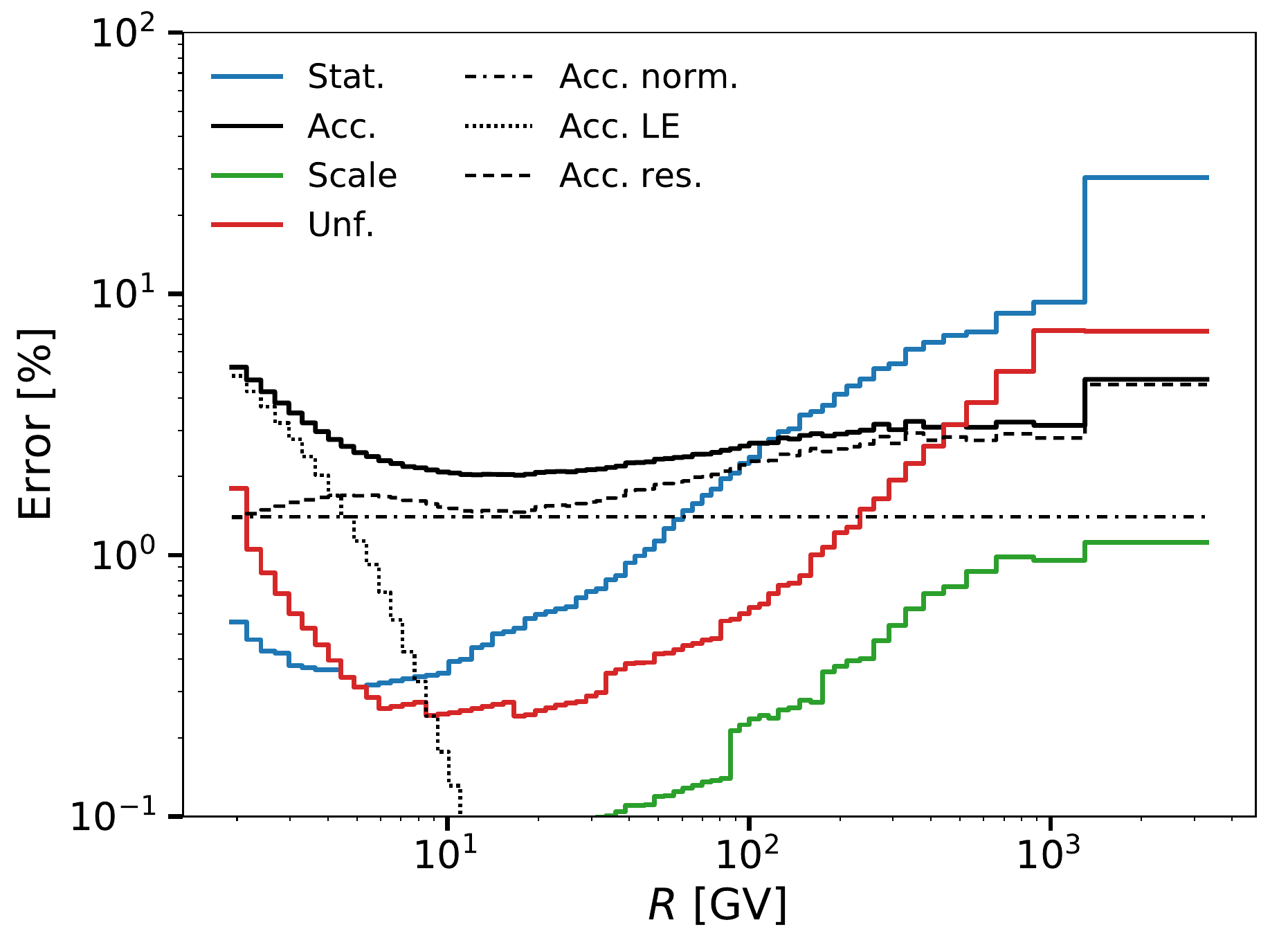}
\caption{AMS-02 errors for B/C data. Solid lines correspond to the errors provided in \citetads{2018PhRvL.120b1101A}, namely statistical, acceptance, scale, and unfolding (the step-like evolution is artificial and related to the rounding of the values provided in the table). The orange lines correspond to a further split of the acceptance errors: normalisation (norm.), low energy (LE), and residual (res.). See text for details.}
\label{fig:AMS_errors}
\end{center}
\end{figure}

The errors on the B/C ratio measured by AMS-02 are described in
\citetads{2018PhRvL.120b1101A}. The different contributions obtained from
table~VI of the Supplemental Material of \citetads{2018PhRvL.120b1101A} are shown in Fig.~\ref{fig:AMS_errors} as thick solid lines.

As explained in~\citetads{2018PhRvL.120b1101A}, the unfolding error (Unf.)
corresponds to the contribution coming from the uncertainty on the rigidity
resolution function and the unfolding procedure.  The rigidity scale error
(Scale) is the sum of the contribution from residual tracker misalignment and
from the uncertainty on the magnetic field map measurement and its temperature
time-dependent correction.  The acceptance error (Acc.) is the sum of different
contributions: survival probability of Boron and Carbon in the detector,
Boron contamination from heavier nuclei fragmentation (mainly carbon), and
uncertainty on the `data/Monte Carlo' corrections to the Boron and Carbon acceptances.

\subsection{Building the covariance matrix: correlation length}

To properly take into account AMS-02 data uncertainties,
one needs to define the covariance matrices $C_\alpha$ for $\alpha$=(Stat., Unf., Scale, Acc.),
and minimise the $\chi^2$ defined by Eqs~(\ref{eq:chi2}--\ref{eq:cov}). As these covariance 
matrices are not provided explicitly in \citetads{2018PhRvL.120b1101A}, we start from the covariance matrices of relative errors ${\cal C}_{\rm rel}^\alpha$, estimated from the following expression:
\begin{equation}
 (C_{\rm rel}^\alpha)_{ij} = \sigma^\alpha_i \sigma^\alpha_j \exp\left(-\frac{1}{2}
\frac{(\log (R_i/R_j)^2} {(l_\rho^\alpha)^2} \right)\,,
\end{equation}
with $(C_{\rm rel}^\alpha)_{ij}$ the $i\!j$-th element built from the relative errors $\sigma^\alpha_i$ and $\sigma^\alpha_j$ at rigidity bins $R_i$ and $R_j$, and where the parameter $l_\rho^\alpha$ is the correlation length associated with the error $\alpha$ (in unit of decade of rigidity).

For this study, we set the covariance matrix to be $({\cal C}^\alpha)_{ij} = ({\cal C}^\alpha_{\rm rel})_{ij}\times {\rm model}_i\times{\rm model_j}$ (see App.~\ref{app:chi2} for a justification), and we set the correlation lengths $l_\rho^\alpha$ to the following values:
\begin{itemize}
    \item $l_{\rho}^\textrm{Stat.} = 0$ because the number of events on each bin are independent;

    \item $l_{\rho}^\textrm{Scale} = \infty$ since the uncertainty on the rigidity
        scale affects all rigidities similarly;

    \item $l_{\rho}^\textrm{Unf.} = 0.5$ because errors from the unfolding procedure
        and from the rigidity response function affect intermediate
        scales. As seen on Fig.~\ref{fig:AMS_errors}, this error is sub-dominant compared
        to Stat. and Acc. errors, and we checked that the results are not affected by our
        choice for $l_{\rho}^\textrm{Unf.}$.

    \item The value of the
        correlation length for the Acc. error is more critical, because this
        error dominates the systematic error and it cannot be easily defined.
        The dependence of $\chi^2_{\rm min}/$dof and of the fitted parameters with this correlation
        is studied below for different values $l_{\rho}^\textrm{Acc.} =
        0.01\ldots 3$, which cover the range from lower than the bin size (fully
        uncorrelated) to the full range (fully correlated).
\end{itemize}

As the acceptance error is a combination of errors which are expected to have a
rather small correlation length (`data/Monte Carlo' corrections)  and others which are
expected to have a large correlation length (cross-section normalisation), one
can try to decompose this error into different contributions with different correlation
lengths. In particular, the rise of the acceptance error at low rigidity is not
expected to be correlated with larger rigidities: it is related to the rapid
change of the acceptance at low energy mostly because of energy losses in the detector.
One can therefore construct a  better description of the covariance matrix by
splitting acceptance errors in three independent parts:
\begin{itemize}
  \item a normalisation error, {\em Acc.\,norm.} (dash-dotted orange line in
  Fig~\ref{fig:AMS_errors}), with a large correlation length ($l_\rho \sim 1.0$);

  \item a rise at low rigidity, {\em Acc.\,LE} (dotted orange line), with an
  intermediate correlation length ($l_\rho \sim 0.3$);

  \item a residual error, {\em Acc.\,res.} (dashed orange line), defined so that
  the quadratic sum of the three contributions equals the full acceptance error.
  This last part corresponds mainly to `data/Monte Carlo' corrections and the rigidity-dependent parts of
  other acceptance errors. Its correlation length is not well defined and left free in the following.
\end{itemize}
Near the completion of this article, we found out that \citetads{PhysRevD.99.103014} also
proposed to use a correlation matrix of errors to analyse AMS-02 data. However, while these authors
focus on \pbar{} and use a single correlation length fit on the data, our analysis
relies on several correlations lengths whose values are motivated by physics processes in
the AMS-02 detectors.

\begin{figure}[t]
\begin{tabular}{l}
{\tiny \hspace{2.cm} Model A \hspace{2.8cm} Model B}\\
\includegraphics[width=0.44\columnwidth]{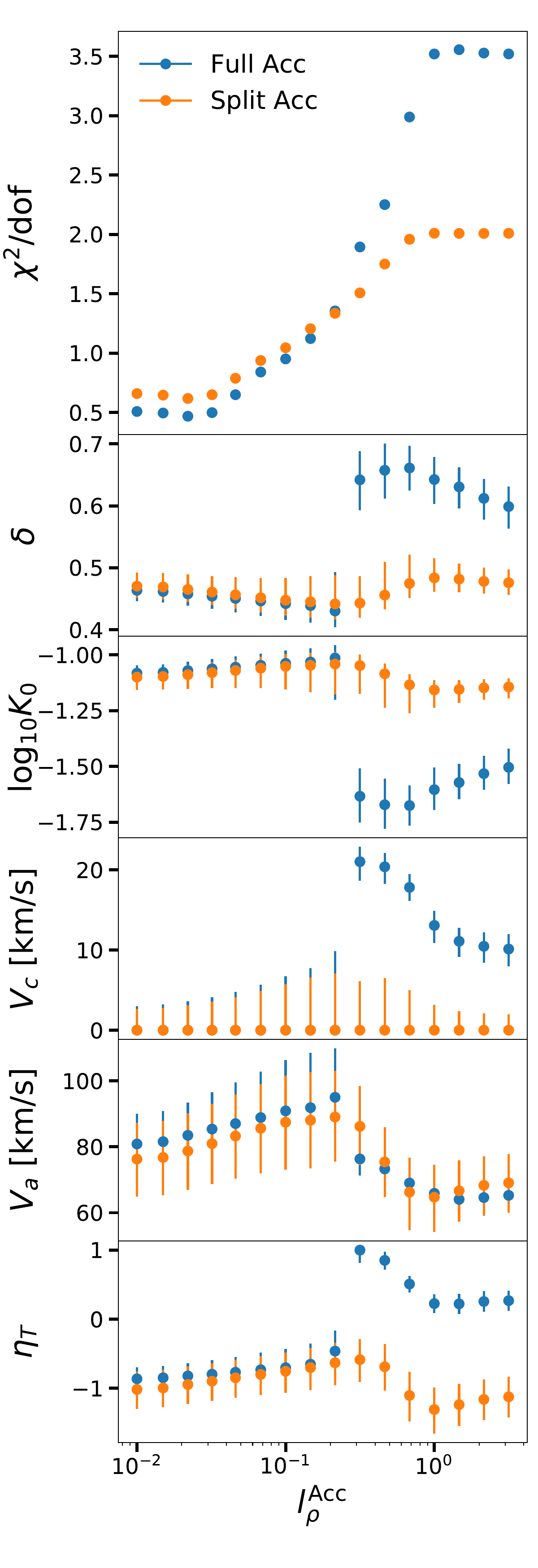}
\includegraphics[width=0.44\columnwidth]{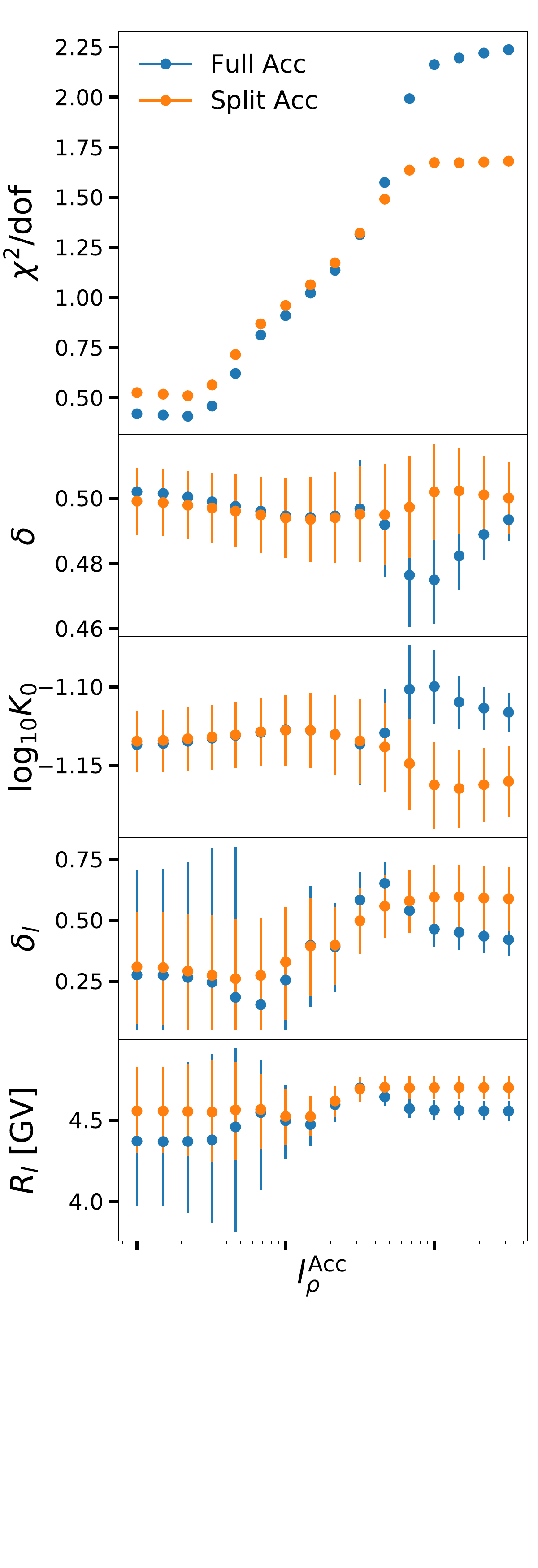}\\[-0.25cm]
{\tiny \hspace{2.cm} Model A \hspace{2.8cm} Model B}\\
\hspace{0.3mm}
\includegraphics[width=0.44\columnwidth]{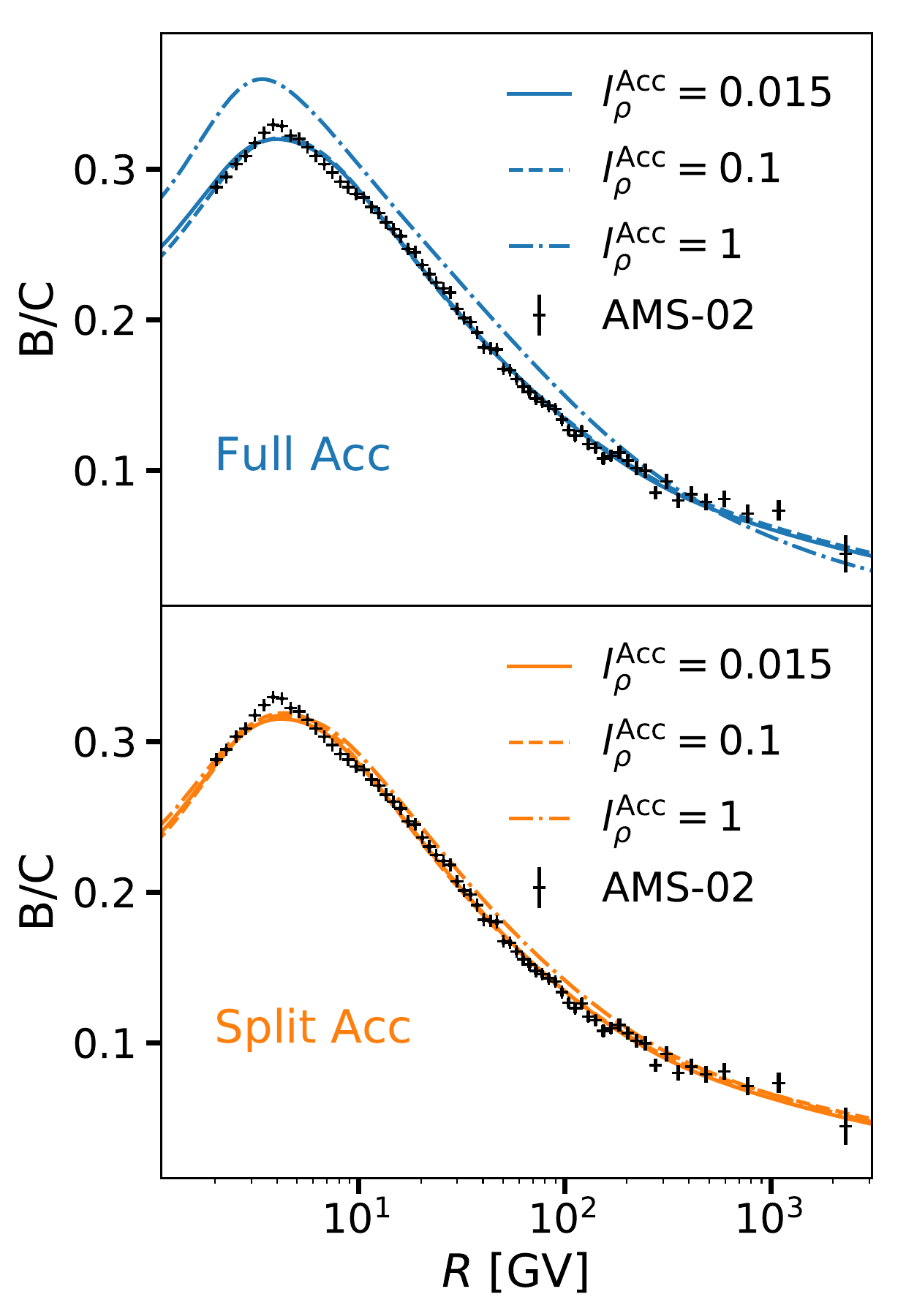}
\includegraphics[width=0.44\columnwidth]{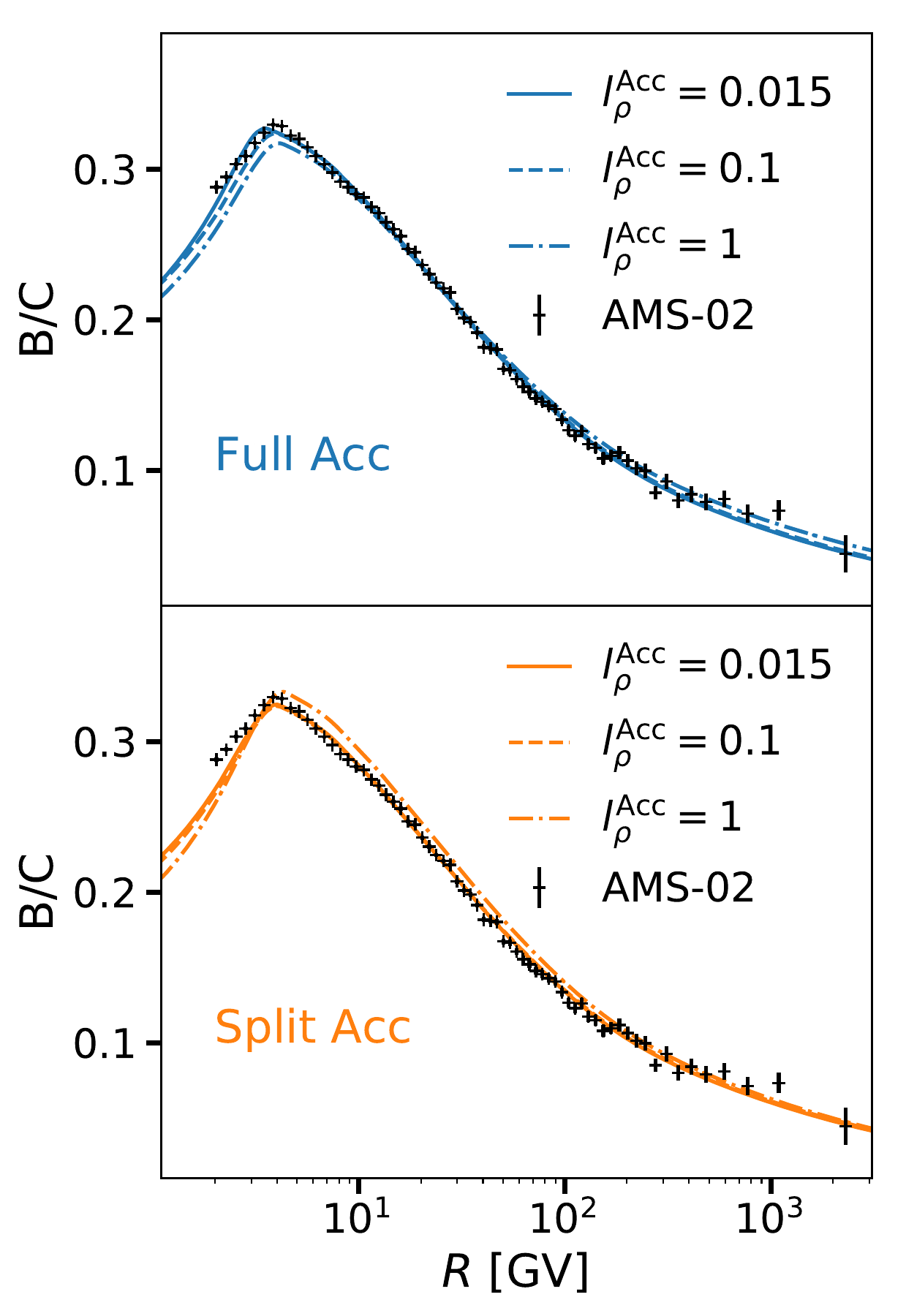}
\end{tabular}
\caption{{\bf Top panels:} values obtained for the $\chi^2_{\rm min}/$dof and the fit parameters (and uncertainties) as a function of $l_{\rho}^\textrm{Acc.}$ for propagation models A (left) and B (right) and for the full acceptance error (blue) and the split acceptance error (orange). In the case of split acceptance error,  $l_{\rho}^\textrm{Acc.}$ corresponds to the correlation length of the {\em Acc.\,res.} contribution only. {\bf Bottom panels:} Comparison of best-fit B/C (lines) and AMS-02 data (symbols) for the above models and configurations, i.e. full (blue lines, top) vs split (orange lines, bottom) acceptance errors. Only a sample of correlation lengths are shown (0.015, 0.1, and 1 decade).}
\label{fig:models_vs_corr}
\end{figure}

\subsection{Parameter and goodness-of-fit dependence on correlation length}

We have built two different covariance matrices, which partly depend on an unknown
correlation length $l_{\rho}^\textrm{Acc.}$. We can now perform a $\chi^2$ analysis
with \minos{} to extract robust errors on the parameters. The analysis is repeated on the two
propagation model configurations A and B introduced in Sect.~\ref{sec:prop_model}.

The top panels of Fig.~\ref{fig:models_vs_corr} show the values obtained for $\chi^2_{\rm min}/$dof
and best-fit parameters as a function of $l_{\rho}^\textrm{Acc.}$ for models A
(left) and B (right) and for the full acceptance error (blue circles) and the split
acceptance error (orange circles). The B/C from the best-fit model along with AMS-02 data are shown on the bottom panels of Fig.~\ref{fig:models_vs_corr} for the same models, that is for the full (blue lines, top)
and split (orange lines, bottom) acceptance errors. As expected, $\chi^2_{\rm min}/$dof strongly
depends on $l_{\rho}^\textrm{Acc.}$ for both models. The best-fit parameters are stable
(i.e. fluctuate within errors estimated from the fit) for low and large $l_{\rho}^\textrm{Acc.}$ but undergo a rapid jump around
$l_{\rho}^\textrm{Acc.} = 2$ for model A when one uses the full acceptance error
description. These features are problematic since it means that
the best-fit parameters are very sensitive to the choice of
$l_{\rho}^\textrm{Acc.}$. In addition, with the full acceptance error,
the best-fit obtained for model A and $l_{\rho}^\textrm{Acc.}\approx1$
does not pass through the data points as featured by the upper-left plot of the bottom panels in Fig.~\ref{fig:models_vs_corr}. This is explained as follows:
with a large correlation, the cost on the $\chi^2$ for a global deviation between data and
model is moderate and thus accepted, and would correspond to a global bias in the measured B/C.
Although correct from a mathematical standpoint, this interpretation is disputable given our
crude modelling of the systematic errors. This unwanted behaviour is absent when we use the
split acceptance error modelling.

From the above results, we conclude that the best way to handle
the systematic errors is to use the split acceptance errors approach.
Indeed, not only does it provide a more realistic description of the acceptance systematic error,
but it also leads to more stable results w.r.t. the values taken for 
$l_{\rho}^\textrm{Acc.}$. In this approach, $l_{\rho}^\textrm{Acc.}=0.1$ is a
reasonable choice which gives a $\chi^2_{\rm min}/$dof $\sim 1$ and
conservative errors for the fit parameters.

\begin{figure}[t]
\begin{tabular}{c}
{\tiny Model A}\\
\includegraphics[trim={.7cm 0 0.2cm 0.2cm},clip,width=0.9\columnwidth]{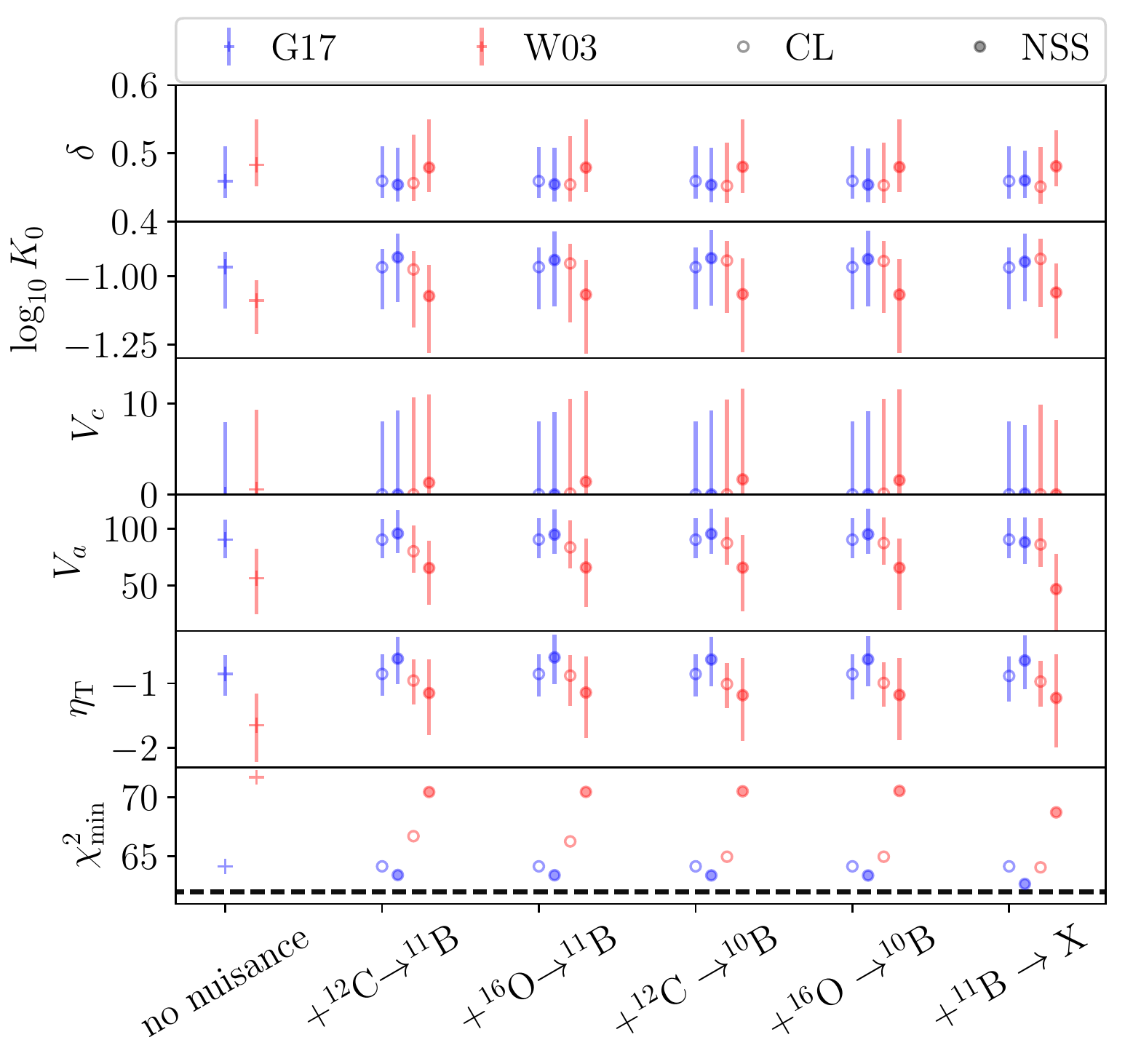}\\
{\tiny Model B}\\
\includegraphics[trim={.7cm 0 0.2cm 0.2cm},clip,width=0.9\columnwidth]{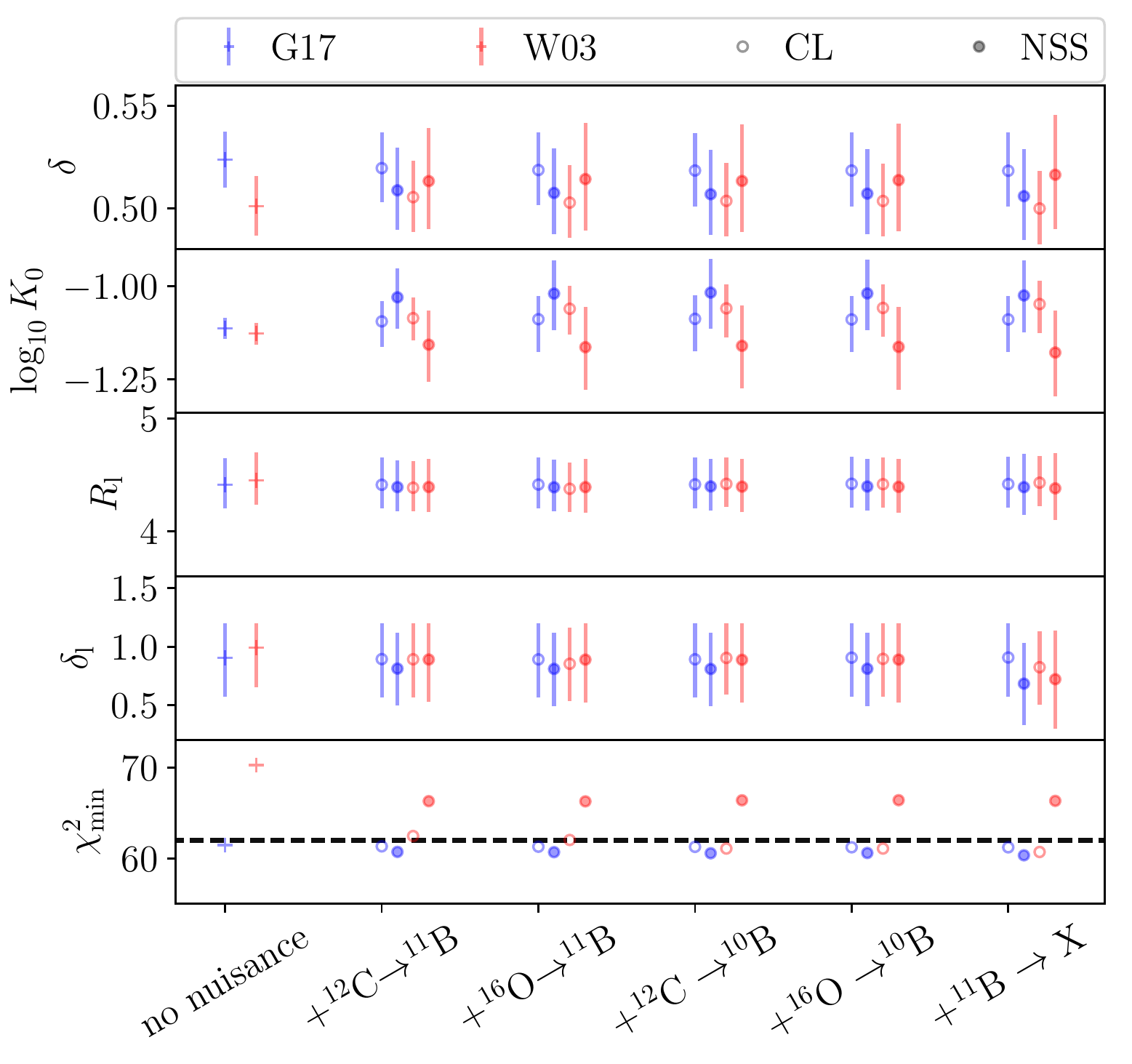}
\end{tabular}
\caption{Evolution of the best-fit parameters and asymmetric $1\sigma$ errors (extracted from \minos{}), increasing the number of nuclear cross-section reactions used as nuisance parameters (from left to right on the $x$-axis); Model A (top) and B (bottom) are shown. For every additional reaction as nuisance, four different fits are performed: starting either with G17) or W03 parametrisations for the production cross sections; using either CL or NSS for the implementation of nuisance parameters. The black dashed line in the $\chi^2_{\rm min}$ panel corresponds to the threshold $\chi^2_{\rm min}/{\rm dof}=1$. See text for discussion.}
\label{fig:nuis_and_cov}
\end{figure}

\section{Joint impact of cross-section uncertainties and data systematics}
\label{sec:xs_and_covmatrix}

The previous section provides us with a realistic treatment of the data errors. In this light, we can revisit the impact of cross-section uncertainties on the best-fit parameters and errors (Sect.~\ref{sec:xs_nuis}). Indeed, we showed in the most challenging case (statistical only in the data) that NSS and LC nuisance parameters enable to recover the correct transport parameters when starting from the wrong cross sections, whereas systematic errors dominate over a wide dynamical range (see Fig.~\ref{fig:AMS_errors}). Nevertheless, because neither NSS nor the LC approach is perfect (see discussion in Sect.~\ref{sec:xs_concl}), it is important to test both to ensure, as a minimal consistency check, that consistent values of the transport parameters (within their uncertainties) are obtained in both approaches.
For this purpose, we analyse the `real' B/C data using the covariance matrix of errors (see previous section), with both NSS and LC approaches, and varying the number of reactions used as nuisance parameters. This allows us to validate our strategy for the actual data, as used in the companion paper \citepads{2019arXiv190408917G} for the B/C physics analysis. It also exemplifies how an analysis should be carried out in the methodology we propose.

Our results are presented in Fig.~\ref{fig:nuis_and_cov}, in which we display the best-fit parameters and errors of Model A (top) and Model B (bottom) for an increasing number of reactions used as nuisance (from left to right). The four production cross sections are introduced by order of importance with respect to their contribution to the secondary boron. We then introduce the inelastic cross sections of $^{11}$B; we have checked that the impact of other inelastic cross sections is negligible. We recall that we can start from different parametrisations of the full network of cross sections: for production cross sections, we either start from G17) or W03), whereas we only consider T99 for inelastic ones (their impact is negligible, see below). Several comments are in order about these results:
\begin{itemize}
  \item For both Model A and B, even without nuisance parameters (\textit{no nuisance} on $x$-axis), the best-fit parameters for the two cross-section cases considered (W03 and G17) are consistent within errors. This means that the covariance matrix mitigates most of the errors coming from using the wrong cross sections (see discussion in Sect.~\ref{sec:xs_mock}). This would not have been the case (not shown) using simply the statistical and systematic uncertainties in quadrature. This further demonstrates the importance of having the correct covariance matrix of errors. It also emphasises the fact that energy correlations on intermediate scales (one decade) in data uncertainties could relax, to some extent, the need for very accurate cross sections.

  \item When adding nuisance parameters (NSS or LC), the consistency between the parameters is improved, in particular for $\delta$ (in both models A and B). With more degrees of freedom, this translates into better $\chi^2_{\rm min}$ values and larger uncertainties on the parameters. The latter is only significant for Model~B, and the lack of increase in Model~A probably comes from the fact that it has many degenerate `low-energy' degrees of freedom ($V_a$, $V_c$, $\eta_T$) which already make the uncertainties maximal. We stress that the improvement depends on the initial set of cross sections used and method, but overall, for this specific analysis, with the covariance matrix of error dominating the error budget, only a few reactions need to be taken into account.

  \item Focusing on the fit quality ($\chi^2_{\rm min}$ values), we see that G17 systematically gives a better fit than W03. In the LC approach, the productions cross section initially set to W03 also choose to go to G17. This is probably not too surprising as G17 is expected to better match existing cross-section data, whereas W03 is based on a global semi-empirical fit to these same data (see Sect.~\ref{sec:XS} and \citealtads{2018PhRvC..98c4611G}). 

\end{itemize}

Based on these results, the recommendation for the B/C analysis, is to start from G17 production cross sections, and to use the NSS nuisance parameters. The latter allow for more freedom than LC ones, so that slight improvements are still possible when including several cross-section reactions as nuisance (though these improvements are not very important statistically in this case). These recommendations are followed to define benchmark models in our companion paper \citepads{2019arXiv190408917G}, in which the values of the parameters shown in Fig.~\ref{fig:nuis_and_cov} are discussed and interpreted.

\section{Recommendations and conclusions}
\label{sec:conclusions}

Faced with the challenges of interpreting cosmic-ray data of unprecedented accuracy, we have refined the methodology to properly account for all uncertainties (model, ingredient, and data) in the fit to the data. The proposed methodology was exemplified on the analysis of the AMS-02 B/C ratio.

The first step was to ensure a model precision higher than the data uncertainty: we inspected in detail the numerical stability of the model and the impact of energy boundary conditions. Some low-energy boundary conditions fare better than others, but setting any of them to MeV values ensure a good precision of the model calculation above a few hundreds of MeV/n. When using a Crank-Nicholson approach to solve the second-order differential diffusion equation on energy, we have checked its precision and provided a criterion to numerically converge to the correct solution when $V_a\rightarrow0$ (i.e. if no reacceleration, corresponding to a first-order differential equation).  We have also quantified the systematics from using the point-estimate calculation of a flux or ratio compared to the correct calculation integrating the model over the energy bin. While the discussion is partly specific to the model used and the species inspected, our considerations are generic. Obviously, the precision tests should always be repeated and compared to the data uncertainty for other species, models, and data considered.

The second step was to handle properly cross-section uncertainties. We detailed the impact of the most important reactions on the B/C ratio ($\lesssim 3\%$ for inelastic cross sections and $\sim10\%$ for production cross sections). We then proposed an approach to account for these uncertainties via Gaussian distributed nuisance parameters, based on a combination of Normalisation, Scale, and low-energy Slope cross-section modifications (NSS) or based on linear combinations (LC) of existing parametrisations. We validated this choice on simulated data, showing that the degrees of freedom enabled by these nuisance parameters allow to recover the true parameters (when starting from a different set of cross sections simulated data were generated with). Simulated data also show that starting from the wrong cross-section values, valid propagation models would be excluded based on statistical criterion ($\chi^2_{\rm min}/{\rm dof}\gg 1$). The nuisance parameters we proposed also cure this problem.

The third step was to handle as best as possible data uncertainties. We accounted for possible energy correlations in the AMS-02 data via the covariance matrix of errors. As the AMS-02 collaboration does not provide such a matrix, we proposed a best-guess model, based on the information available in \citetads{2018PhRvL.120b1101A} and its supplemental material. The crucial parameters are the correlation length associated with various systematics, correlating more or less strongly various energy bins. The dominant effect is from the acceptance systematics, and we showed an unphysical dependence of the transport parameters with the acceptance correlation length. After a more careful inspection of the systematics, we discussed the fact that the acceptance systematics is actually a mix of several systematics with very different correlation lengths. Splitting the acceptance systematics in three parts stabilised the dependence of the transport parameters with the correlation length. To fully solve this issue, the publication, along with the data, of the correlation matrix by the AMS-02 collaboration is necessary. This is likely a very difficult task, and waiting for its completion, further informations on the various systematics, further splits and indications on each of the systematics correlation length would already be extremely useful. Indeed, not only does it possibly biases the transport parameters fit to the data, but it also has a huge impact on the statistical interpretation of the model inspected: depending on the correlation length assumed, we can either conclude on a perfect fit to the data $\chi^2_{\rm min}/{\rm dof}\sim 1$ or exclude the model ($\chi^2_{\rm min}/{\rm dof}\gtrsim 2-3$).

The fourth and last step was to consider a realistic analysis, applying the method developed to handle cross-section uncertainties ($2^{\rm nd}$ step) with the full data uncertainties, that is accounting for the covariance matrix of errors ($3^{\rm rd}$ step). Because of energy correlations in the systematics, the impact of cross-section uncertainties can be lessened. In the context of the analysis of the AMS-02 B/C data, the impact of systematics was found to be dominant over that of cross-section uncertainties. In any case, for any analysis, we recommend to implement the dominant nuclear reactions as nuisance parameters, checking the results against various choices of production cross sections and the two possible strategies for the nuisance parameters (NSS or LC).

Our methodology can be used for any CR species, but the most important cross sections and their uncertainties depend on the species (e.g. \citealtads{2018PhRvC..98c4611G}), so that specific nuisance parameters need to be changed for both the NSS and LC methods. Then, data uncertainties have generally different origins for different species, with different sub-detectors and selections cuts applied. For these reasons, the conclusions that can be drawn concerning the most impacting effect (cross sections or data systematics) can be different from one species to another, and so should be carefully inspected for each species and data considered.

Further results based on this methodology are presented in two companion papers: \citetads{2019arXiv190408917G} present the interpretation of the B/C ratio data and constraints on the transport parameters. \citetads{2019arXiv190607119B} use these transport parameters and their uncertainties to calculate the astrophysical flux of antiprotons. We emphasise that all the results presented here and in the companion papers are based on \usine{} v3.5 \citepads{2018arXiv180702968M}, available at \url{https://lpsc.in2p3.fr/usine}\footnote{The current release is v3.4, but v3.5, specifically developed for this analysis, will be online soon.}.

\begin{acknowledgements}
We thank our `cosmic-ray' colleagues at LAPTh, LAPP, LUPM, and Universidade de S{\~a}o Paulo, and in particular P. Serpico for stimulating discussions and feedback that helped us clarify and improve the presentation of the methodology presented in this paper. This work has been supported by the `Investissements d'avenir, Labex ENIGMASS' and by the French ANR, Project DMAstro-LHC, ANR-12-BS05-0006. The work of Y.G. is supported by the IISN, the FNRS-FRS and a ULB ARC. The work of M.B. is supported by the European Research Council ({ERC}) under the EU Seventh Framework Programme (FP7/2007-2013) / {ERC} Starting Grant (Agreement No 278234 -- {"NewDark"} project). 
\end{acknowledgements}

\appendix

\section{Systematics from $R$ to $E_{k/n}$ approximate conversion}
\label{app:RtoEkn}

CR data are mostly published and analysed as a function of the kinetic energy per nucleon (see, e.g. the data collected in the Cosmic-Ray Data Base\footnote{\url{https://lpsc.in2p3.fr/crdb/}}, \citealtads{2014A&A...569A..32M}). The latter quantity is conserved in nuclear reactions (in the straight-ahead approximation) and propagation codes usually solve the transport equation per $E_{k/n}$. However, it is not the quantity CR detectors measure; for instance, hadronic calorimeters provide the total energy, whereas spectrometers like AMS-02 provide the rigidity. Conversion from one energy unit to another is only exact if the nucleus ($m$, $A$, $Z$) is identified. For elements, unless the isotopic content is known, the conversion is approximate.

The uncertainty brought from energy unit conversions was neglected in the past because of larger uncertainties, but this is no longer possible for modern data. For instance, the conversion from $R$ to $E_{k/n}$ in an experimental context in which only elemental fluxes are measured is discussed by the  PAMELA collaboration in App.~B of \citetads{2014ApJ...791...93A}, and by the AMS collaboration for the B/C ratio in the Supplemental Material of \citetads{2016PhRvL.117w1102A}. As the practice remains in the field to fit data as a function of $E_{k/n}$, we argue below that this is not a good procedure.

\begin{figure}[t]
\begin{center}
\includegraphics[width=\columnwidth]{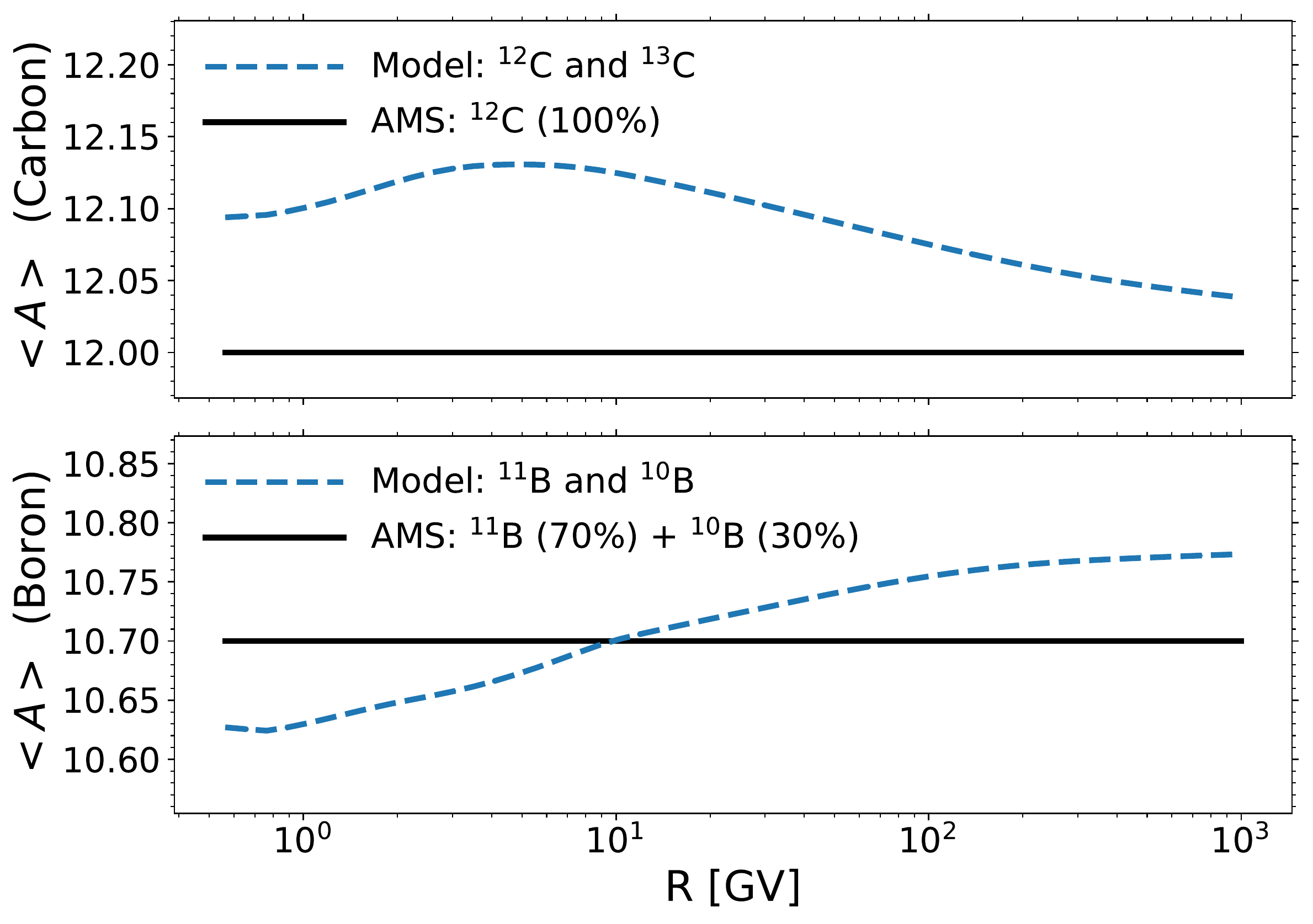}
\caption{Average mass number for carbon (top) and Boron (bottom) as a function of rigidity. The dashed blue lines correspond to the value calculated from the model (with the isotopic content known), whereas the solid black lines correspond to the choice made in \citetads{2016PhRvL.117w1102A}. See text for details.}
\label{fig:RvsEkn_Amean}
\end{center}
\end{figure}

To convert B/C data from $R$ to $E_{k/n}$, the AMS-02 collaboration relies on an average mass number,
\[
   \langle A \rangle_Z = \displaystyle \frac{\sum_{i\in Z} (A_i F_i)}{F_Z},
\]
of 12 for Carbon and 10.7 for Boron. Figure~\ref{fig:RvsEkn_Amean} shows these values compared to the theoretical calculation from a model reproducing B/C data. In the top panel, the varying $\langle A \rangle_C$ for the model results from the fact that Carbon contains mostly primary $^{12}$C and a small fraction of secondary $^{13}$C ($\lesssim15\%$ at $\lesssim 1$~GV and steadily vanishing at higher rigidities). For the Boron (bottom panel), both $^{10}$B and $^{11}$B are of secondary origin, but their rigidity evolution is related to two subtle effects: (i) $\approx 15\%$ of $^{10}$B comes from the decay of $^{10}$Be \citepads{2018PhRvC..98c4611G}, and as the effective lifetime increases with energy, the fraction of $^{10}$B  with respect to that of $^{11}$B decreases with rigidity; (ii) $^{10}$B has a larger fraction originating from 2-step reactions (w.r.t. direct `1-step' production) than $^{11}$B has \citepads[see Table~1 and Fig.~3 of][]{2018PhRvC..98c4611G}, and as 2-step reactions have a steeper rigidity dependence than 1-step ones, again the fraction of $^{10}$B  decreases with rigidity (w.r.t. that of $^{11}$B).

\begin{figure}[t]
\begin{center}
\includegraphics[width=\columnwidth]{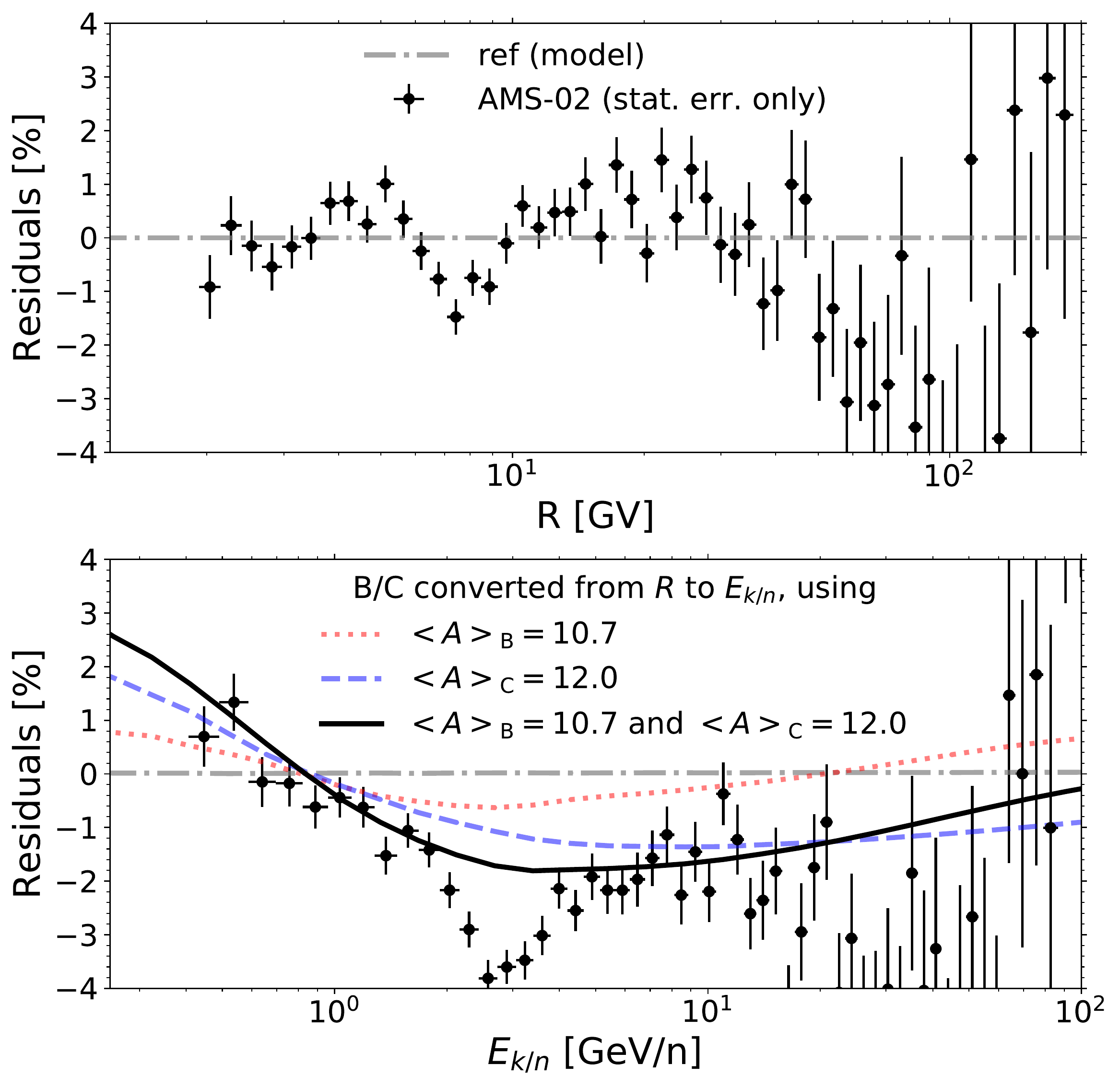}
\caption{{\em Top panel}: B/C residual (vs $R$) calculated from the difference between AMS-02 data and our reference model (best-fit model). {\em Bottom panel}: B/C residual (vs $E_{k/n}$) calculated from the difference between various $R$-to-$E_{k/n}$-converted B/C (red, blue and black lines) and the `exact' B/C (reference model converted with correct isotopic content). The AMS-02 data vs $E_{k/n}$ are taken from \citetads{2016PhRvL.117w1102A}, and result from the same conversion as done for the solid black line. See text for discussion}
\label{fig:RvsEkn_convert}
\end{center}
\end{figure}
In Fig.~\ref{fig:RvsEkn_convert}, we show the impact of these two different choices when converting B/C from $R$ to $E_{k/n}$. The top panel shows the residual of B/C data vs $R$ (w.r.t. our best-fit model), in order to give a visual reference for the difference between the model and data. The bottom panel shows residuals of B/C vs $E_{k/n}$ for different conversions (w.r.t. our best-fit model converted without approximation). The black curve shows the conversion bias when assuming a constant $\langle A \rangle$ for Boron and Carbon, where part of the bias comes from assuming $\langle A \rangle_B=10.7$ (red dotted line), and part from assuming $\langle A \rangle_C=12$ (blue dashed line): the bias is positive and the largest at the lowest $E_{k/n}$ ($\sim 3\%$) and null at 1~GeV/n; it is negative, reaches $\sim 2\%$ at $\sim 4$~GeV/n, and then decreases. The pattern of AMS-02 data (converted using the same approximation) with respect to the black solid line is similar to the one seen in the top panel, indicating that the origin of the discrepancy with the exact model calculation is the wrong assumption made for $\langle A \rangle$.

We can summarise the above subtle discussion as follows: if we fit a model on AMS-02 B/C data as a function of rigidity, this model will be offset from the converted AMS-02 B/C data as a function of $E_{k/n}$. The offset is not a simple scaling, it is instead energy dependent because the isotopic content of B and C elements is energy dependent (in a non-trivial way). Whereas the maximum bias is `only' $\sim 3\%$, this is already significant compared to other AMS-02 uncertainties. Moreover, in AMS publication, the uncertainty associated with the conversion is estimated to range from $1\%$ (lowest energy) to $4\%$ (highest energy), which does not reflect our results. For all these reasons, we conclude that AMS-02 data, and in general all data, should be fit in their native energy scale in order to avoid a non-necessary bias introduced by converting the data to another energy scale. We stress that the \usine{} package allows one to fit any combination of data in their native energy scale.

\section{$\chi^2$ with covariance or nuisance}
\label{app:chi2}

To characterise the impact of the uncertainties on the model parameters, we rely on the $\chi^2$ analysis implemented in \usine{} and described in \citetads{2018arXiv180702968M}, using the \minuit{} package \citepads{1975CoPhC..10..343J} for minimisation. In particular, the \minos{} option in \minuit{} allows to reliably reconstruct asymmetric error bars on the parameters, taking into account both parameter correlations and non-linearities. The generic form of the $\chi^2$ we use is
\begin{equation}
  \chi^2 = \sum_{t}\left( \sum_{q} \left( {\cal D}_{\rm cov}^{t,q} + {\cal N}^{t,q}\right) + {\cal N}^t\right) + {\cal N},
  \label{eq:chi2}
\end{equation}
where we loop over all time periods $t$ (corresponding to different modulation levels) and all quantities $q$ selected in the minimisation. The quantities ${\cal D}_{\rm cov}$ and ${\cal N}$ are detailed below.

\paragraph{Covariance}
The quadratic distance ${\cal D}_{\rm cov}$ measures the distance between the data and the model, accounting for a covariance matrix ${\cal C}$,\footnote{If several systematics $\alpha$ are present, the global covariance matrix is given by the sum of all associated covariance matrices, ${\cal C} = \sum_{\alpha}{\cal C}^\alpha$.}
\begin{equation}
 {\cal D}_{\rm cov} \!=\! \sum_{i,j=1}^{n_E,n_E}({\rm data}_i-{\rm model}_i) \; ({\cal C}^{-1})_{ij} \; ({\rm data}_j-{\rm model}_j),
     \label{eq:cov}
\end{equation}
which correlates ${ij}$ energy bins ($n_E$ bins in total). Without correlations, ${\cal C}$ is diagonal (systematic errors $\sigma_k$ on data), and we recover the standard expression
\begin{equation}
  {\cal D}_{\rm no-cov} = \sum_{k=1}^{n_E} \frac{\left({\rm data}_k-{\rm model}_k\right)^2}{\sigma_k^2} {\rm \quad (no~covariance)}.
\end{equation}

\paragraph{Covariance from relative errors}

As discussed in Sect.~\ref{sec:cov}, we built for each AMS-02 systematics $\alpha$ the covariance matrix of relative errors ${\cal C}_{\rm rel}^\alpha$. The latter can be related to the covariance ${\cal C}_\alpha$ require for Eq.~(\ref{eq:cov}) in two different ways:
\begin{eqnarray}
({\cal C}_{\rm model}^\alpha)_{ij} &=& ({\cal C}^\alpha_{\rm rel})_{ij} \times {\rm model}_i \times {\rm model_j}\,, \label{eq:Cmodel}\\
({\cal C}_{\rm data}^\alpha)_{ij} &=& ({\cal C}^\alpha_{\rm rel})_{ij} \times {\rm data}_i \times {\rm data_j}\,. \label{eq:Cdata}
\end{eqnarray}
While using ${\cal C}_{\rm data}$ may seem more natural, \citet{Blobel2006} showed that if an overall normalisation factor is present in the data, including it in the fit should be done via a factor in the model, not in the data; otherwise, the reconstructed model parameters are biased (see also \citealtads{1994NIMPA.346..306D}). For this reason we decided to use  ${\cal C}_{\rm model}$ in our analysis. We note however that a global normalisation factor corresponds to a situation in which the correlation length is infinite, which is not the case for
the data we consider (see Sect.~\ref{sec:cov}). To ensure that either using Eq.~(\ref{eq:Cmodel}) or (\ref{eq:Cdata}) does not affect our conclusions, we fit 10\,000 mock B/C data under these two assumptions, and checked that (i) the input model parameters were recovered in both cases, and (ii) up to the level of precision reached, potential biases were much smaller than the $1\sigma$ uncertainties on the reconstructed parameters.

\paragraph{Nuisance parameters}
Nuisance parameters are parameters contingent to the analysis performed, but whose value can affect the result of the analysis. An example is given by CR cross sections, that are instrumental for the model calculation, but whose values and uncertainties were determined by `external' experiments. Nuisance parameters can appear at various levels of the modelling: (i) global nuisance parameters ${\cal N}$ related to the model and thus independent of the data (e.g. cross sections), (ii) time-dependent nuisance parameters ${\cal N}^t$ (e.g. modulation parameter for a specific data-taking period), (iii) data-dependent nuisance parameters ${\cal N}^{t,q}$ (e.g. systematic errors on data as an alternative to using a covariance matrix).

In principle, any probability distribution function is possible, and it is determined from the auxiliary experiment. However, in \usine{}, only Gaussian-distributed nuisance parameters are enabled, so that each adds in the $\chi^2 $, Eq.~(\ref{eq:chi2}), a contribution
\begin{equation}
   {\cal N} = \frac{(y-\bar{y})^2}{\sigma_y^2},
   \label{eq:chi2_nuis}
\end{equation}
where $\bar{y}$ and $\sigma_y^2$ are the mean and variance of the parameter, and $y$ the tested value.

\section{Coefficients for boundary conditions}
\label{app:boundary}

\begin{table*}
\centering
\caption{Matrix coefficients for several boundary conditions of the transport equation. Our prescriptions are highlighted in boldface. {\bf Top}: first matrix coefficients $a_0$ and $b_0$ (Eqs.~\ref{eq:coeff_a} and \ref{eq:coeff_b}) for low-energy boundary conditions (prescription {\bf L3}). {\bf Bottom}: last matrix coefficients $b_K$ and $c_K$ (Eqs.~\ref{eq:coeff_b} and \ref{eq:coeff_c}) for high-energy boundary conditions (prescription {\bf H4}).}
\label{tab:coeff}
\begin{tabular}{l c c }
\hline
\hline
Low-energy conditions & $b_0$  & $c_0$ \\
\hline
\\[-0.2cm]
\#L1: No energy flow $J_E=0$
& $1+\displaystyle\frac{\alpha_0}{\Delta x}\left(\displaystyle\frac{\beta_{0}}{2}+\displaystyle\frac{\gamma_{\frac{1}{2}}}{\Delta x}\right)$
& $\displaystyle\frac{\alpha_0}{\Delta x}\left(\displaystyle\frac{\beta_{1}}{2}-\displaystyle\frac{\gamma_{\frac{1}{2}}}{\Delta x} \right)$
\\[0.5cm]
\#L2: $\left. {\partial^2 u}/{\partial x^2} \right|_{\displaystyle x_{0}}=0$  (1st order)
& $1+\displaystyle\frac{\alpha_0}{\Delta x}\left\{-\beta_{0}+\left(\displaystyle\frac{\gamma_{\frac{1}{2}}-\gamma_{-\frac{1}{2}}}{\Delta x}\right)\right\}$
& $\displaystyle\frac{\alpha_0}{\Delta x}\left\{\beta_{1}-\left(\displaystyle\frac{\gamma_{\frac{1}{2}}-\gamma_{-\frac{1}{2}}}{\Delta x}\right)\right\}$
\\[0.5cm]
{\bf \#L3: $\left. {\partial^2 u}/{\partial x^2} \right|_{\displaystyle x_{0}}=0$ (2nd order)}
& $1+\displaystyle\frac{\alpha_0}{\Delta x}\left\{-\beta_{-1}+\left(\displaystyle\frac{\gamma_{\frac{1}{2}}-\gamma_{-\frac{1}{2}}}{\Delta x}\right)\right\}$
& $\displaystyle\frac{\alpha_0}{\Delta x}\left\{\left(\frac{\beta_{1}+\beta_{-1}}{2}\right)-\left(\displaystyle\frac{\gamma_{\frac{1}{2}}-\gamma_{-\frac{1}{2}}}{\Delta x}\right)\right\}$
\\[0.5cm]
\#L4: $\left. {\partial f}/{\partial p\,} \right|_{\displaystyle x_{0}}=0$
& $1+\displaystyle\frac{\alpha_0}{\Delta x}\left\{\beta_{-1}\Delta x\,\delta_0+\left(\displaystyle\frac{\gamma_{\frac{1}{2}}+\gamma_{-\frac{1}{2}}+\gamma_{-\frac{1}{2}}2\Delta x\,\delta_0}{\Delta x}\right)\right\}$
& $\displaystyle\frac{\alpha_0}{\Delta x}\left\{\left(\frac{\beta_{1}-\beta_{-1}}{2}\right)-\left(\displaystyle\frac{\gamma_{\frac{1}{2}}+\gamma_{-\frac{1}{2}}}{\Delta x}\right)\right\}$
\\[0.5cm]
\hline\hline
High-energy conditions & $a_K$ & $b_K$ \\
\hline
\\[-0.2cm]
\#H1: No energy flow $J_E=0$
& $\displaystyle-\frac{\alpha_K}{\Delta x}\left(\displaystyle\frac{\beta_{K-1}}{2}+\displaystyle\frac{\gamma_{K-\frac{1}{2}}}{\Delta x} \right)$
& $1+\displaystyle\frac{\alpha_K}{\Delta x}\left(\displaystyle-\frac{\beta_{K}}{2}+\displaystyle\frac{\gamma_{K-\frac{1}{2}}}{\Delta x}\right)$
\\[0.5cm]
\#H2: $\left. {\partial^2 u}/{\partial x^2} \right|_{\displaystyle x_{K}}\!=0$  (1st order)
& $\displaystyle\frac{\alpha_K}{\Delta x}
\left\{-\beta_{K-1}+\left(\displaystyle\frac{\gamma_{K+\frac{1}{2}}-\gamma_{K-\frac{1}{2}}}{\Delta x}\right)\right\}$
& $1+\displaystyle\frac{\alpha_K}{\Delta x}
\left\{\beta_{K}-\left(\displaystyle\frac{\gamma_{K+\frac{1}{2}}-\gamma_{K-\frac{1}{2}}}{\Delta x}\right)\right\}$
\\[0.5cm]
\#H3: $\left. {\partial^2 u}/{\partial x^2} \right|_{\displaystyle x_{K}}\!=0$ (2nd order)
& $\displaystyle\frac{\alpha_K}{\Delta x}\left\{-\left(\frac{\beta_{K+1}+\beta_{K-1}}{2}\right)+\left(\displaystyle\frac{\gamma_{K+\frac{1}{2}}-\gamma_{K-\frac{1}{2}}}{\Delta x}\right)\right\}$
& $1+\displaystyle\frac{\alpha_K}{\Delta x}\left\{\beta_{K+1}-\left(\displaystyle\frac{\gamma_{K+\frac{1}{2}}-\gamma_{K-\frac{1}{2}}}{\Delta x}\right)\right\}$
\\[0.5cm]
{\bf \#H4: Pure diffusive limit} $u=u^{0}$
& 0
& 1
\\[0.1cm]
\hline
\end{tabular}
\end{table*}

The discretisation of Eq.~(\ref{eq:crank_0}) on a grid over $x \equiv \ln E_{k/n}$ gives
\beq
 u_k+\frac{\alpha_k}{\Delta x}\left(J_{k+\frac{1}{2}}-J_{k-\frac{1}{2}}\right)=u^{0}_k\;,
\label{eq:crank_1}
\eeq
with the current $J_{k+\frac{1}{2}}$ defined to be
\beq
J_{k+\frac{1}{2}}=\frac{1}{2}\left(_{\,}\beta_{k+1}u_{k+1}+\beta_{k}u_{k\,}\right)-\frac{\gamma_{k+\frac{1}{2}}}{\Delta x}\left(_{\,}u_{k+1}-u_{k\,}\right)\;.\label{eq:current}
\eeq
This equation can readily be written as a matrix equation
\beq
\mathbb{M}\,\mathbb{U}=\mathbb{U}^{0}, \quad \text{with}\quad \mathbb{U}=\begin{pmatrix} u_0 \\ \vdots \\ u_k \\ \vdots\\ u_K \end{pmatrix}\quad ,
\label{eq:matrix_MU_U0}
\eeq
\beq
\text{and}\quad \mathbb{M}=\begin{bmatrix} b_0 & c_0 &  &  &  &  \\ a_1 & b_1 & c_1 &  &  & \\  & \ddots & \ddots & \ddots & & \\ & & \ddots & \ddots & \ddots &   \\  &  & & a_{K-1} & b_{K-1} &c_{K-1} \\  &  &  &  & a_K & b_K\end{bmatrix}\;.
\label{eq:def_matrix_M}
\eeq
$\mathbb{M}$ is a tridiagonal matrix defined by its coefficients
\begin{align}
&a_k=\frac{\alpha_k}{\Delta x}\left(-\frac{\beta_{k-1}}{2}-\frac{\gamma_{k-\frac{1}{2}}}{\Delta x} \right)\;,\label{eq:coeff_a}\\ \nonumber\\
&b_k=1+\frac{\alpha_k}{\Delta x^2}\left(\gamma_{k+\frac{1}{2}}+\gamma_{k-\frac{1}{2}}\right)\;,\label{eq:coeff_b}\\ \nonumber\\
&c_k=\frac{\alpha_k}{\Delta x}\left(\frac{\beta_{k+1}}{2}-\frac{\gamma_{k+\frac{1}{2}}}{\Delta x} \right)\;.\label{eq:coeff_c}
\end{align}
Solving this system requires to fix the boundary conditions. Several possibilities have been tested in Appendix~\ref{app:stability}.

As regards the low-energy boundary at $x_{\rm min} \equiv x_{0}$,  we list several suitable prescriptions, and report results for the corresponding $b_0$ and $c_0$ in the half-upper part of Table~\ref{tab:coeff}.
\begin{itemize}
\item \textit{No energy flow (L1)}.
The condition $J_E=0$ at $x_{0}$ translates into $J_{-\frac{1}{2}}=0$ with the defined grid steps.
\item \textit{No curvature in the spectrum (L2 and L3)}.
Using the prescription of \cite{leveque1998finite} for a second order accurate method yields
\beq
\left.\frac{\partial^2 u}{\partial x^2}\right|_{x_{0}} =
\displaystyle\frac{\left(\displaystyle\frac{u_1-u_0}{\Delta x}\right) - \left(\displaystyle\frac{u_0-u_{-1}}{\Delta x}\right)}{\Delta x}=0 \;,
\eeq
which implies that $u_{-1}=2_{\,}u_0-u_1$. We compute the coefficients $b_0$ and $c_0$ for a first (L2) and second (L3) order accurate method; the former was used in \citetads{2001ApJ...563..172D}.
\item \textit{No phase space density gradient (L4)}.
The phase space distribution $f$ is flat for a CR momentum $p=0$. We implement this condition at $x_{0}$. Requiring that ${\partial f}/{\partial p}$ vanishes translates into
\beq
\left. {\displaystyle \frac{\partial (u/pE)}{\partial p}} \right|_{x_{0}} = 0 \;.
\eeq
If written in term of $x=\ln{E_{k}}$ it reads
\beq
\left. \frac{\partial u}{\partial x} \right|_{x_{0}} = \left\{\frac{E_{k,0}}{E_0} \left(1+\frac{E_0^2}{p_0^2}\right)\right\} u_0 \equiv \delta_{0\,} u_0 \;.
\eeq
If we discretise this condition, we are led to
\beq
\left. \frac{\partial u}{\partial x} \right|_{x_{0}} = \frac{u_1-u_{-1}}{2 \Delta x} = \delta_{0\,} u_0 \;,
\eeq
and get, according to \citep{leveque1998finite}, a second order accurate method. Injecting this condition into the differentiation scheme let us define $b_0$ and $c_0$.
\end{itemize}

At high-energy (i.e. at the highest point $x_{\rm max} \equiv x_{K}$ of the grid), several conditions can be implemented along the same lines. The resulting coefficients $a_K$ and $b_K$ are listed in the half-bottom part of Table~\ref{tab:coeff}.
\begin{itemize}
\item \textit{No energy flow (H1)}.
The condition $J_E=0$ at $x_{K}$ translates into $J_{K+\frac{1}{2}}=0$.
\item \textit{No curvature in the spectrum (H2 and H3)}.
We require that
\beq
\left.\frac{\partial^2 u}{\partial x^2}\right|_{x_{K}} =
\displaystyle\frac{\left(\displaystyle\frac{u_{K+1}-u_{K}}{\Delta x}\right) - \left(\displaystyle\frac{u_{K}-u_{K-1}}{\Delta x}\right)}{\Delta x}=0 \;,
\eeq
which implies that $u_{K+1}=2_{\,}u_{K}-u_{K-1}$.
\item \textit{No energy losses nor diffusive reacceleration  (H4)}.
As discussed in Appendix~\ref{app:stability}, the CR density $u$ is given by $u^{0}$ at high energy insofar as energy losses and diffusive reacceleration do not play any role in this regime. We require that $u(x_{K})=u^0(x_{K})$, hence the condition $a_K=0$ and $b_K=1$ of Table~\ref{tab:coeff}.
\end{itemize}

\section{Stability of the numerical solution}
\label{app:stability}

The numerical solution of Eq.~(\ref{eq:crank_0}) might exhibit instabilities when diffusive reacceleration vanishes with the Alfv\'enic speed $V_a$. To explore the onset of these instabilities and to remedy them, we consider the same 1D geometry as discussed in Sect.~\ref{sec:prop_model}, that is a thin disc of half-thickness $h$ and a magnetic halo of half-thickness $L$. We focus on Model A, presented in Sect.~\ref{sec:model_params}, that is a standard diffusion/convection/reacceleration transport model.
For definitiveness, all figures presented in this Appendix are based on the following values for the transport parameters: $K_{0} = 0.030 \, {\rm kpc^{2} \, Myr^{-1}}$, $\delta = 0.65$, $\eta_t = - 0.49$, $L = 10 \, {\rm kpc}$, $V_{c} = 15.1$ km/s and $V_{a} = 74.6$ km/s. These values are among those that give a good fit to the B/C ratio and as such are sufficient to illustrate our discussion.

\subsection{Simplified 1D model and solutions}

\paragraph{Transport equation}
The CR density in space and energy $u(z,E)$ fulfils the transport equation
\beq
- K_{\,}{\displaystyle \frac{\partial^{2} u}{\partial z^{2}}} +
{\displaystyle \frac{\partial}{\partial z}} \{ V_{c}(z) \, u \} +
{\displaystyle \frac{\partial}{\partial E}}\left\{b_{\,}u - K_{EE\,}{\displaystyle \frac{\partial_{}u}{\partial E}}\right\} =
q_{\rm acc} \;,
\label{eq:master_CR_2}
\eeq
where $z$ is the vertical co-ordinate. The source term $q_{\rm acc}$ denotes the rate with which CRs are injected. CRs diffuse with a spatial diffusion coefficient $K$ that follows Eq.~(\ref{eq:K_std}), are convected away from the disc at velocity $V_{c}$, suffer from energy losses at a rate $b(E) < 0$ (ionisation, Coulomb friction, and adiabatic expansion in the wind), or gain energy via an energy diffusion coefficient\footnote{The following definition, taken from \citetads{2001ApJ...555..585M}, is similar to that of Eq.~(\ref{eq:kpp}) since $K_{EE} = \beta^{2}_{\,}K_{pp}$, albeit with a slightly larger coefficient of $2/9 \sim 0.22$ instead of $0.17$.} $K_{EE} = (2/9) \times V_a^2 \beta^4 E^2/K$. CR injection, as well as energy losses and energy diffusion are localised in the disc. 

From now on, the Galactic disc is treated as in the infinitely thin limit, and all nuclear interactions on the interstellar medium (ISM) have been discarded. Requiring that the CR density vanishes at the boundaries $z = \pm L$ of the magnetic halo allows to straightforwardly express $u(z,E)$ as a function of its value inside the Galactic disc $u(E,0)$.

\begin{figure}[t]
\begin{center}
\includegraphics[width=0.85\columnwidth]{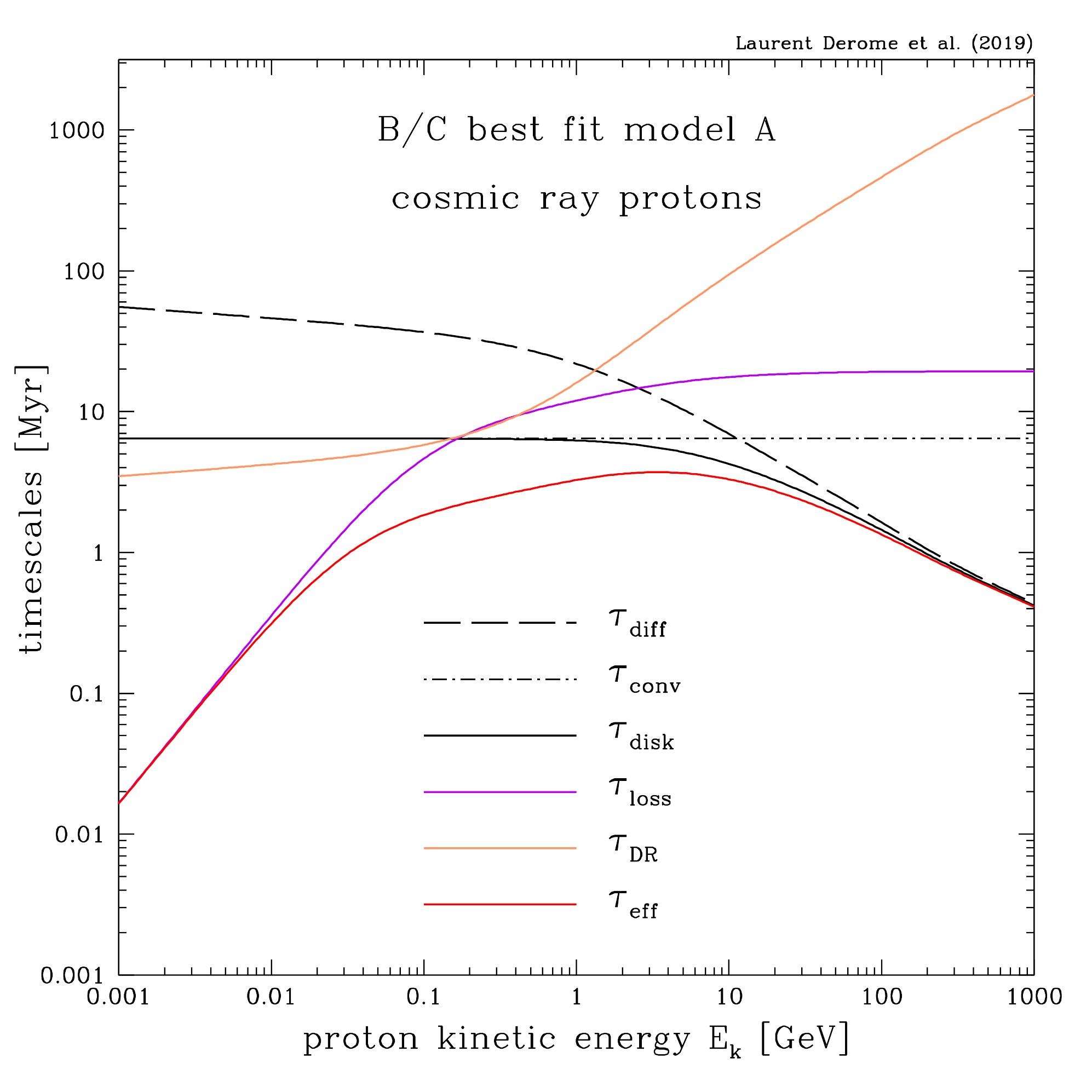}
\caption{The timescales associated to the various processes at play in the Galactic transport of CR protons are plotted as a function of kinetic energy $E_{k}$. The diffusion $\tau_{\rm diff}$ and convection $\tau_{\rm conv}$ timescales are respectively featured by the long dashed and short dashed-dotted black curves. They are combined into the disc residence time $\tau_{\rm disc}$ plotted as the solid black line. The energy loss and diffusive reacceleration timescales correspond to the solid purple and orange curves. The red solid line features the combined timescale $\tau_{\rm eff}$ as given in Eq.~(\ref{eq:tau_eff}).}
\label{fig:taux_H_NEW_model}
\end{center}
\end{figure}

\paragraph{Timescales}
Transport inside the magnetic halo, energy losses and diffusive reacceleration are the three processes at play in the transport of CR nuclei. To determine which of these processes is dominant and inside which energy range it prevails, we can calculate the associated timescales. Galactic diffusion and convection can be combined into the residence time of CRs inside the Galactic disc
\beq
{\tau_{\rm disc}} = {\tau_{\rm conv}} \left\{ 1 - e^{\displaystyle - {\tau_{\rm diff}}/{\tau_{\rm conv}}} \right\} \;,
\eeq
where $\tau_{\rm conv} = {h}/{V_{c}}$ and $\tau_{\rm diff} = {h L}/{K}$ are the convection and diffusion timescales.
For energy losses, we define $\tau_{\rm loss}$ as the ratio $- {E_{k}}/{b(E)}$, where $E_{k}$ is CR kinetic energy.
Diffusive reacceleration occurs over a timescale $\tau_{\rm DR} \equiv {E_{k}^{2}}/{K_{EE}}$.
These timescales are plotted as a function of kinetic energy $E_{k}$ in Fig.~\ref{fig:taux_H_NEW_model}. The behaviour is fairly generic and the same trends appear for heavier nuclei as well as for secondary species.

All these processes can be combined through the effective timescale $\tau_{\rm eff}$ which we define as
\beq
{\displaystyle \frac{1}{\tau_{\rm eff}}} = {\displaystyle \frac{1}{\tau_{\rm disc}}} +
{\displaystyle \frac{1}{\tau_{\rm loss}}} + {\displaystyle \frac{1}{\tau_{\rm DR}}} \;.
\label{eq:tau_eff}
\eeq

\paragraph{High- and low-energy analytical solutions}

We are only interested in the solution in the disc, $u(z=0,E)\equiv u$, and using the above timescales
the transport equation (\ref{eq:master_CR_2}) boils down into the PDE
\beq
{\displaystyle \frac{u}{\tau_{\rm disc}}} +
{\displaystyle \frac{d}{d E}}\!\left\{b_{}u - K_{EE}{\displaystyle \frac{d{}u}{d E}}\right\} =
q_{\rm acc} \;.
\label{eq:master_CR_3}
\eeq

\begin{itemize}
   \item High-energy limit: space diffusion dominates over the other processes. As energy losses and diffusive reacceleration do not play any role in this regime, the solution to the CR transport equation~(\ref{eq:master_CR_3}) is
      \beq
         u_{\rm diff}(E_{k}) \equiv u^{0} = \tau_{\rm disc} \, q_{\rm acc}.
      \eeq

   \item Low-energy limit: energy losses dominate and diffusive reacceleration can be neglected. The transport equation~(\ref{eq:master_CR_3}) has an analytic solution which can be cast into the form
      \begin{eqnarray}
         u_{\rm loss}(E_{k}) \!\!\! & \!\! = \!\!\!\! &
         \left\{ {\displaystyle \frac{b(E_{k,{\rm max}})}{b(E_{k})}} \right\} u^{0}(E_{k,{\rm max}}) \, e^{\displaystyle - \tilde{t}(E_{k})} \nonumber \\
         & \!\! + \!\!\!\!&
         {\displaystyle \frac{1}{|b(E_{k})|}}
         {\displaystyle \int_{E_{k}}^{E_{k,{\rm max}}}} \!\!\!\!\! q_{\rm acc}(E_{k}') \;
         e^{\displaystyle - \{ \tilde{t}(E_{k}) - \tilde{t}(E_{k}') \}} \, dE_{k}' \;,
         \label{eq:psi_loss_2}
      \end{eqnarray}
   where the pseudo-time $\tilde{t}$ is defined as
      \beq
         \tilde{t}(E_{k}) =
         {\displaystyle \int_{E_{k}}^{E_{k,{\rm max}}}} \,
         {\displaystyle \frac{\tau_{\rm loss}}{\tau_{\rm disc}}} \; {\displaystyle \frac{dE_{k}'}{E_{k}'}} \;.
      \eeq
   For above-mentioned reasons, the analytic solutions $u^{0}$ and $u_{\rm loss}$ become equal in the high-energy limit. We have set them equal at the highest energy point $E_{k,{\rm max}}$ of our analysis, for which a value of $10^{3}$ GeV is assumed.
\end{itemize}
These solutions are used below to check the precision of the numerical calculation.

\paragraph{Numerical solution}

The numerical solution of Eq.~(\ref{eq:master_CR_2}) is obtained as follows. First, the equation can be recast into Eq.~(\ref{eq:crank_0}), where the coefficients $\alpha$, $\beta$, and $\gamma$, of our simplified model are
\beq
\alpha = {\displaystyle \frac{\tau_{\rm disc}}{E_{k}}}
\;,\;\;
\beta = {\displaystyle \frac{- E_{k}}{\tau_{\rm loss}}} = b(E)
\;\;{\rm and}\;\;
\gamma = {\displaystyle \frac{E_{k}}{\tau_{\rm DR}}} = {\displaystyle \frac{K_{EE}}{E_{k}}} \;.
\eeq
As explained in Appendix~\ref{app:boundary}, this Eq.~(\ref{eq:crank_0}) can be put on a grid in $x\equiv \ln E_{k/n}$ to numerically solve the equation. The solution depends on the boundary conditions implemented, the impact of which is studied below.

\subsection{Impact of boundary conditions} 
\begin{figure}[t]
\begin{center}
\includegraphics[width=0.85\columnwidth]{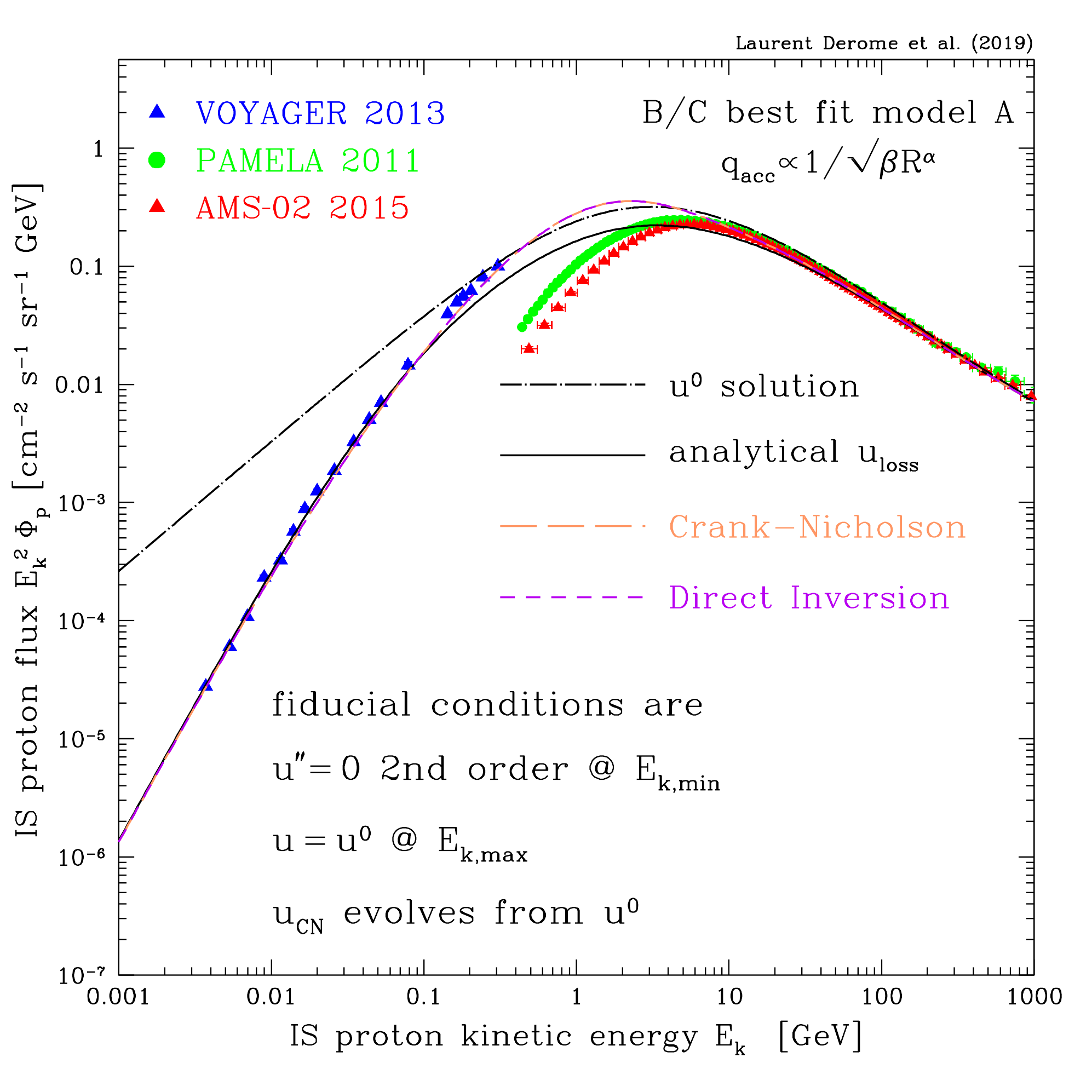}
\caption{The CR proton flux $\Phi_{\rm p}$ is plotted as a function of kinetic energy $E_{k}$. The long dashed-dotted and solid black curves stand respectively for the approximations $u^{0}$ and $u_{\rm loss}$. The exact solution $u$ is derived numerically assuming boundary conditions L3 and H4. The long dashed orange and short dashed purple curves correspond to different methods used to solve transport Eq.~(\ref{eq:crank_0}).
Protons are injected with a rate $q_{\rm acc} = {N_{\rm p}}/{\sqrt{\beta}\,R^{\alpha}}$ where $N_{\rm p} = 5.8 \times 10^{-11}$ protons cm$^{-3}$ GeV$^{-1}$ Myr$^{-1}$ and $\alpha = 2.3$. With these values, we get $\Phi_{\rm p}$ in rough agreement with the Voyager~1 \citepads{2013Sci...341..150S}, PAMELA \citepads{2011Sci...332...69A} and AMS-02 \citepads{2015PhRvL.114q1103A} data.}
\label{fig:fiducial_case}
\end{center}
\end{figure}

Various boundary conditions can be implemented at the lowest $x_{\rm min} \equiv x_{0}$ and highest $x_{\rm max} \equiv x_{K}$ energy points of the x-grid as shown in Table~\ref{tab:coeff}.

\paragraph{Reference boundary conditions vs  analytical solutions:}
At $x_{\rm max}$, the four boundary conditions H of Table~\ref{tab:coeff} yield the same CR proton flux and we always find that $u$ is close to $u^{0}$ above a few tens of GeV. We have decided to implement prescription H4, which is the most natural condition given that $\tau_{\rm eff} \simeq \tau_{\rm disc}$ at high energy.
At $x_{\rm min}$, the prescription that yields the most stable behaviour is L3 with ${\partial^{2}u}/{\partial x^{2}}= 0$ up to second order. Our fiducial conditions are therefore L3 and H4.

For illustrative purpose, Fig.~\ref{fig:fiducial_case} shows a comparison of the analytical and numerical solutions\footnote{The source term has been set proportional to ${1}/{\sqrt{\beta}\,R^{\alpha}}$ to grossly match the proton data. Notice that we have not performed a fit since Fig.~\ref{fig:fiducial_case} is meant to be an illustration of how the various solutions $u^{0}$, $u_{\rm loss}$ and $u$ behave with kinetic energy $E_{k}$.}.
The long dashed-dotted black curve stands for the approximation $u^{0}$, for which energy losses and diffusive reacceleration are switched off. The solid black curve features the solution $u_{\rm loss}$,
where diffusive reacceleration alone is suppressed.
The exact solution $u$ is featured by the short dashed purple and long dashed orange curves. The former is obtained through the direct inversion of Eq.~(\ref{eq:matrix_MU_U0}) while for the latter, a Crank-Nicholson recursion is used to get $u_{\rm CN}$ converging from $u^{0}$ to the exact solution $u$. Both results agree with a precision better than $10^{-10}$.

\begin{figure}[t]
\begin{center}
\includegraphics[width=0.85\columnwidth]{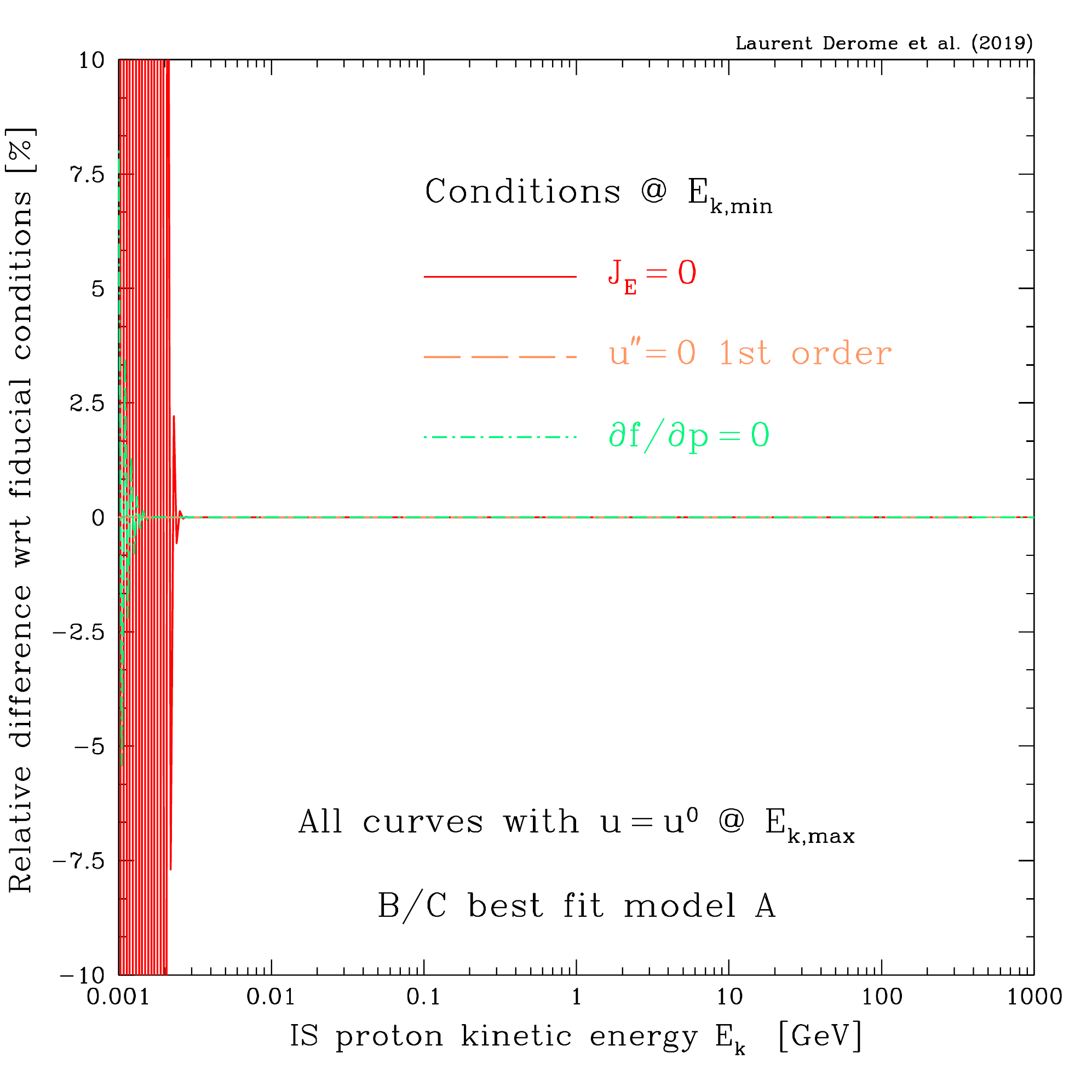}
\caption{Changing the low-energy boundary conditions of Table~\ref{tab:coeff} modifies the numerical result obtained for $u$. In this plot, the variations of the proton flux relative to the fiducial case of Fig.~\ref{fig:fiducial_case} are displayed as a function of kinetic energy $E_{k}$. Notice that all conditions yield the same flux above 3~MeV. Close to the boundary, prescriptions L1 and L4 generate wiggles and the flux becomes inaccurate. Prescriptions L2 and L3 (fiducial) yield the same result.}
\label{fig:COMP_wrt_fiducial_HEBC_0}
\end{center}
\end{figure}

\paragraph{Varying low-energy boundary conditions:}
In Fig.~\ref{fig:COMP_wrt_fiducial_HEBC_0}, we plot the relative difference induced on the fiducial flux of Fig.~\ref{fig:fiducial_case} when prescription L3 is respectively replaced by conditions L1 (solid red), L2 (long dashed orange) and L4 (short dashed-dotted green) of Table~\ref{tab:coeff}. Above 3~MeV, all fluxes agree up to double precision. Below that energy, some differences appear. Condition L2 always yields a flux that agrees with the fiducial result with a precision better than $10^{-4}$. Condition L4 is associated with moderate wiggles, with a relative difference that nevertheless reaches 10\% at 1~MeV. The worst prescription is L1 which generates very large instabilities exceeding 100\% below 2~MeV. It is remarkable that in spite of these, the fiducial result is obtained above 3~MeV.

\begin{figure}[t]
\begin{center}
\includegraphics[width=0.85\columnwidth]{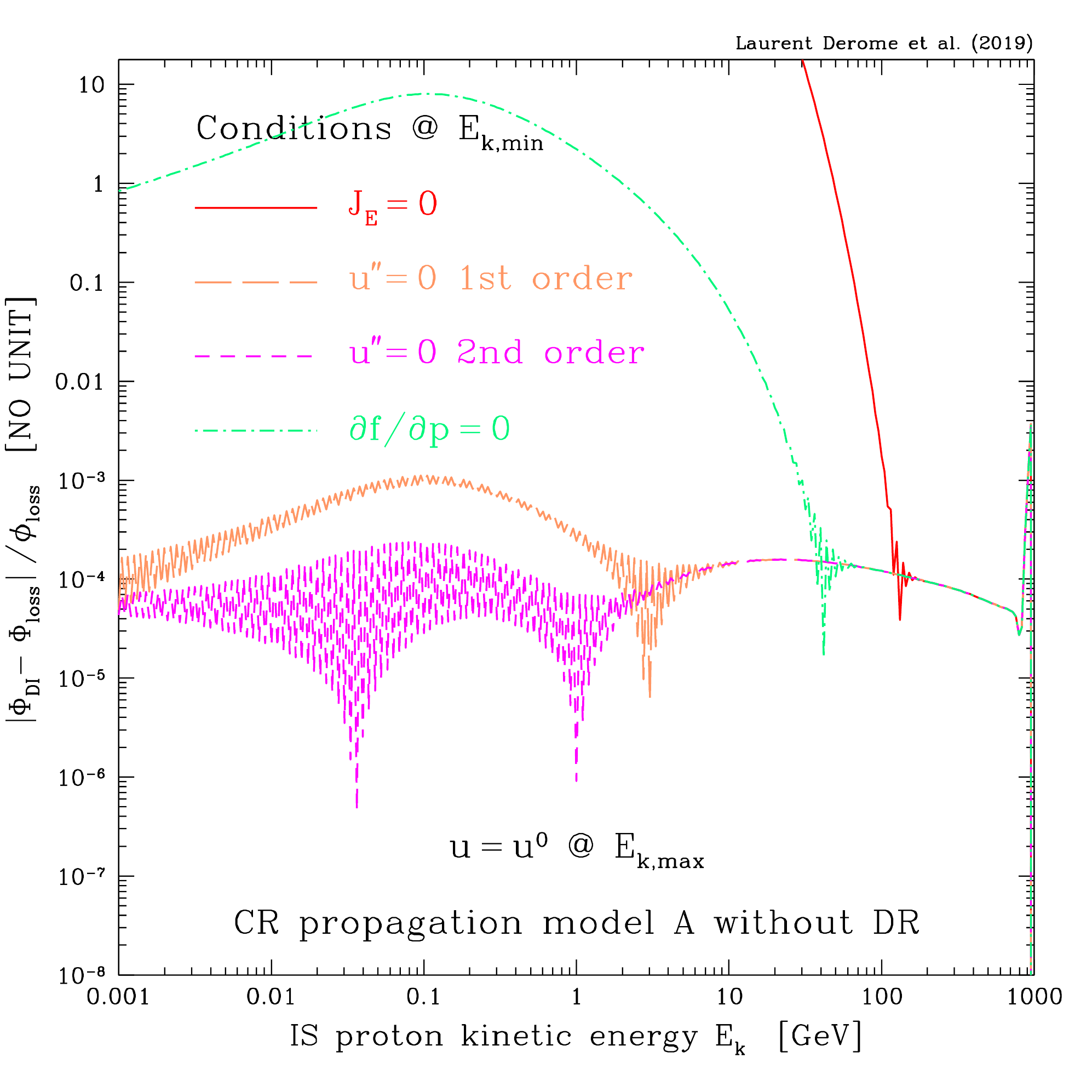}
\caption{The cosmic-ray transport parameters of Model~A are used, except for the Alfv\'enic speed $V_a$ set equal to $1$~m/s. In this regime, the numerical result $u$ of Eq.~(\ref{eq:master_CR_3}) is equal to the analytic solution $u_{\rm loss}$. The relative difference between $u$ and $u_{\rm loss}$ is plotted as a function of kinetic energy $E_{k}$.
The low-energy boundary prescriptions of Table~\ref{tab:coeff} are respectively featured by the solid red (L1), long dashed orange (L2), short dashed purple (L3) and short dashed-dotted green (L4) curves. The most precise condition is L3 while the worst one is L1.
}
\label{fig:model_a_no_DR_COMP_HEBC_0_log}
\end{center}
\end{figure}

In the regime where diffusive reacceleration is switched off, the numerical solution $u$ of Eq.~(\ref{eq:master_CR_3}) is given by the analytic solution $u_{\rm loss}$ displayed in relation~(\ref{eq:psi_loss_2}). This situation offers a unique opportunity to investigate how low-energy boundary conditions affect the stability and precision of the numerical solution. To this purpose, we have used the cosmic-ray transport parameters of Model~A with the exception of a vanishing Alfv\'enic speed $V_a$.
The relative difference between $u$ and $u_{\rm loss}$ is plotted as a function of proton kinetic energy in Fig.~\ref{fig:model_a_no_DR_COMP_HEBC_0_log}. The numerical solution is derived by direct inversion of Eq.~(\ref{eq:matrix_MU_U0}).
Depending on the prescription used at $x_{\rm min}$, $u$ can be very close to the actual result $u_{\rm loss}$ or completely out of range.
As featured by the short dashed purple curve, the most precise condition is L3 with a level of precision of $10^{-4}$. Condition L2 yields also a very accurate solution $u$ with a relative error of at most $10^{-3}$ at 100~MeV.
As could have been anticipated from Fig.~(\ref{fig:COMP_wrt_fiducial_HEBC_0}), setting $J_{E} = 0$ yields the worst numerical result which is orders of magnitude larger than the correct solution below 30~GeV.
Finally, condition  L4 only yields the correct result above 70~GeV. We find that the relative difference between $u$ and $u_{\rm loss}$ even exceeds 100\% below 2~GeV, as exhibited by the short dashed-dotted green curve of Fig.~\ref{fig:model_a_no_DR_COMP_HEBC_0_log}.

\begin{figure}[t]
\begin{center}
\includegraphics[width=0.85\columnwidth]{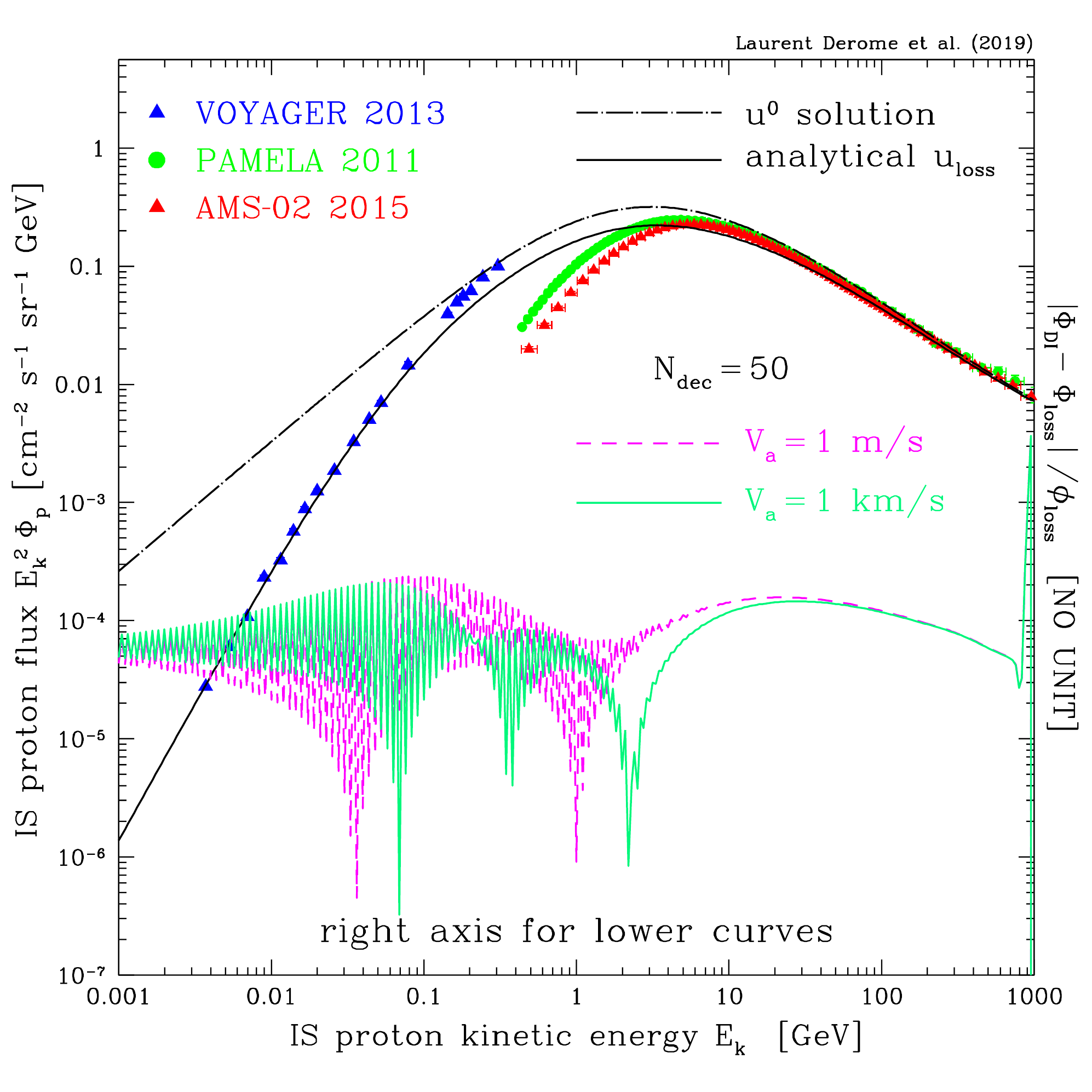}
\vskip -0.3cm
\includegraphics[width=0.85\columnwidth]{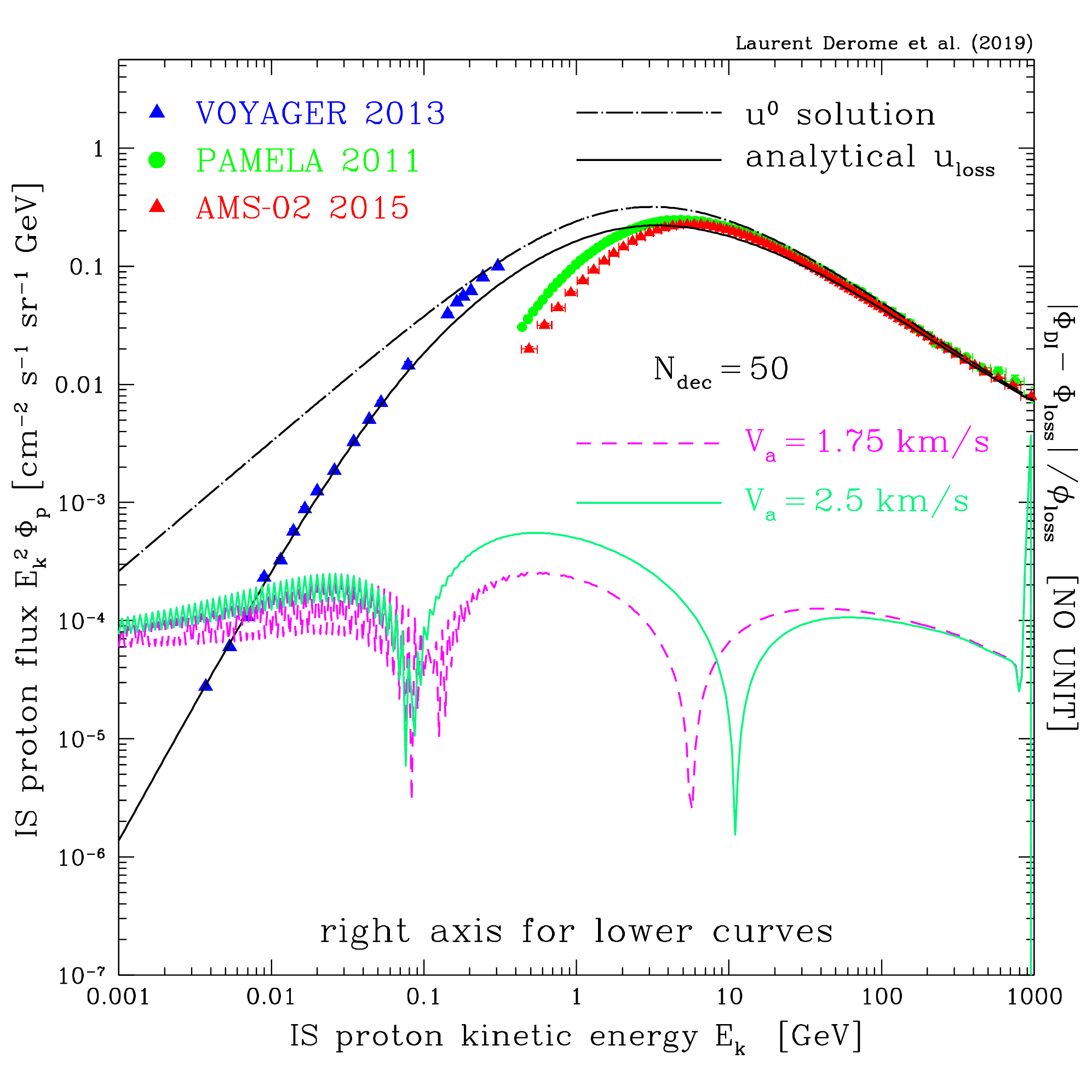}
\caption{In these two panels, the proton flux is plotted as in Fig.~\ref{fig:fiducial_case} with the exception of the Alfv\'enic speed $V_{a}$ for which different values are assumed as indicated. When diffusive reacceleration vanishes, we expect the proton flux to be given by the analytic solution $u_{\rm loss}$. The relative difference between the numerical result $u$ (direct inversion) and $u_{\rm loss}$ is calculated with the low-energy boundary prescription L3. As $V_{a}$ increases from 1~m/s to 2.5~km/s, $u$ becomes more stable and wiggles disappear. The transition occurs for a critical value of $V_{a} \sim 1.75$~km/s.
The number of bins per decade of energy is $N_{\rm dec} = 50$.}
\label{fig:NEW_fiducial_various_DR}
\end{center}
\end{figure}

\subsection{Numerical stability when $V_a\rightarrow0$}
Although condition L3 yields the most precise solution, the short dashed purple curve of Fig.~\ref{fig:model_a_no_DR_COMP_HEBC_0_log} exhibits wiggles for vanishing Alfv\'enic speed. In our example, these instabilities never exceed a level of $10^{-4}$ and have no effect on the numerical solution $u$. In some configurations though, in particular those with a less fine-grained x-grid, they could reach a level where they would impair the capability of the fitting routine used in the B/C analysis. It is then important to understand the reason for these instabilities and to remedy them.

\paragraph{Is it specific to the solver used?}
To commence, we have investigated if these instabilities are related to the method used to derive the numerical solution $u$. In the direct inversion procedure, the matrix $\mathbb{M}$ defined in Eq.~(\ref{eq:def_matrix_M}) is inverted through a fast recursion that takes advantage of its tridiagonal nature. One may wonder if this procedure does not generate numerical errors insofar matrix $\mathbb{M}$ could be far from the unity matrix $\mathbb{I}$. Complementarily, the Crank-Nicholson procedure makes use of the matrix $\mathbb{I} + {\mathbb{M}_{\,} \Delta t}/{2}$ which is arbitrarily close to unity if the time step $\Delta t$ of the recursion is small enough. We find that deriving $u$ through a Crank-Nicholson recursion yields a relative change with respect to the direct inversion result of Fig.~\ref{fig:model_a_no_DR_COMP_HEBC_0_log} (short dashed purple curve) which is always less than $10^{-9}$ above 10~MeV while reaching a maximum of $4 \times 10^{-6}$ at $E_{k} = 2$~MeV. In these calculations, the Alfv\'enic speed is $V_{a} = 1$~m/s. Increasing the speed alleviates further the discrepancy between the two numerical results. The difference is at most $2 \times 10^{-6}$ for a velocity of 1~km/s and decreases below $4 \times 10^{-7}$ at 1.75~km/s. We conclude that the same wiggles appear should $u$ be derived by directly inverting Eq.~(\ref{eq:matrix_MU_U0}) or by letting $u$ evolve \`a la Crank-Nicholson. The origin of instabilities has to be looked elsewhere.

\paragraph{Failure of $2^{\rm nd}$-order schemes to solve $1^{\rm st}$-order equations}
While performing the previous check, we have serendipitously observed that the wiggles disappear as $V_{a}$ increases. In the panels of Fig.~\ref{fig:NEW_fiducial_various_DR}, four different values have been assumed for the Alfv\'enic speed and the relative difference ${|u - u_{\rm loss}|}/{u_{\rm loss}}$ is plotted as a function of $E_{k}$ for each of them. In the upper panel, the numerical solution $u$ exhibits instabilities which are no longer visible in the lower panel. The solution becomes smooth for a critical value of order 1.75~km/s. Above that speed, $u$ slowly departs from $u_{\rm loss}$ as diffusive reacceleration starts to be felt by the proton flux.

\begin{figure}[t]
\begin{center}
\includegraphics[width=0.85\columnwidth]{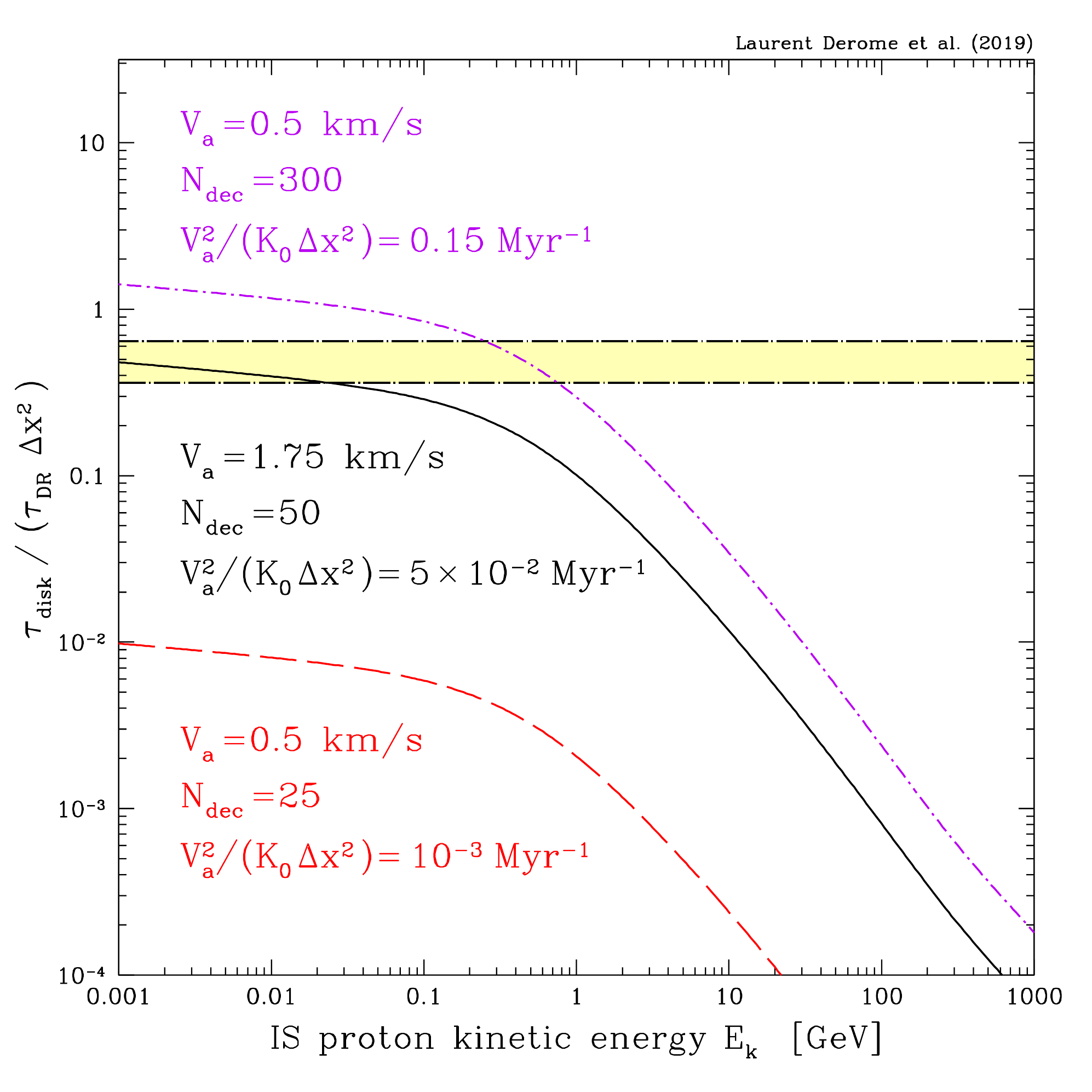}
\caption{The ratio ${\tau_{\rm disc}}/{\tau_{\rm DR\,}\Delta x^{2}}$ is plotted as a function of kinetic energy $E_{k}$ for three different configurations. In all cases, the parameters of Model~A have been assumed with the exception of the Alfv\'enic speed.
The solid black curve corresponds to the critical value $V_{a} = 1.75$~km/s above which the instabilities which affect the numerical solution $u$ in Fig.~\ref{fig:NEW_fiducial_various_DR} are noticeably reduced. For that velocity, the ratio ${\tau_{\rm disc}}/{\tau_{\rm DR\,}\Delta x^{2}}$ reaches a maximum of $0.5$. Enlarging the range of Alfv\'enic speeds for which the onset of stability occurs yields the yellow band.
The long dashed red and short dashed-dotted purple curves both correspond  to $V_{a} = 0.5$~km/s. In the former case, the spacing $N_{\rm dec}$ of the energy grid is $25$, whereas it is $300$ in the latter.
}
\label{fig:mathieu_criterion}
\end{center}
\end{figure}

A tentative explanation for the presence of instabilities is that when $V_{a}$ is vanishingly small, the numerical result $u$ is not derived with the appropriate method. The transport equation~(\ref{eq:master_CR_3}) becomes first order in this regime and only one boundary condition suffices to determine its solution. We have actually obtained $u_{\rm loss}$ by requiring that it should be equal to $u^{0}$ at the highest energy point. Strictly speaking, a low-energy boundary condition is  no longer necessary.
Of course, this is not as simple as that since we proceed numerically through the inversion of matrix $\mathbb{M}$ whose elements $a_{k}$, $b_{k}$ and $c_{k}$ depend on the functions $\alpha$, $\beta$ and $\gamma$ as discussed in Appendix~\ref{app:boundary}. Energy losses and diffusive reacceleration respectively enter in the definition of matrix $\mathbb{M}$ through the combinations ${\alpha_{\,}\beta}/{\Delta x}$ and  ${\alpha_{\,}\gamma}/{\Delta x^{2}}$, with ${\Delta x}$ the spacing of the energy grid. 

\paragraph{Regularisation of the $2^{\rm nd}$-order scheme}
From a numerical perspective, the Alfv\'enic speed vanishes when ${\alpha_{\,}\gamma}/{\Delta x^{2}}$ is negligible with respect to 1, as is clear from Eq.~(\ref{eq:coeff_b}). To explore this regime, we define the numerical strength of diffusive reacceleration through the ratio
\beq
\left| {\displaystyle \frac{\alpha_{\,}\gamma}{\Delta x^{2}}} \right| \equiv {\displaystyle \frac{\tau_{\rm disc}}{\tau_{\rm DR\,}\Delta x^{2}}} \;,
\eeq
which we have plotted as a function of kinetic energy $E_{k}$ in Fig.~\ref{fig:mathieu_criterion}. Model~A has been assumed with the exception of $V_{a}$. The solid black curve corresponds to an Alfv\'enic speed of 1.75~km/s. At that critical value, the instabilities of the numerical result $u$ start to recede, as featured in the lower panel of Fig.~\ref{fig:NEW_fiducial_various_DR} by the smoothness of the short dashed purple curve above 300~MeV. Concomitantly the ratio ${\tau_{\rm disc}}/{\tau_{\rm DR\,}\Delta x^{2}}$ reaches a maximum of $0.5$. Notice that the onset of stability is an ill-defined process which takes place for Alfv\'enic speeds between 1.5 and 2~km/s, hence the yellow strip of Fig.~\ref{fig:mathieu_criterion}.

The prescription which we propose for getting rid of the numerical perturbations of $u$ is to require the ratio ${\tau_{\rm disc}}/{\tau_{\rm DR\,}\Delta x^{2}}$ to overshoot that band, with a maximum exceeding a benchmark value of $0.7$.
%
\begin{figure}[t]
\begin{center}
\includegraphics[width=0.85\columnwidth]{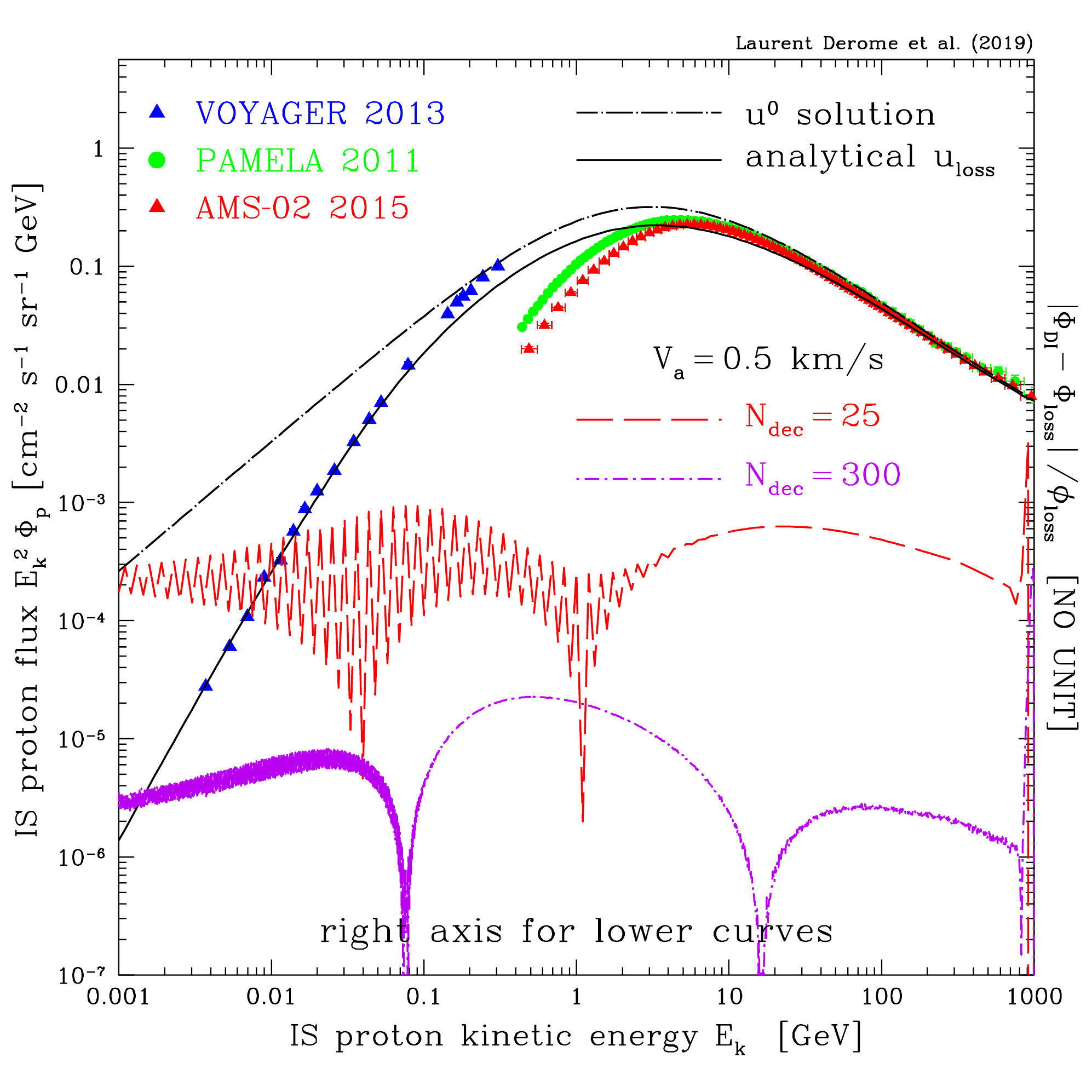}
\caption{Same as in Fig.~\ref{fig:NEW_fiducial_various_DR} but now with an Alfv\'enic speed $V_{a}$ of 0.5~km/s and two different values for $N_{\rm dec}$, the number of bins per decade of energy. The line patterns and colour codes of the two lower curves are the same as in Fig.~\ref{fig:mathieu_criterion}, i.e. the long dashed red and short dashed-dotted purple lines respectively stand for $N_{\rm dec} = 25$ and $300$. Only in the latter case is the criterion for stability fulfilled.}
\label{fig:NEW_various_N_dec_0_5_km_s}
\end{center}
\end{figure}
%
Because the energy grid is made coarser with fewer bins per decade, the interval $\Delta x$ increases and we expect the above criterion to be fulfilled for larger values of $V_{a}$. This is actually what we observe. So far, all the results presented in Appendix~\ref{app:stability} are based on 50 bins per decade. If we degrade the resolution of the x-grid and use 25 bins per decade, we find that wiggles start to recede when the Alfv\'enic speed exceeds a critical value of $1.75 \times 2 = 3.5$~km/s. Conversely, if the x-grid is refined with 100 bins per decade, the transition occurs at ${1.75}/{2} = 0.875$~km/s.

As a final consistency check, we have derived $u$ with an Alfv\'enic speed $V_{a}$ of 0.5~km/s and two different values for the energy spacing $N_{\rm dec}$. Our results are featured in Fig.~\ref{fig:NEW_various_N_dec_0_5_km_s} where the same line patterns and colour codes as in Fig.~\ref{fig:mathieu_criterion} have been used for the two lower curves. The long dashed red line has been derived with 25 bins per decade. Notice how the corresponding curve of Fig.~\ref{fig:mathieu_criterion does not overshoot the yellow band and hence does not fulfil the stability criterion}. We expect $u$ to exhibit numerical instabilities which we actually observe in Fig.~\ref{fig:NEW_various_N_dec_0_5_km_s}. In the case of the short dashed-dotted purple curve, a more refined grid is used with $N_{\rm dec} = 300$ and the numerical solution $u$ is smooth. The stability condition is now satisfied as shown by the corresponding curve of Fig.~\ref{fig:mathieu_criterion}.

\paragraph{Recommendation and conclusion}
To conclude this Appendix, let us discuss how to implement practically the criterion which we have found. To ensure the stability of the numerical result $u$, we require the ratio ${\tau_{\rm disc}}/{\tau_{\rm DR\,}\Delta x^{2}}$ to reach a maximum in excess of $0.7$. In the regime where the Alfv\'enic speed is very small, this implies a very fine structure of the energy grid. In Fig.~\ref{fig:boundary_conditions} for instance, the reference B/C case is derived with 5000 energy bins per decade. Our study confirms that this procedure leads to the correct CR flux.
However, the CPU time significantly increases with $N_{\rm dec}$ and the B/C fit may become impracticable. The procedure which we suggest is to fix $N_{\rm dec}$ at a benchmark value of 50 for instance. For a given set of CR propagation parameters, the critical value $V_{a}^{S}$ of the Alfv\'enic speed at which $u$ becomes stable can be determined from the ratio ${\tau_{\rm disc}}/{\tau_{\rm DR\,}\Delta x^{2}}$ calculated, say, at the lowest energy point. In practice, $\tau_{\rm disc}$ and $\tau_{\rm DR}$ are respectively decreasing and increasing functions of $E_{k}$. This procedure ensures that $V_{a}^{S}$ is always much smaller than the Alfv\'enic speed at which diffusive reacceleration becomes as important as disc transport. With $N_{\rm dec} = 50$, the former is only 3.8\% of the latter. During the B/C fit, $V_{a}$ is forced to be larger than the lower limit $V_{a}^{S}$ which we find, in the case of Model~A, to be $1.75 \pm 0.25$~km/s.
In this configuration, a very rough criterion is to require that the ratio ${V_{a}^{2}}/{K_{0\,}{\Delta}x^{2}}$ exceeds $5 \times 10^{-2}$~Myr$^{-1}$ or, alternatively, that ${V_{a}^{2}}/{K_{0}}$ is larger than $10^{-4}$~Myr$^{-1}$.

\section{Impact of selected cross-section uncertainties on B/C}
\label{app:xs_impact}

\begin{figure}[t]
\begin{center}
\begin{tabular}{l}
{\small \hspace{2.3cm} Impact of inelastic reactions}\\
{\small \hspace{1.8cm} Model~A \hspace{2.1cm} Model~B }\\
\includegraphics[width=0.87\columnwidth]{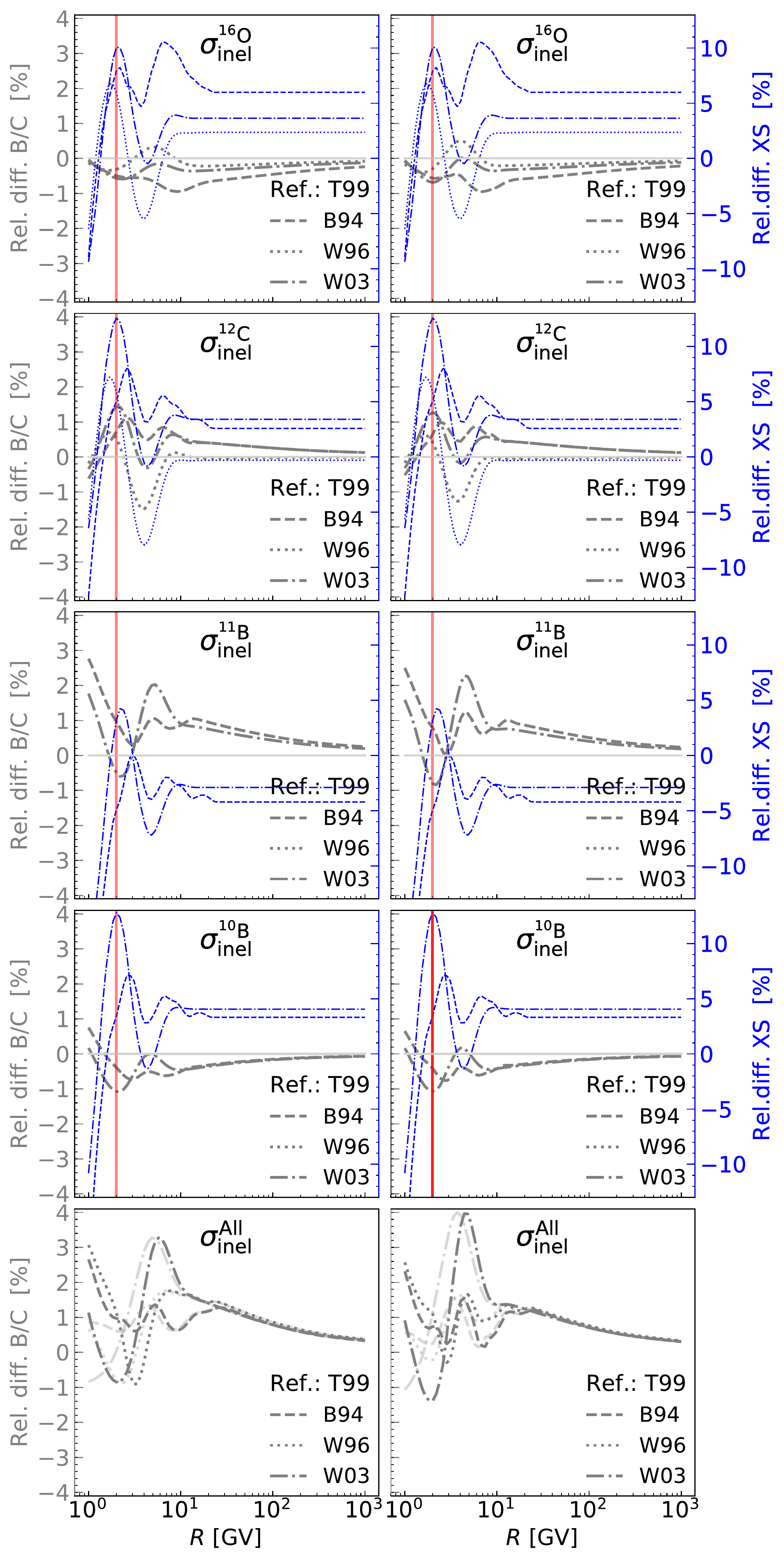}
\end{tabular}
\caption{Impact of inelastic cross-section uncertainties on B/C ratio for specific reactions as a function of rigidity, shown for Model~A and Model~B discussed in Sect.~\ref{sec:prop_model}. In each panel, two quantities are shown: the relative difference between a cross section (of a given reaction) and a reference parametrisation T99 in blue (the associated $y$-axis is on the right-hand side of the plot, also in blue), and the associated impact on the B/C ratio for this cross-section parametrisation w.r.t. to the reference B/C (the associated $y$-axis is on the left-hand side in black). To guide the eye, the vertical red line indicates the rigidity of the first AMS-02 data point. The bottom panels show the overall impact when all reactions (i.e. all nuclei in the network) are replaced. For the latter panel, dark and grey curves correspond to IS and modulated ($\phi_{FF}=0.8$~GV) B/C ratio to emphasise that results are independent of the modulation level.
}
\label{fig:xs_var_inel}
\vspace{-0.5cm}
\end{center}
\end{figure}

\begin{figure}[t]
\begin{center}
\begin{tabular}{l}
{\small \hspace{2.3cm} Impact of production reactions}\\
{\small \hspace{1.8cm} Model~A \hspace{2.cm} Model~B }\\
\includegraphics[width=0.87\columnwidth]{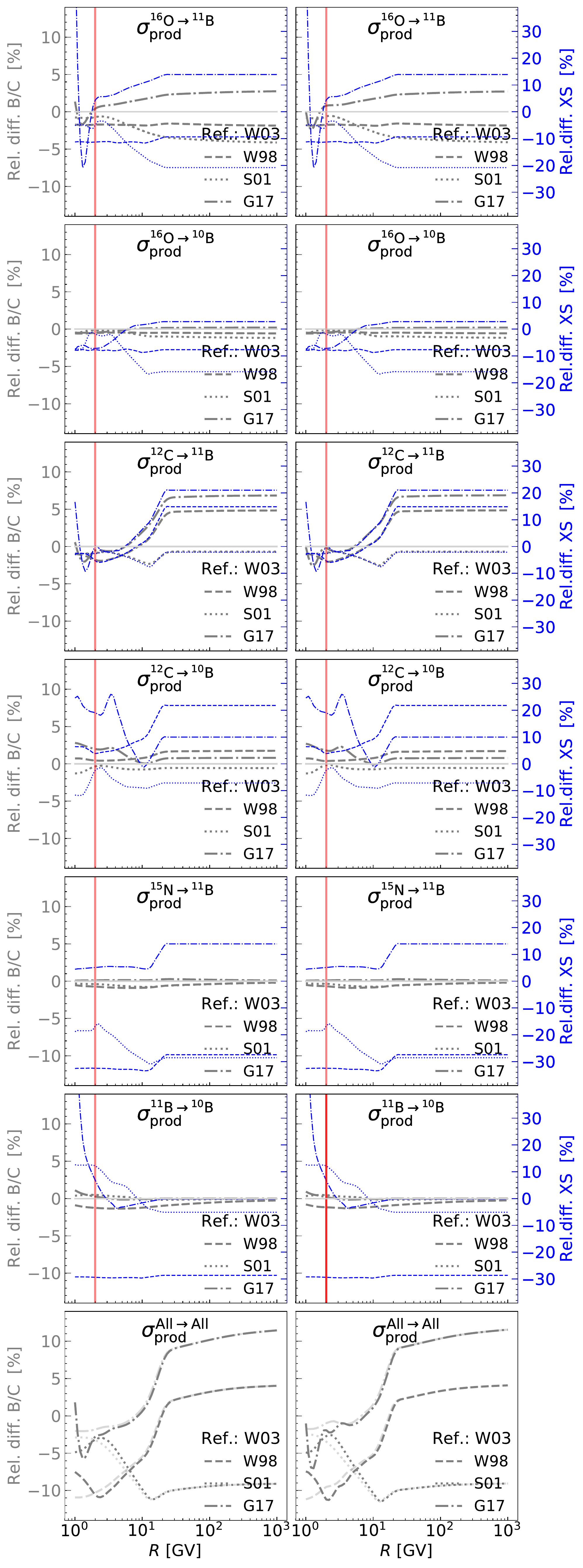}
\end{tabular}
\caption{Same as Fig.~\ref{fig:xs_var_inel} but for production and reference W03.}
\label{fig:xs_var_prod}
\vspace{-0.8cm}
\end{center}
\end{figure}

We detail the impact of the most dominant reactions (see Sect.~\ref{sec:XS}) on the B/C ratio. Figures~\ref{fig:xs_var_inel} and \ref{fig:xs_var_prod} show, in black lines, the impact of changing a single inelastic or production cross section (w.r.t. to the calculation with a reference cross-section file) and, for comparison purpose, blue lines show the relative difference between the reference cross section and the new one on the same plot: the B/C variation is always smaller than that of the cross sections (see Sect.~\ref{sec:XS}).

For inelastic cross sections (Fig.~\ref{fig:xs_var_inel}), the maximum variation between different parametrisations is $\sim 10\%$---related to the different positions and level of the peaks and dips of the cross sections (shown in blue in Fig.~\ref{fig:xs_nuis})---, leading to a $\lesssim3\%$ impact on B/C (black curves). For $^{16}$O, $^{11}$B, and $^{10}$B, the cross section and B/C variations are anti-correlated, while correlated for $^{12}$C. This is explained as follows: for $^{16}$O, a larger destruction cross section means less Oxygen at low energy, and so less Boron production (and less B/C); an increased destruction of $^{10}$B and $^{11}$B also leads to less B/C; the case of $^{12}$C is a trade-off between the fact that an increased destruction means less Carbon (increase in B/C), which means less Boron production (decrease in B/C). If we compare the variation on the B/C relative differences when all cross sections are changed (last panel) to the case in which only one reaction is changed, we see that $^{11}$B and $^{12}$C combined are responsible for almost the whole variability. The impact of the destruction of $^{10}$B is sub-dominant, and at the same level as $^{16}$O.

For production cross sections (Fig.~\ref{fig:xs_var_prod}), the maximum variation between different parametrisations is $\pm 20\%$ for $^{16}{\rm O}\rightarrow ^{11}$B. There is more structure in the G17 cross sections than in all other ones, because they are normalised to data whenever existing. As for inelastic cross sections, the variation in B/C is smaller than the variation in production cross sections, but the relative variations are now correlated for all reactions, because more or less production directly reflects on the B/C ratio. The most impacting reactions also directly reflect the ranking established in \citetads{2018PhRvC..98c4611G}, in which $^{12}{\rm C}\rightarrow ^{11}$B and then $^{16}{\rm O}\rightarrow ^{11}$B have the strongest effect. The sum of these two accounts for most of the variation seen in the bottom panel, in which all reactions in the network were changed at once. We also plot the impact of two reactions involved in `two-step' production of Boron, where $^{15}$N and $^{11}$B are intermediate steps. As discussed in \citetads{2018PhRvC..98c4611G}, the `two-step' reactions can contribute up to $\sim 25\%$ of the total production, and they do not have the same energy-dependence as `one-step' (or direct) production. As can be seen in the two bottom panels, these reactions are suppressed at high energy compared to the other shown, and overall, their impact on B/C is $\lesssim 3\%$. In Sect.~\ref{sec:xs_nuis}, we did some checks adding these dominant channels as nuisance parameters to give extra degrees of freedom w.r.t. the energy dependence, but no gain on the results was found, so they were not considered in the main analysis.

\bibliographystyle{aa} 
\bibliography{methodology}
\end{document}